\documentclass[11pt,a4paper]{article}
\pdfoutput=1
\usepackage{jheppub}
\usepackage{graphicx}
\usepackage{amsmath}
\usepackage{amssymb}
\usepackage{array}
\usepackage{color}

\title{\boldmath  Supersymmetric SO(10) Grand Unification at the LHC and Beyond}

\author[a]{D.~J.~ Miller}
\author[a,b]{and A.~P.~Morais}
\affiliation[a]{SUPA, School of Physics and Astronomy, University of Glasgow,\\ Glasgow, G12 8QQ, UK}
\affiliation[b]{Departamento de F\'isica da Universidade de Aveiro and I3N, \\ Campus de Santiago, 3810-183 Aveiro, Portugal}

\emailAdd{David.J.Miller@Glasgow.ac.uk}
\emailAdd{aapmorais@ua.pt}

\abstract{We study models of supersymmetric grand unification based on the SO(10) gauge group. We investigate scenarios of non-universal gaugino masses including models containing a mixture of two representations of hidden sector chiral superfields. We analyse the effect of excluding $\mu$ from the fine-tuning measure, and confront the results with low energy constraints, including the Higgs boson mass, dark matter relic density and supersymmetry bounds. We also determine high scale Yukawa coupling ratios and confront the results with theoretical predictions. Finally, we present two additional benchmarks that should be explored at the LHC and future colliders.}

\begin{document} 
\maketitle
\flushbottom

\setcounter{footnote}{0}

\section{Introduction} \label{sec:int}

While the discovery of a Higgs boson with mass around $125\,{\rm GeV}$ has been a triumph for the LHC and its detectors {\cite{ATLAS:2012ae, Chatrchyan:2012tx, Aad:2012tfa, Chatrchyan:2012ufa}}, the lack of new physics beyond the Standard Model (SM) is troubling. We are left with a theory that is plagued by extreme fine tuning (the hierarchy problem), no particle suitable for describing dark matter, and no prospect of a unified framework for the gauge interactions. In the days before the LHC, many hoped supersymmetry would cure these ills (and more!) but experimental limits~\cite{Aad:2014wea,ATLAS_newsusy,CMS_newsusy} now push the theory to much higher energies where it itself appears to be fine-tuned. It is important to stress that this ``little hierarchy problem'' afflicting supersymmetry is much much less severe than that of the SM, and supersymmetry provides many additional advantages. This has led some theorists to throw away the desire for naturalness altogether and embrace anthropic arguments for why nature is ``simply un-natural'' (see for example \cite{ArkaniHamed:2012gw}).

However, it is possible that the fine-tuning of supersymmetry is simply a measure of the physics that we do not know. One would expect any ultraviolet completion of the model to provide relations between the high scale parameters, constraining the theory to lie on a surface within the naive larger parameter space. This surface need not be a trivial one so it is important to explore theories with non-trivial relations between the parameters, such as scenarios with non-universal masses. Furthermore, the correct fine-tuning measure in such a theory would only allow the parameters to vary within these constraints (i.e.\ on the surface) so some parameters, or combinations of parameters, may need to be left out of the true fine-tuning measure. While we do not know the ultraviolet completion we cannot predict which parameters or combination thereof this should be. However, we can turn the argument around and use the fine-tuning of the incomplete model to inspire construction of the ultraviolet completion. 

In a recent publication \cite{Miller:2013jra}, we investigated supersymmetric Grand Unified Theories (GUTs) with an SU(5) gauge group. We considered models with both universal and non-universal gaugino masses and confronted our spectra with low energy constraints from the newly discovered Higgs boson {\cite{ATLAS:2012ae, Chatrchyan:2012tx, Aad:2012tfa, Chatrchyan:2012ufa}} and the dark matter (DM) relic density~\cite{Hinshaw:2012fq, Ade:2013xsa}. We also discussed fine-tuning and the effect of {\em not} including the Higgs-higgsino mass parameter $\mu$ into fine-tuning determinations. This latter choice was motivated by the difficulty in avoiding fine-tuning arising from $\mu$ and the soft supersymmetry breaking parameters simultaneously. We showed that if one does not include $\mu$ in the fine-tuning measure, scenarios with low fine-tuning can be found for particular choices of non-universal gaugino mass ratios that have a spectrum heavy enough to evade experimental constraints. These include two orbifold inspired models and a model with a gauge-kinetic function embedded in a $\mathbf{200}$ of SU(5). In this paper we extend our analysis to supersymmetric Grand Unified models based on the SO(10) gauge symmetry, following the same philosophy as in~\cite{Miller:2013jra}. However we will develop our studies further for non-universal gauginos  by considering combinations of hidden sector fields in distinct SO(10) representations. We examine the possible breaking chains following the SU(5) and Pati-Salam routes and investigate constraints on the parameter space. Natural supersymmetric models with a heavy spectrum (though not in a GUT context) have also recently been investigated in~\cite{Cheng:2014taa}.

The rank five SO(10) group is the smallest group capable of accommodating all the SM fermions (and accompanying sfermions) of each generation in a single anomaly-free sixteen dimensional irreducible representation (irrep) $\mathbf{16}$. Compared to SU(5), this leaves room for an extra SM singlet which is identified with a right-handed neutrino (and sneutrino) field. An SO(10) GUT makes it possible to build the mass matrices of the Dirac and heavy Majorana neutrinos  \cite{Babu:1995uu, Albright:2000dk}, permits the determination of the charged fermion masses and mixing angles, provides a seesaw mechanism \cite{Minkowski:1977sc, Mohapatra:1979ia,Schechter:1980gr,Schechter:1981cv} and predicts neutrino oscillations. In the minimal version, the two Higgs doublets of the Minimal Supersymmetric Standard Model (MSSM) are embedded in a single fundamental representation $\mathbf{10}$.  This prescription allows a single Yukawa term in the superpotential of the form,
\begin{eqnarray}
W_y = \mathbf{y}_{10} \mathbf{16} \cdot \mathbf{16} \cdot \mathbf{10}
\label{eq:Wy10}.
\end{eqnarray}
If the third generation charged fermions obtain their masses entirely from this Yukawa interaction it is possible to achieve top-bottom-tau Yukawa Unification (\textit{tb}$\tau$ YU) \cite{Ajaib:2013uda,Ajaib:2013zha,Badziak:2013eda,Shafi:2012yy,Shafi:2012zz,Anandakrishnan:2014nea}. However, this prescription fails to predict the correct fermion masses and mixings, so the Higgs sector must be extended.

Consider the tensor product of two $\mathbf{16}$s, as found in the required mass term,
\begin{eqnarray}
\mathbf{16} \otimes \mathbf{16} = \mathbf{10}_s \oplus \mathbf{120}_a \oplus \mathbf{126}
\label{eq:16x16}.
\end{eqnarray}
Since the mass term itself must be a singlet, the only suitable additional Higgs representations are $\mathbf{\overline{126}}$ and $\mathbf{120}$ Higgs. Although \textit{tb}$\tau$ YU may be spoiled by the addition of extra fields, realistic charged fermions and neutrino masses can be obtained if we assume that the MSSM Higgs doublets $H_{u,d}$ are a superposition of the components that reside in distinct SO(10) representations \cite{Babu:1992ia,Matsuda:2000zp,Matsuda:2001bg,Fukuyama:2002ch}. To preserve \textit{tb}$\tau$ YU alongside with correct fermion masses, the contributions to $H_{u,d}$ from the extra Higgs fields must be small (see \cite{Joshipura:2012sr} and references therein). However, if those contributions are sizable we can still have \textit{tb}$\tau$ {\em Quasi}-Yukawa unification (QYU)~\cite{Dar:2011sj}.

In general, the presence of SO(10) GUT representations introduces colour triplet fields and higher dimensional operators contributing to proton decay (see e.g.~\cite{Fukuyama:2004xs}). Here we will assume that this is solved by some unknown mechanism at the GUT scale, for example embedding the model in higher dimensions \cite{higher_dim}.

The remainder of the paper is organised as follows. In Section~\ref{sec:SO10GUT} we review the possible routes for breaking the symmetry from SO(10) down the SM gauge group. We then describe the model to be examined in Section~\ref{sec:so10pars}, including our choices for input parameters at the GUT scale. In Section~\ref{sec:scansSO10} we briefly describe the constraints on the particle spectrum, both from experiment and theory. Our first analysis on universal gaugino masses is described in Section~\ref{sec:ugmSO10}, and we extend this to non-universal gaugino masses with no constraint on the gaugino masses in Section~\ref{sec:nugm}. In Section~\ref{sec:sfgmrSO10} we discuss models that that predict particular (non-universal) gaugino mass ratios, including those resulting from combinations of hidden sector fields in diferent SO(10) representations. This also includes an in-depth analysis of the Pati-Salam breaking route and two benchmark scenarios. Finally in Section~\ref{sec:conc} we will summarise our results and draw some conclusions. 


\section{Breaking Chains}
\label{sec:SO10GUT}

SO(10) may be broken to the SM gauge group, $G_{SM}$, through either $SU(5) \times U(1)$ (normal \cite{Georgi:1974sy} or flipped \cite{DeRujula:1980qc, Barr:1988yj, Barr:1981qv} embedding), or through Pati-Salam (PS) $SU(4) \times SU(2)_L \times SU(2)_R$ \cite{Pati:1974yy}. Motivated by the convergence of the gauge couplings in the MSSM, we assume here that this occurs entirely at the GUT scale or very close to it. First we consider the breaking via SU(5),
\begin{equation}
SO(10)\to SU(5) \times U(1)_X \to 
SU(3) \times SU(2) \times U(1)_Z \times U(1)_X \to G_{SM}. \label{eq:GG}
\end{equation}
The branching rules at the first breaking for the $\mathbf{16}$- and $\mathbf{10}$-plets are,
\begin{eqnarray}
\mathbf{16} &\rightarrow& \mathbf{1}_{-5} \oplus \mathbf{\overline{5}}_3 \oplus \mathbf{10}_{-1}, \\
\mathbf{10} &\rightarrow& \mathbf{5}_2 \oplus \mathbf{\overline{5}}_{-2},
\label{eq:BrMG}
\end{eqnarray}
where the subscripts denote $X$, the $U(1)_X$ charge, and at the second breaking,
\begin{eqnarray}
\mathbf{1} &\rightarrow& \left( \mathbf{1},\mathbf{1} \right)_0, \\
\mathbf{5} &\rightarrow& \left( \mathbf{1},\mathbf{2} \right)_3 \oplus \left( \mathbf{3},\mathbf{1} \right)_{-2} , \\
\mathbf{\overline{5}} &\rightarrow& \left( \mathbf{1},\mathbf{2} \right)_{-3} \oplus \left( \mathbf{\overline{3}},\mathbf{1} \right)_2 , \\
\mathbf{10} &\rightarrow& \left( \mathbf{1},\mathbf{1} \right)_6 \oplus \left( \mathbf{\overline{3}},\mathbf{1} \right)_{-4} \oplus  \left( \mathbf{3},\mathbf{2} \right)_1,
\label{eq:BrM1}
\end{eqnarray}
where the subscripts denote $Z$, the $U(1)_Z$ charge. The normalisations of $X$ and $Z$ are those adopted in Slansky \cite{Slansky:1981yr} (see also \cite{Fukuyama:2004ps}). To relate them to the original generators of the unbroken groups one must normalise them according to,
\begin{equation}
\hat X = \frac{1}{\sqrt{40}} X, \qquad \hat Z = \frac{1}{6}\sqrt{\frac{3}{5}} Z.
\end{equation}

Since we now have two Abelian $U(1)$ symmetries, the weak hypercharge generator $Y$ is a linear combination of the $U(1)_X$ and $U(1)_Z$ generators, and we also have an orthogonal generator $Y^{\perp}$. For SU(5) there are two possible ways to arrange the SM fields in the SU(5) multiplets while maintaining the SM quantum numbers, Georgi-Gashow (GG) \cite{Georgi:1974sy} or flipped (FL) \cite{DeRujula:1980qc, Barr:1988yj, Barr:1981qv} embedding. 

For the GG embedding, the fields are identified in the same way as we did for SU(5) in Ref.~\cite{Miller:2013jra} with the addition of a right-handed neutrino field as $( \mathbf{1},\mathbf{1} )_0$. The SM hypercharge generator and its orthogonal partner is then given by,
\begin{eqnarray}
Y &=& Z/3, \label{eq:YGG}\\
Y^{\perp} &=& -X,
\label{eq:YGGperp}
\end{eqnarray}
where $Y$ is normalised so the electromagnetic charge is $Q_{\rm em} = I_3 +Y/2$. The normalisation of $Y^{\perp}$ is arbitrary. 

For FL SU(5) \cite{DeRujula:1980qc, Barr:1988yj}, the fields correspond to swapping $\hat{u}^{\dagger}_R$ with $\hat{d}^{\dagger}_R$ and $\hat{e}^{\dagger}_R$ with $\hat{N}^{\dagger}_R$ in the GG identification. The hypercharge $Y$ and its orthogonal partner are,
\begin{eqnarray}
Y &=& -\frac{1}{15} \left( 6X + Z \right) = - \frac{2}{5} \sqrt{\frac{5}{3}} \left( \sqrt{24} \hat X + \hat Z \right), 
\label{eq:YFL}\\
Y^{\perp} &=& -\frac{1}{15} \left( -X + 4 Z \right) = \frac{\sqrt{40}}{15} \left( \hat X - \sqrt{24} \hat Z \right).
\label{eq:YFLperp}
\end{eqnarray}

For Pati-Salam (PS) breaking \cite{Pati:1974yy},
\begin{equation}
SO(10)\to  SU(4) \times SU(2)_L \times SU(2)_R \to SU(3) \times SU(2)_L \times SU(2)_R \times U(1)_W \to G_{SM},
\label{eq:PS}
\end{equation}
the decomposition of the $\mathbf{16}$ and $\mathbf{10}$ plets after the breaking of SO(10) are,
\begin{eqnarray}
\mathbf{16} &\rightarrow&  \left( \mathbf{4},\mathbf{2}, \mathbf{1} \right) \oplus \left( \mathbf{\overline{4}},\mathbf{1}, \mathbf{2} \right) ,\label{eq:BrPS16}\\[2mm]
\mathbf{10} &\rightarrow&  \left( \mathbf{1},\mathbf{2}, \mathbf{2} \right) \oplus \left( \mathbf{6},\mathbf{1}, \mathbf{1} \right) , \label{eq:BrPS10}
\end{eqnarray}
and for the second breaking
\begin{eqnarray}
\left( \mathbf{4},\mathbf{2}, \mathbf{1} \right) &\rightarrow& \left( \mathbf{1},\mathbf{2}, \mathbf{1} \right)_{3} \oplus \left( \mathbf{3},\mathbf{2}, \mathbf{1} \right)_{-1} , \label{eq:4BR}\\[2mm]
\left( \mathbf{\overline{4}},\mathbf{1}, \mathbf{2} \right) &\rightarrow& \left( \mathbf{1},\mathbf{1}, \mathbf{2} \right)_{-3} \oplus \left( \mathbf{3},\mathbf{1}, \mathbf{2} \right)_{1} , \label{eq:bar4BR}\\[2mm]
 \left( \mathbf{1},\mathbf{2}, \mathbf{2} \right) &\rightarrow&  \left( \mathbf{1},\mathbf{2}, \mathbf{2} \right)_0 ,
\label{eq:BrM2}
\end{eqnarray}
where the right-handed fields are grouped in $SU(2)_R$ doublets with right-isospin \mbox{$I_R = 1/2$} and eigenvalues $I_{R\,3} = \pm 1/2$. Once again we have two possible assignments of SM particles to these multiplets in order to reproduce the correct quantum numbers. 
The hypercharge generator is distinct from the $U(1)_{W}$ generator and orthogonal to both the $SU(3)$ and $SU(2)_L$ generators. Therefore $Y$ is a linear combination of the $U(1)_W$ and $SU(2)_R$ generators. Freedom remains for $SU(2)_R$ rotations in a plane perpendicular to the $U(1)_W$ axes, but fixing $Y$ also constrains the $SU(2)_R$ axes. 

The first possibility is to identify,
\begin{equation}
 \left( \mathbf{4},\mathbf{2}, \mathbf{1} \right) =
\begin{pmatrix}
\hat u^x & \hat \nu \\
\hat d^x & \hat e
\end{pmatrix}_L,
\qquad
 \left( \mathbf{\overline{4}},\mathbf{1}, \mathbf{2} \right) =
\begin{pmatrix}
\hat u^{\dagger}_x & \hat N^{\dagger} \\
\hat d^{\dagger}_x & \hat e^{\dagger}
\end{pmatrix}_R,
\label{eq:RH}
\end{equation}
for matter fields, where $x$ is a colour index, and,
\begin{eqnarray}
 \left( \mathbf{1},\mathbf{2}, \mathbf{2} \right)  &=&
\begin{pmatrix}
\hat h^{+}_u & \hat h^{0}_d \\
\hat h^{0}_u & \hat h^{-}_d
\end{pmatrix},
\label{eq:Higgs}
\end{eqnarray}
for the Higgs fields. Leptons are interpreted as part of a four colour quark, unified in $\mathbf{4}$-plets of $SU(4)$. This leads to the hypercharge assignments,
\begin{eqnarray}
Y &=& -2 I_{R\,3} - W/3 , \label{eq:YPS}\\
Y^{\perp} &=& 4 I_{R\,3} - W . \label{eq:YPSperp}
\end{eqnarray}
Alternatively we can ``flip'' the assignments again with a $\pi$ rotation in $SU(2)_R$. Then the hypercharge assignments are instead
\begin{eqnarray}
Y &=& 2 I_{R\,3} - W/3 ,
\label{eq:YFLPS} \\
Y^{\perp} &=& 4 I_{R\,3} + W .
\label{eq:YFLPSperp}
\end{eqnarray}
In the above the normalisation of $W$ is that adopted by Slansky; the normalisation of $Y$ is again set by the SM charges, but that of $Y^\perp$ is arbitrary. Note that $W$ is related to $B-L = -W/3$.


\section{The SO(10) GUT Model}\label{sec:so10pars}

\hspace{5mm} We consider the minimal realistic SO(10) GUT model \cite{Aulakh:2003kg,Bajc:2004xe} with the superpotential given by,
\begin{eqnarray}
W_{SO(10)} &=& \left(y_{\mathbf{10}}\right)_{ij}\mathbf{16}_{i a} \left( \mathbf{C} \Gamma_{\alpha} \right)^{a b} \mathbf{16}_{j b} \mathbf{10}_{\alpha} \nonumber \\ &+&  \frac{1}{5 !} \left(y_{\mathbf{126}}\right)_{ij}\mathbf{16}_{i a} \left( \mathbf{C} \Gamma_{[\alpha} \Gamma_{\beta} \Gamma_{\rho} \Gamma_{\sigma} \Gamma_{\lambda]} \right)^{a b} \mathbf{16}_{j b} \mathbf{\overline{126}}_{\alpha \beta \rho \sigma \lambda}\nonumber\\
 &+& \mu_1 \mathbf{10}_{\alpha} \mathbf{10}_{\alpha} + \mu_2 \mathbf{126}_{\alpha \beta \rho \sigma \lambda} \mathbf{\overline{126}}_{\alpha \beta \rho \sigma \lambda}  + W_{\mathbf{X_{\mathbf{R}}}}.
\label{eq:W10}
\end{eqnarray}
The $\Gamma_{\mu}$ matrices satisfy a rank 10 Clifford algebra, and $\mathbf{C}$ is an SO(10) charge conjugation matrix. $W_{\mathbf{X_{\mathbf{R}}}}$ includes the chiral superfields $X_{\mathbf{R}}$ belonging to an SO(10) symmetric representation $\mathbf{R}$, contained in the product of two adjoint representations, $\mathbf{45} \times \mathbf{45}$, and whose scalar components are responsible for the breaking of the GUT symmetry at the high scale. $\left\{i, j\right\} = 1,2,3$ are generation indices and $\left\{a, b\right\} = 1, \ldots ,16$ are spinor indices. All the MSSM quark and lepton superfields as well as the right-handed neutrino superfield,  $\hat{Q}_L,~\hat{u}^{\dagger}_R,~\hat{e}^{\dagger}_R,~\hat{L}_L,~\hat{d}^{\dagger}_R,~ \rm{and}~\hat{N}^{\dagger}_R$, are embedded in the $\mathbf{16}$ representation. The Higgs superfields $\hat{H}_u$ and $\hat{H}_d$ belong to a superposition of the  $\mathbf{10}$ and of the $\overline{\mathbf{126}}$ representations in order to generate the correct fermion masses and mixings. In addition, we allow terms involving a $\mathbf{126}$-plet that do not couple directly to the ordinary matter in the $\mathbf{16}$-plet to also be present in $W_{\mathbf{X_{\mathbf{R}}}}$. D-terms arising from the expectation values of the $\overline{\mathbf{126}}$ are canceled by those of the $\mathbf{126}$, and the D-term mass splittings at the GUT scale become identical to a model with a single Higgs $\mathbf{10}$-plet. We assume that the MSSM $\mu$-term is a combination of the bilinear coefficients $\mu_1$ and $\mu_2$. Finally $y_{\mathbf{10}}$ and $y_{\mathbf{126}}$ are Yukawa coupling matrices and typically $y_{\mathbf{126}}$ has entries much smaller than those in $y_{\mathbf{10}}$.

\subsection{Soft Scalar Masses}

The effective Lagrangian for the Higgs and sfermion masses is given by,
\begin{eqnarray}
-\mathcal{L}_{mass} &=&
 m^2_{H_d}\left|H_d\right|^2 +  m^2_{H_u}\left|H_u\right|^2 + \tilde{Q}^{~\alpha x}_{Li} \left( m^2_{\tilde{Q}_L} \right)^i_{~j} \tilde{Q}^{\ast~j}_{L \alpha x} + \tilde{L}^{~\alpha}_{L i} \left( m^2_{\tilde{L}_L} \right)^i_{~j}\tilde{L}^{\ast~j}_{L \alpha}\nonumber\\
 &+& \tilde{u}^{\ast~x}_{R i} \left( m^2_{\tilde{u}_R} \right)^i_{~j}\tilde{u}^{~j}_{R x} + \tilde{d}^{\ast~x}_{R i} \left( m^2_{\tilde{d}_R} \right)^i_{~j}\tilde{d}^{~j}_{R x} + \tilde{e}^{\ast}_{R i} \left( m^2_{\tilde{e}_R} \right)^i_{~j}\tilde{e}^{~j}_{R }.
\label{eq:Lmass}
\end{eqnarray}
When the symmetry is broken, the sfermions from the single $\mathbf{16}$ take a common soft mass $m_{\mathbf{16}}$, whereas the $\mathbf{10}\oplus \mathbf{\overline{126}}$ Higgs fields take a mass $m_{\mathbf{10}+\mathbf{126}}$, arising from the individual masses of the $\mathbf{10}$ and $\mathbf{\overline{126}}$. Additionally, D-term splittings should be included in the scalar masses due to rank reduction. The boundary conditions for the GG embedding follow from (\ref{eq:YGGperp}) yielding,
\begin{eqnarray}
m^2_{Q_{ ij}}\left( 0 \right) \:\: =\:\: m^2_{u_{ij}}\left( 0 \right)  \:\: =\:\:   m^2_{e_{ij}}\left( 0 \right) &=& 
\begin{pmatrix}
K_{\mathbf{16}} & 0 & 0\\
0 & K_{\mathbf{16}} & 0\\
0 & 0 & 1
\end{pmatrix}\left( m^2_{\mathbf{16}} + g^2_{10} D\right),
\label{eq:10scalarBCGG}\\[2mm]
m^2_{L_{ ij}}\left( 0 \right)  \:\: =\:\:  m^2_{d_{ij}}\left( 0 \right) &=& 
\begin{pmatrix}
K_{\mathbf{16}} & 0 & 0\\
0 & K_{\mathbf{16}} & 0\\
0 & 0 & 1
\end{pmatrix}\left( m^2_{\mathbf{16}} -  3g^2_{10} D\right),
\label{eq:5scalarBCGG}\\[2mm]
m^2_{N_{ij}}\left( 0 \right) &=& 
\begin{pmatrix}
K_{\mathbf{16}} & 0 & 0\\
0 & K_{\mathbf{16}} & 0\\
0 & 0 & 1
\end{pmatrix}\left( m^2_{\mathbf{16}} + 5 g^2_{10} D\right),
\label{eq:1scalarBCGG}\\[2mm]
m^2_{H_u}\left( 0 \right) &=& m^2_{\mathbf{10} + \mathbf{126}} - 2 g^2_{10} D, \label{eq:HuBCGG}\\
m^2_{H_d}\left( 0 \right) &=&  m^2_{\mathbf{10} + \mathbf{126}} + 2 g^2_{10} D, \label{eq:HdBCGG}
\end{eqnarray}
where $g^2_{10} D$ is the D-term contribution for the mass splittings and $g_{10}$ is the unified gauge coupling of SO(10). We allow an hierarchy between the third and first two generations, but keep the first two degenerate in order to avoid dangerous Flavour-Changing Neutral-Currents (FCNC)~\cite{Baer:2004xx}. Consequently, this model has one extra parameter, $K_{\mathbf{16}} > 0$, which accounts for the third generation's non-universality at the GUT scale. To be consistent with a type-I seesaw mechanism, we add a large Majorana mass $\mathcal{M}_{ij}$ to the right-handed sneutrino field boundary condition. This term may emerge when a neutral component of the $\mathbf{\overline{126}}$ Higgs acquires an expectation value at the high scale. Since $N_R$ is a SM singlet, it retains a large mass dominated by the Majorana contribution so does not become a dark matter candidate as it could if its mass was purely Dirac.

For the FL embedding, when we apply the charge assignments (\ref{eq:YFL}) and (\ref{eq:YFLperp}), the boundary conditions at the GUT scale have the same form as the GG ones, but with opposite sign D-term splittings. Since we consider both positive and negative D-term contributions, there is no practical difference between the GG and FL embeddings in the scalar sector, and the boundary conditions take the same form as in (\ref{eq:10scalarBCGG}-\ref{eq:HdBCGG}).

For the PS breaking route, the charge assignments in eqs. (\ref{eq:YPSperp}) and (\ref{eq:YFLPSperp}), yield exactly the same D-term splittings as for the GG and FL embeddings respectively. Since we assume that the breaking to $G_{SM}$ is entirely accomplished at the GUT scale or very close to it, once again we have boundary conditions of (\ref{eq:10scalarBCGG}-\ref{eq:HdBCGG}).

\subsection{Soft Trilinear Couplings}

The explicit soft supersymmetry-breaking terms that contain scalar trilinear couplings are given by,
\begin{eqnarray}
-\mathcal{L}_{trilinear} &=&
 \varepsilon_{\alpha \beta} \left[ a_{u ij}H^{\alpha}_u \tilde{u}_{R i x}\tilde{Q}^{\beta x}_{L j} - a_{d ij}H^{\alpha}_d \tilde{d}_{R i x}\tilde{Q}^{\beta x}_{L j} - a_{e ij}H^{\alpha}_d \tilde{e}_{R i}\tilde{L}^{\beta}_{L j} +  b H^{\alpha}_d H^{\beta}_u  \right] + {\rm h.c.} \nonumber \\
\label{eq:Ltri}
\end{eqnarray}
Since $y_{\mathbf{126}} \ll y_{\mathbf{10}}$, we consider contributions only from $ y_{\mathbf{10}}$ and we impose the simplified boundary condition,
\begin{eqnarray}
a_{t}\left( 0 \right) = a_b\left( 0 \right) = a_{\tau}\left( 0 \right) = a_{\mathbf{10}}. \label{eq:AdeBC}
\end{eqnarray}
where $a_{\mathbf{10}}$ is a single unified trilinear coupling at the GUT scale.

\subsection{Gaugino Masses} \label{subsec:gauginoSO10}

The hidden sector auxiliary fields $\hat X_i$ are now in a representation (or combination of representations) belonging to the symmetric product $\left(\mathbf{45} \times \mathbf{45}\right)_{symm} = \mathbf{1} + \mathbf{54} + \mathbf{210} + \mathbf{770}$. The coefficient of the gaugino mass term~\cite{Bhattacharya:2009wv},
\begin{eqnarray}
\frac{1}{2} \frac{\langle F^j_X \rangle}{\langle Re f_{\alpha \beta} \rangle} \left\langle \frac{\partial f^{*}_{\alpha \beta}}{\partial \varphi^{j *}} \right\rangle\tilde \lambda^{\alpha} \tilde \lambda^{\beta},
\label{eq:LM}
\end{eqnarray}
will only generate universal masses when the F-term $F_X$ is a trivial representation. In the above, $f_{\alpha \beta}$ is the gauge kinetic function, $\tilde \lambda^\alpha$ is a gaugino fermion and $\varphi_i$ is the scalar component of $\hat X_i$. It is in this sector where the GUT scale constraints arising from the GG, FL and the two PS embeddings will differ. In particular, the transformation properties of the $F$-terms under the full SO(10) symmetry as well as under its maximal proper subgroups, fixes distinct coefficients in (\ref{eq:LM}). A detailed description with all possible coefficients can be found in \cite{Martin:2009ad} (see also \cite{Bhattacharya:2009wv,Chakrabortty:2008zk}). The effective soft gaugino mass terms are then
\begin{eqnarray}
-\mathcal{L}_{gaugino} = \frac{1}{2} \left[ M_1 \tilde{B}\cdot \tilde{B} + M_2 \, \tilde{W}^a \cdot \tilde{W}^a+  M_3 \, \tilde{g}^a \cdot \tilde{g}^a  + {\rm h.c.}\right].
\label{eq:Lg}
\end{eqnarray}
As in Ref.~\cite{Miller:2013jra}, we will examine the following sets of boundary conditions at the GUT scale:
\begin{itemize}
\item[I.] universal gaugino masses: $M_1 = M_2 = M_3 \equiv M_{1/2}$,
\item[II.] non-universal gaugino masses: $M_1/\rho_1 = M_2/\rho_2 = M_3 \equiv M_{1/2}$.
\end{itemize}

\subsection{Summary of the Parameter Space}

In addition to the usual SM parameters, our SO(10) model is described by eight high scale parameters, $m_{16}$, $K_{16}$, $m_{10+126}$, $g^2_{10}D$,  $M_{1/2}$,  $\rho_1$, $\rho_2$, $ a_{10}$, as well as $\tan \beta$ and the sign of $\mu$. Despite the common scalar masses, the SO(10) model differs from the constrained MSSM or Non-Universal Higgs Mass (NUHM) models due to the D-term splittings.

\section{Constraints on the Particle Spectrum}
\label{sec:scansSO10}

We allow the GUT scale third generation scalar mass and that of the Higgs multiplets, $m_{16}^{(3)}$ and $m_{10+126}$ respectively, to lie between zero and $4\,{\rm TeV}$. We allow the D-term splittings $\sqrt{g^2_{10}D}$ to vary in the range $\pm 4~\rm{TeV}$. To ensure vacuum stability, we only accept points where the sum of the input scalar masses with the respective D-term splittings is positive. The first and second generation input scalar masses are obtained from multiplying $m_{16}$ by $K_{16}$ which we allow to be between zero and $15$. We require $M_3$ to be less than $4\,{\rm TeV}$; if examining scenarios with  universal gaugino masses, this also sets $M_1$ and $M_2$, but if examining non-universal gauginos, we also vary $\rho_{1,2}$ between $\pm 15$.
Finally the single trilinear coupling, $a_{10}$ is allowed to vary between $\pm 10\,{\rm TeV}$, and our only (non-SM) low energy input $\tan \beta$ is constrained to lie in the range $1\--60$. These parameter ranges are summarised in Table \ref{table:input_range}.

\begin{table}[ht!]
\centering
\begin{tabular}{|c|rcl|}
\hline
Parameter & \multicolumn{3}{c|}{range [TeV]} \\
\hline
$m_{16}^{(3)}$ &  $0$ &--& $4$ \\
$m_{10+126}$ &  $0$ &--& $4$ \\
$\sqrt{g^2_{10}D}$ & $-4$ &--& $4$ \\
$M_{1/2}$ & $0$ &--& $4$\\
$a_{10}$ & $-10$ &--& $10$\\ 
\hline
\end{tabular}
\begin{tabular}{|c|rcl|}
\hline
Parameter & \multicolumn{3}{c|}{range} \\\hline
$K_{16}$ & $0$ &--& $15$ \\
$\rho_{1,2}$ & $-15$ &--& $15$ \\
$\tan \beta$ & $1$ &--& $60$ \\
&&&\\
&&&\\
\hline
\end{tabular}
\caption{\it Input parameter ranges for the scan.}
\label{table:input_range}
\end{table}

We have updated our experimental constraints for direct supersymmetry searches \cite{Aad:2014wea,ATLAS_newsusy,CMS_newsusy}. In particular, we require the first and second generation squarks to have masses greater than $1.7\,{\rm TeV}$ and the gluino to be heavier than $1.2\,{\rm TeV}$. We do not explicitly constrain the third generation squarks since we find scenarios that violate the appropriate searches are already ruled out by other experimental constraints. 
We impose Higgs boson mass bounds~\cite{ATLAS:2012ae,Chatrchyan:2012tx} combined in quadrature with a $2\,{\rm GeV}$ theoretical uncertainty (estimated by the mass difference for the light CP-even Higgs obtained with SOFTSUSY and SUSPECT~\cite{Djouadi:2002ze}, as reported in~\cite{Arbey:2012dq}) to give a (1$\sigma$) uncertainty on our output Higgs boson mass of $125.7 \pm 2.1\,{\rm GeV}$.
We impose direct dark matter production bounds from LUX~\cite{Akerib:2013tjd}. We also discard scenarios with too high dark matter relic density, using the cosmological parameters of the nine year WMAP observations published in \cite{Hinshaw:2012fq}.
We combine in quadrature the experimental standard deviation ($\Omega_c h^2=0.1157 \pm 0.0023$) with a $10\%$ theoretical uncertainty estimated from the LSP mass difference in \mbox{micrOMEGAS 2.4.5}~\cite{Belanger:2004yn} and SOFTSUSY 3.3.0 \cite{Allanach:2001kg}. The resulting bounds for our micrOMEGAS relic density are $\Omega_c h^2 = 0.1157 \pm 0.0118$.
 
We also impose bounds on new physics from $b \rightarrow s \gamma$ \cite{Amhis:2012bh},  $B_s \rightarrow \mu^+ \mu^-$ \cite{:2012ct}, the purely leptonic $B \rightarrow \tau \nu_{\tau}$ decay~\cite{Adachi:2012mm} and the muon anomalous magnetic moment $a_{\mu}$~\cite{Bennett:2006fi}. We confront these with with our micrOMEGAS, again assuming a $10\%$ theoretical error. 
This provides constraints for the branching ratios \mbox{${\rm Br} \left(b \rightarrow s \gamma \right) = \left( 355 \pm 43.8\right) \times 10^{-6}$} and ${\rm Br} \left(B_S \rightarrow \mu^{+} \mu^{-} \right) = \left( 3.2^{+1.5}_{-1.2}\times 10^{-9} \right)$. For ${\rm Br} \left(B \rightarrow \tau \nu_{\tau} \right)$, mircOMEGAS outputs the ratio of the measured value with the SM prediction~\cite{Biancofiore:2013ki}, which we denote as $\mathcal{R}_{\tau \nu_{\tau}}$ and we find the constraint $\mathcal{R}_{\tau \nu_{\tau}}  = 1.42 \pm 0.70$. 
The deviation of the experimental value for the anomalous magnetic moment of the muon, $a_{\mu} = \left( g-2 \right)_{\mu}/2$~\cite{Bennett:2006fi}, from the SM prediction~\cite{Davier:2009zi} may be attributed to supersymmetric contributions~\cite{vonWeitershausen:2010zr}. However it is also possible that some or all of such deviation results from some other additional causes. Therefore, we only require that the supersymmetric contribution does not {\it exceed} \mbox{$\Delta a_{\mu}(exp - SM) = (25.5 \pm 8.0) \times 10^{-10}$}.

For each of the above measurements we compare our prediction with experiment and determine the probability of our scenarios being consistent with observations, assuming Gaussian errors. We combine the individual probabilities into a global one through the relation $P_{\rm tot} = P_{m_h} \cdot P_{\Omega_c h} \cdot P_{b \rightarrow s \gamma} \cdot P_{\mathcal{R}_{\tau \nu_{\tau}}} \cdot P_{B_s \rightarrow \mu \mu} \cdot P_{a_{\mu}}$, requiring that this is always larger than $10^{-3}$. This approach excludes scenarios with multiple predictions in the vicinity of their $\pm 2 \sigma$ bound, that would otherwise be accepted by imposing the constraints on a one-by-one basis. In particular scenarios with values below $\Omega_c h^2 = 0.1157$ are always accepted and we only use $P_{\Omega_c h}$ in the usual way to determine whether the mechanism provides the preferred, to little or an excess of Dark Matter relic density. Similarly, if the additional contribution to $\Delta a_{\mu}$ is less than the theoretical-experimental deviation we set $P_{a_\mu}=1$, otherwise we use the above uncertainty to quantify $P_{a_\mu}$.

We follow the same approach for fine tuning as we did for SU(5) in \cite{Miller:2013jra}.

\section{Universal Gaugino Masses}
\label{sec:ugmSO10}

We begin our analysis by examining scenarios with universal gaugino masses, generating points with inputs in the ranges given in Section~\ref{sec:scansSO10}. Using the full two-loop RGEs, as implemented in SOFTSUSY, we evolve them to the electroweak scale, but do not force {\it exact} gauge coupling unification.

We apply the LHC direct and LUX (2$\sigma$) bounds and discard scenarios with Higgs boson masses outside the range $122.6-127\,{\rm GeV}$ to avoid unnecessary computation. We force vacuum stability including a simplified approach to the CCB-2,3 constraints \cite{Casas:1995pd} by implementing the cuts
\begin{equation}
\left| \frac{a_{10}}{ \sqrt{m^2_{16} + g^2_{10}D}} \right| \lesssim 3, \qquad
\left| \frac{a_{10}}{\sqrt{ m^2_{16} - 3 g^2_{10}D}} \right| \lesssim 3 \label{eq:UFB}. 
\end{equation}
Note that having a single trilinear coupling, rather than the two distinct couplings we see in SU(5), forces these conditions to be considerably more restrictive than those of SU(5). 
At this stage we also discard scenarios with a charged dark matter. Out of 1,000,000 initial scenarios, this leaves approximately 180,000. This is a substantial increase in the number of surviving points in comparison to SU(5); $18\%$ of the initial tries  in comparison to $3\%$ (even though experimental constraints are now more restrictive). 

Using the electroweak scale outputs of SOFTSUSY as inputs for micrOMEGAS, we generate predictions for the remaining experimental observables, such as the dark matter relic density, and derive a probability ${ P}_{\rm tot}$ for each scenario (again using the procedure described in \cite{Miller:2013jra}). We require ${ P}_{\rm tot}>10^{-3}$, reducing the number of viable scenarios to 2151, corresponding to $0.2\%$ of the initial attempts, of which 458 ($0.05\%$ of the initial attempts) have the preferred relic density. In comparison, we previously had approximately $0.02\%$ of scenarios surviving for SU(5) with only $0.002\%$ with the preferred relic density, though these SU(5) results were for a reduced range of $M_{1/2}\le 2\,$TeV.

For some parameter choices we find that SOFTSUSY and microOMEGAS provide rather different values for the masses of the lightest supersymmetric particle (LSP) or next-to-lightest supersymmetric lightest particle (NLSP) due to the reduced loop accuracy of the latter. Since this makes the calculation of the dark matter relic density unreliable, we reject scenarios for which the discrepancy is greater than 5\%.

\begin{figure}[ht!]
\centering
\includegraphics[width=0.48\textwidth]{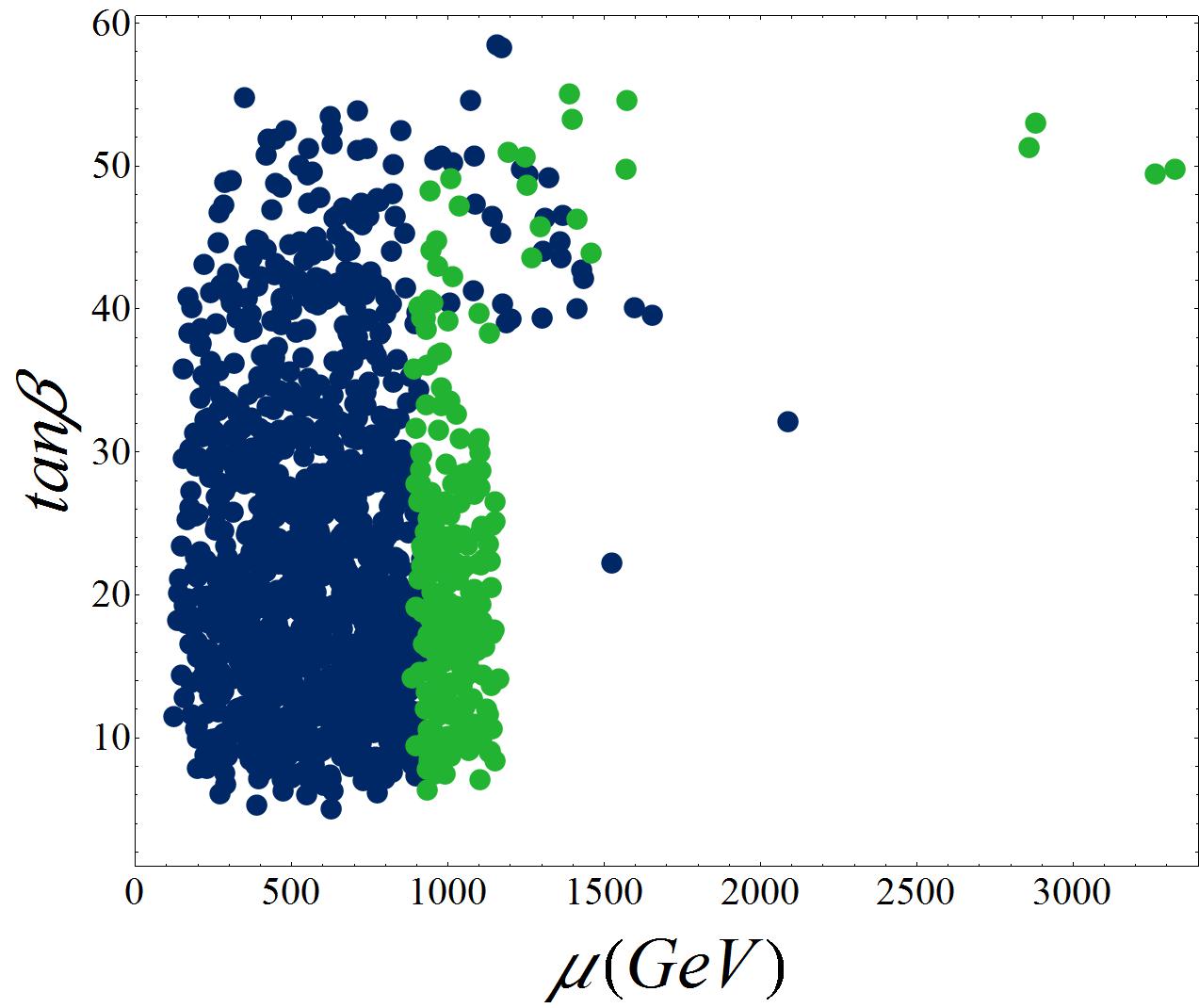} 
\includegraphics[width=0.48\textwidth]{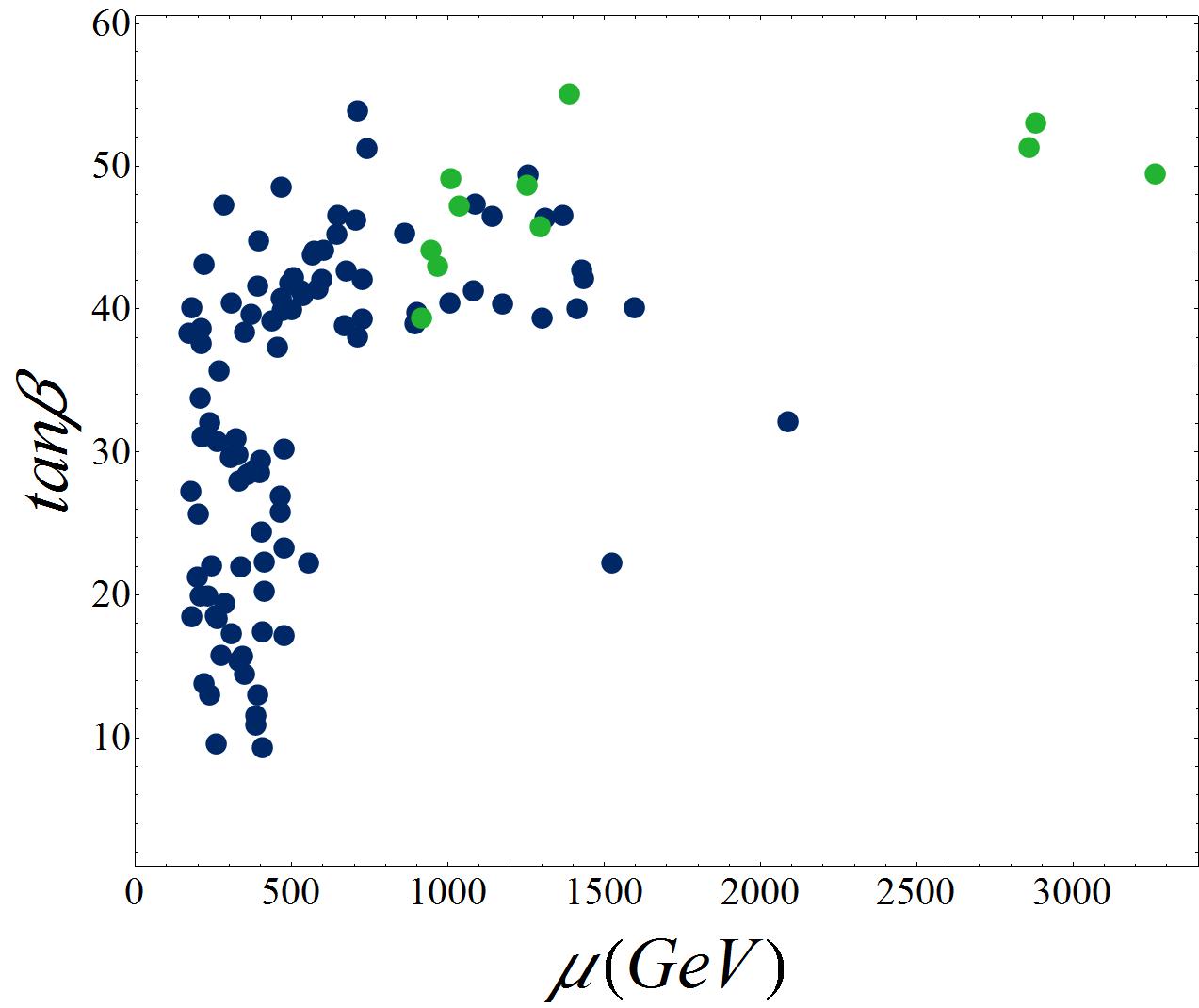} 
\caption[\it Left: Viable universal gaugino mass scenarios in the $\mu$ - $\tan \beta$ plane.]{\it Viable universal gaugino mass scenarios in the $\mu$ - $\tan \beta$ plane. Blue points represent scenarios with a dark matter relic density below $2\sigma$ bounds, while green points have the preferred relic density. Right: As the left hand plot, but with the restricted input $M_{1/2} \le 2~\rm{TeV}$.}
\label{fig:UmutanbGG}
\end{figure}

Fig.~\ref{fig:UmutanbGG} (left) shows the distribution of the surviving points in $\mu$ and $\tan \beta$, where scenarios with dark matter below the 2$\sigma$ relic density bounds are shown in blue, while those with the preferred value are shown in green.
For a better comparison with the SU(5) results, we also show (right) only the points generated with $M_{1/2} \le 2~\rm{TeV}$. For this restricted range we have approximately 500,000 attempts with 46,500 surviving points. Out of these solutions 446 or $0.09\%$ (compared with $0.02\%$ for SU(5)) survived the probability cut, of which 13 or $0.003\%$ (compared with $0.002\%$) have the preferred relic density. While the fraction of accepted solutions is more than four times larger in SO(10), the number of points with the correct relic density is only a factor of $3/2$ larger. It is also interesting to note that the direct supersymmetry searches \cite{Aad:2014wea,ATLAS_newsusy,CMS_newsusy} have very little impact on our scenarios, since scenarios which evade these bounds tend to be ruled out by other constraints.

\begin{figure}[ht!]
\centering
\includegraphics[width=0.48\textwidth]{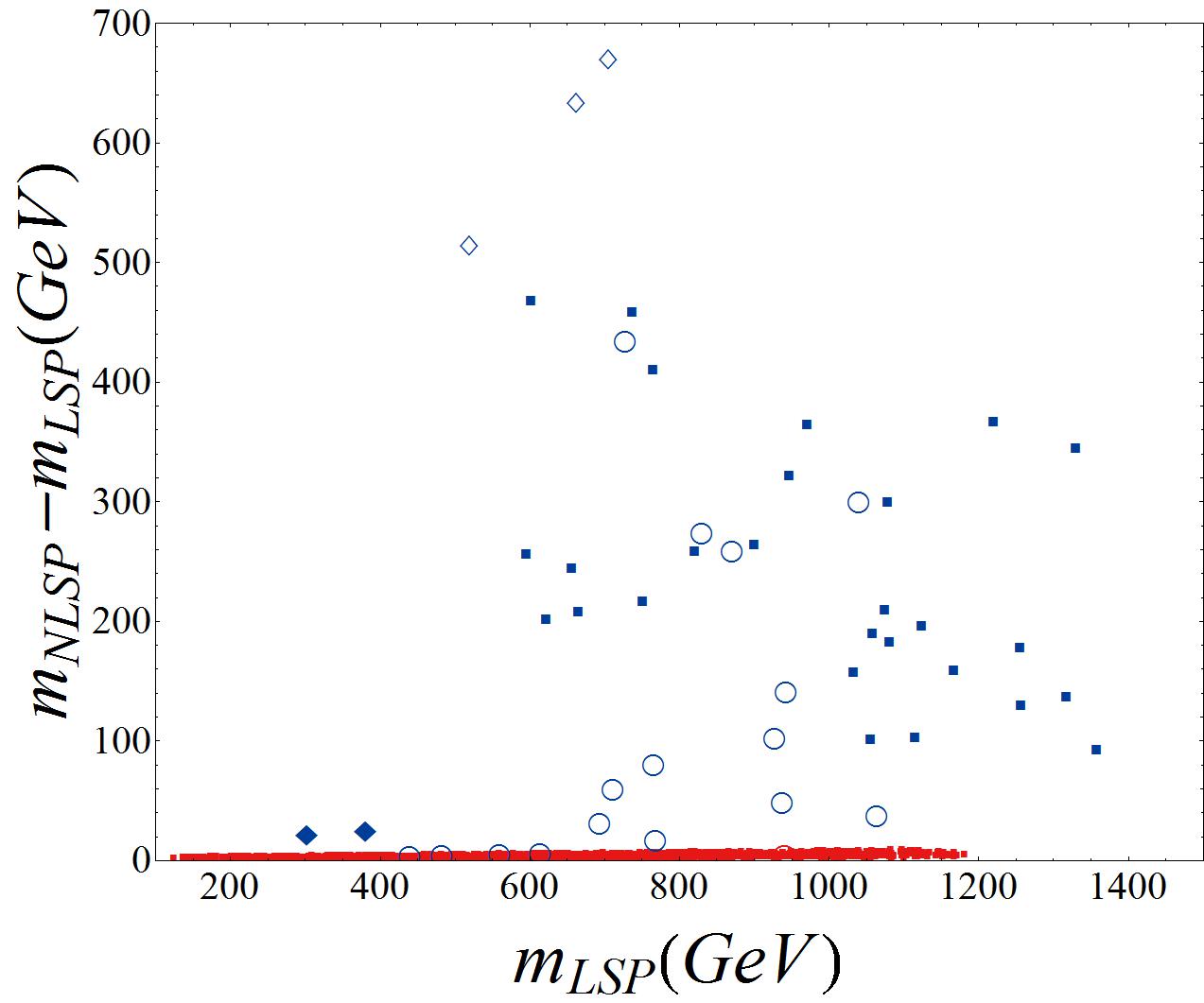}
\includegraphics[width=0.47\textwidth]{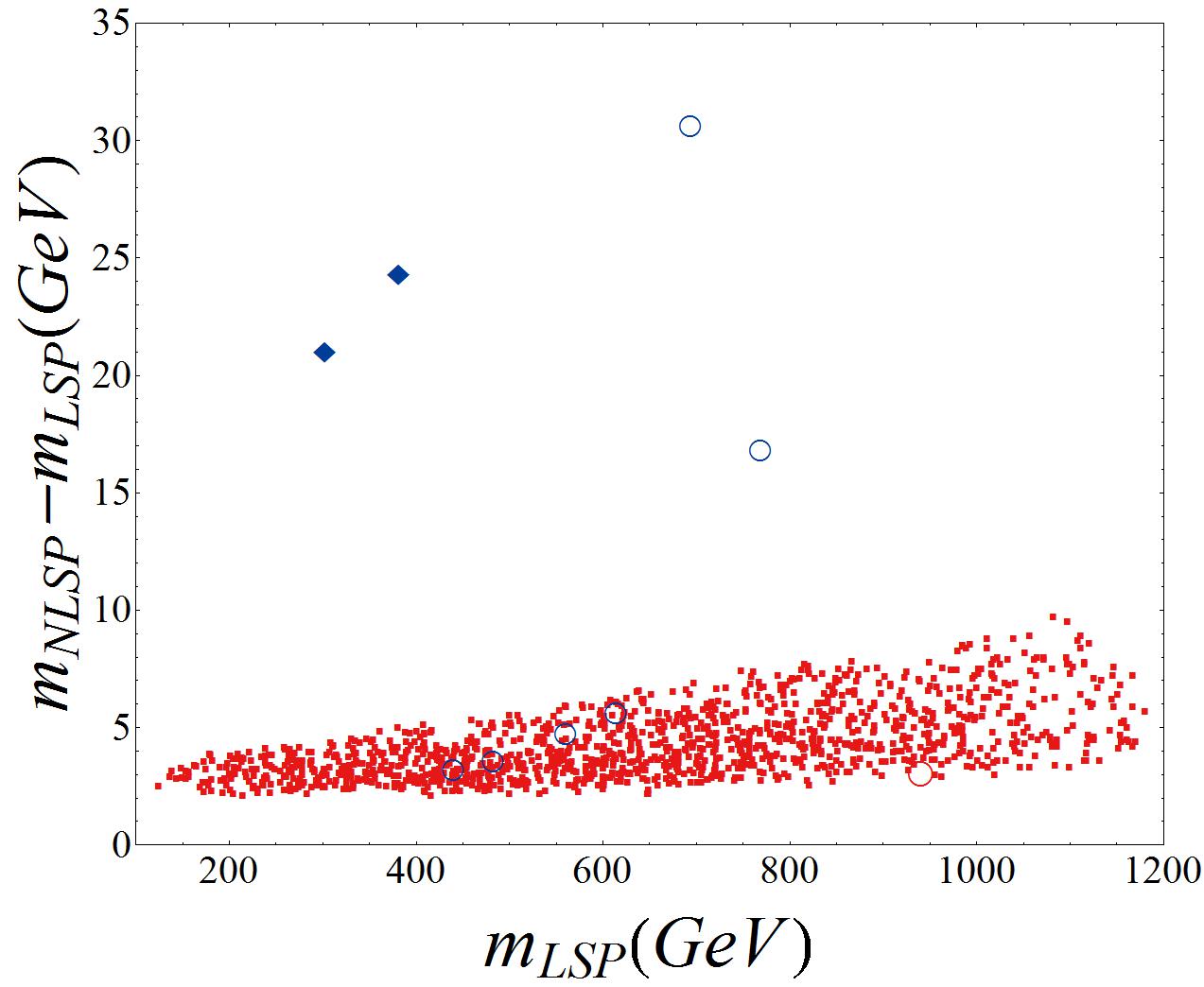}
\label{fig:ULSPtype}
\caption[\it Solutions in the plane of LSP mass vs.~the NLSP-LSP mass splitting for universal gaugino mass scenarios.]{\it Solutions in the plane of LSP mass vs.~the NLSP-LSP mass splitting for universal gaugino mass scenarios. The colour indicates the flavour of LSP, with red and blue denoting higgsino and bino dominated dark matter respectively. The shape indicates the flavour of NLSP; squares, diamonds, triangles, circles and empty diamonds denote chargino, stop, sneutrino, stau and neutralino NLSP respectively. The right-hand plot is a zoomed in version of the left-hand plot.}
\label{fig:UDMGG}
\end{figure}

The majority of the solutions with the preferred dark matter density have $\mu$ close to $1~\rm{TeV}$. In these scenarios, the neutralino LSP and chargino NLSP are both higgsino dominated and very close in mass, allowing them to co-annihilate. However it is also possible to have stops, staus or sneutrinos light enough to favour bino dominated neutralino-sfermion co-annihilation. In particular, one such solution provides the lightest stau $\tilde{\tau}_1$ we found with a mass of $502~\rm{GeV}$. The nature and mass splittings of the LSP and NLSP are shown in Fig.~\ref{fig:UDMGG}.

We find fewer points with stau-neutralino co-annihilation than in SU(5). Such solutions are represented by blue circles in regions where the LSP - NLSP mass splittings are small. This is again a consequence of the new more restrictive stability conditions. In Fig.~\ref{fig:UDMGG} (right) we also see a rare solution where the neutralino is dominated by its Higgsino component, but instead of a chargino, the NLSP is a stau (the red circle). For this scenario, the third generation slepton has mass coincidentally between the almost degenerate neutralino-chargino pair (the neutralino, stau and chargino masses are $940$, $944$ and $947~\rm{GeV}$ respectively). There are also many viable scenarios where the NLSP is considerably heavier than a bino dominated LSP, up to as much as $700~\rm{GeV}$ heavier; these solutions have a heavy Higgs boson approximately twice the mass of the neutralino, allowing dark matter annihilation via a Higgs resonance. 

\begin{figure}[ht!] 
\centering
\includegraphics[width=0.48\textwidth]{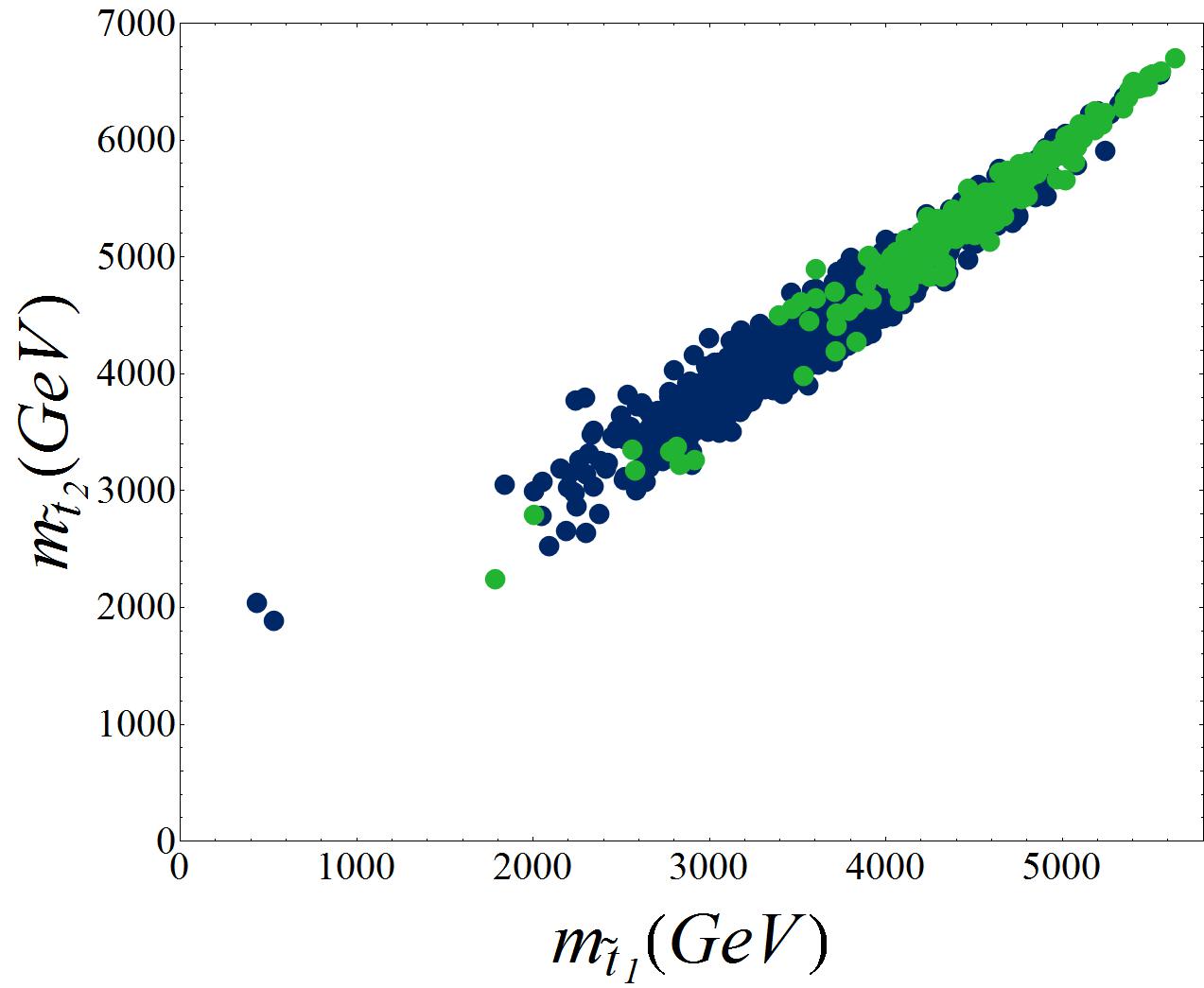}
\includegraphics[width=0.49\textwidth]{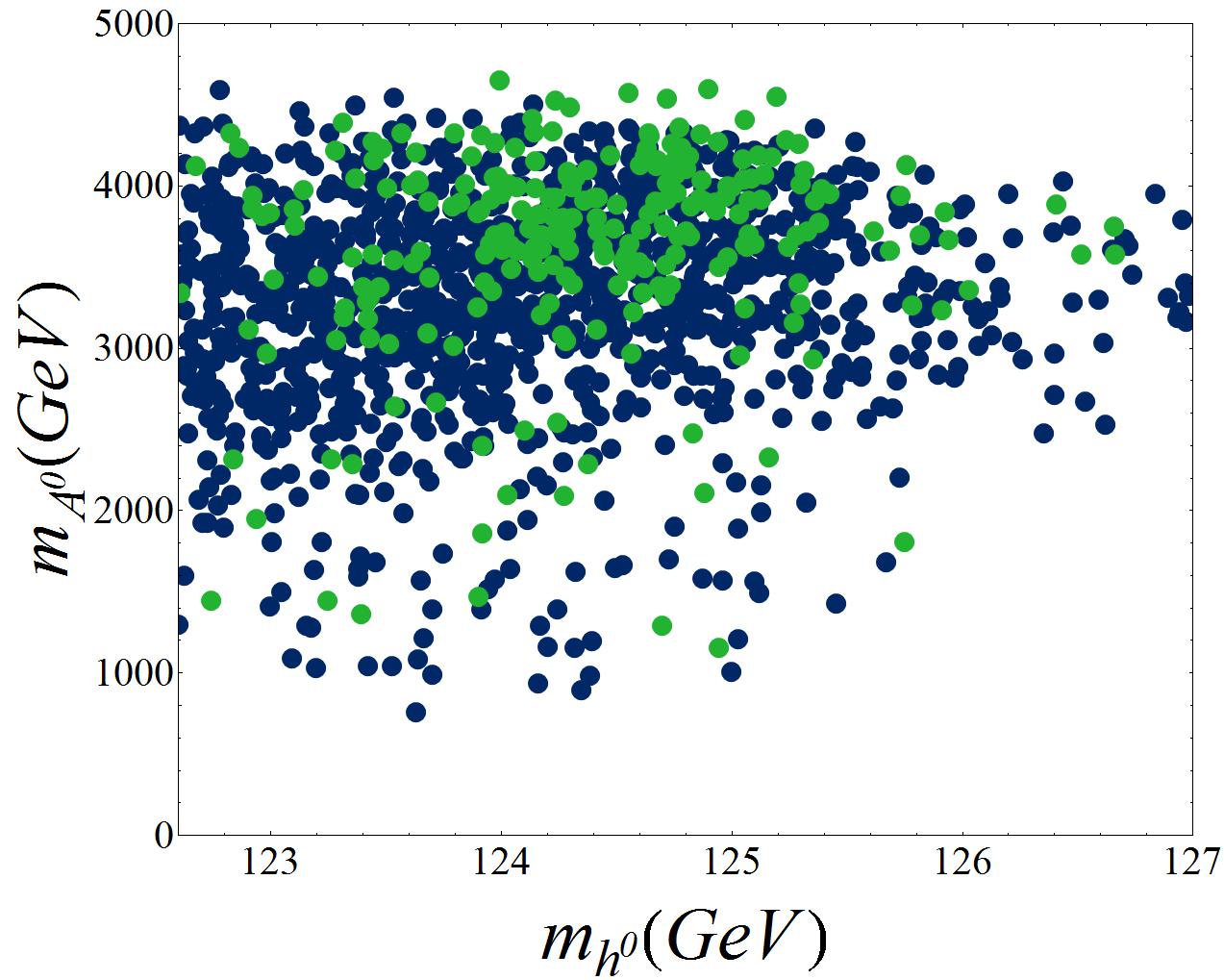}
\caption[\it Viable universal gaugino mass scenarios in the stop mass and the lightest scalar - pseudoscalar mass planes]{\it Viable universal gaugino mass scenarios in the stop mass (left) and the lightest scalar - pseudoscalar mass (right) planes, with colours as in Fig.~\ref{fig:UmutanbGG}.}
\label{fig:UstophiggsGG}
\end{figure}

In Fig.~\ref{fig:UstophiggsGG} we show the physical stop masses for viable scenarios (left) and the Higgs boson and its pseudo-scalar partner (right). The furthest blue point to the left in the $m_{\tilde{t}_1}\-- m_{\tilde{t}_2}$ pane corresponds to the lightest stop found and has a mass of $432\,{\rm GeV}$ but too little dark matter. The lightest stop with the preferred relic density has mass $1785\,{\rm GeV}$ (the furthest left green point). We have no difficulty producing Higgs boson in the correct mass range, but we require a CP-odd Higgs with mass $0.7$--$4.7\,{\rm TeV}$. 

\begin{figure}[ht!]
\centering
\includegraphics[width=0.46\textwidth,height=0.275\textheight]{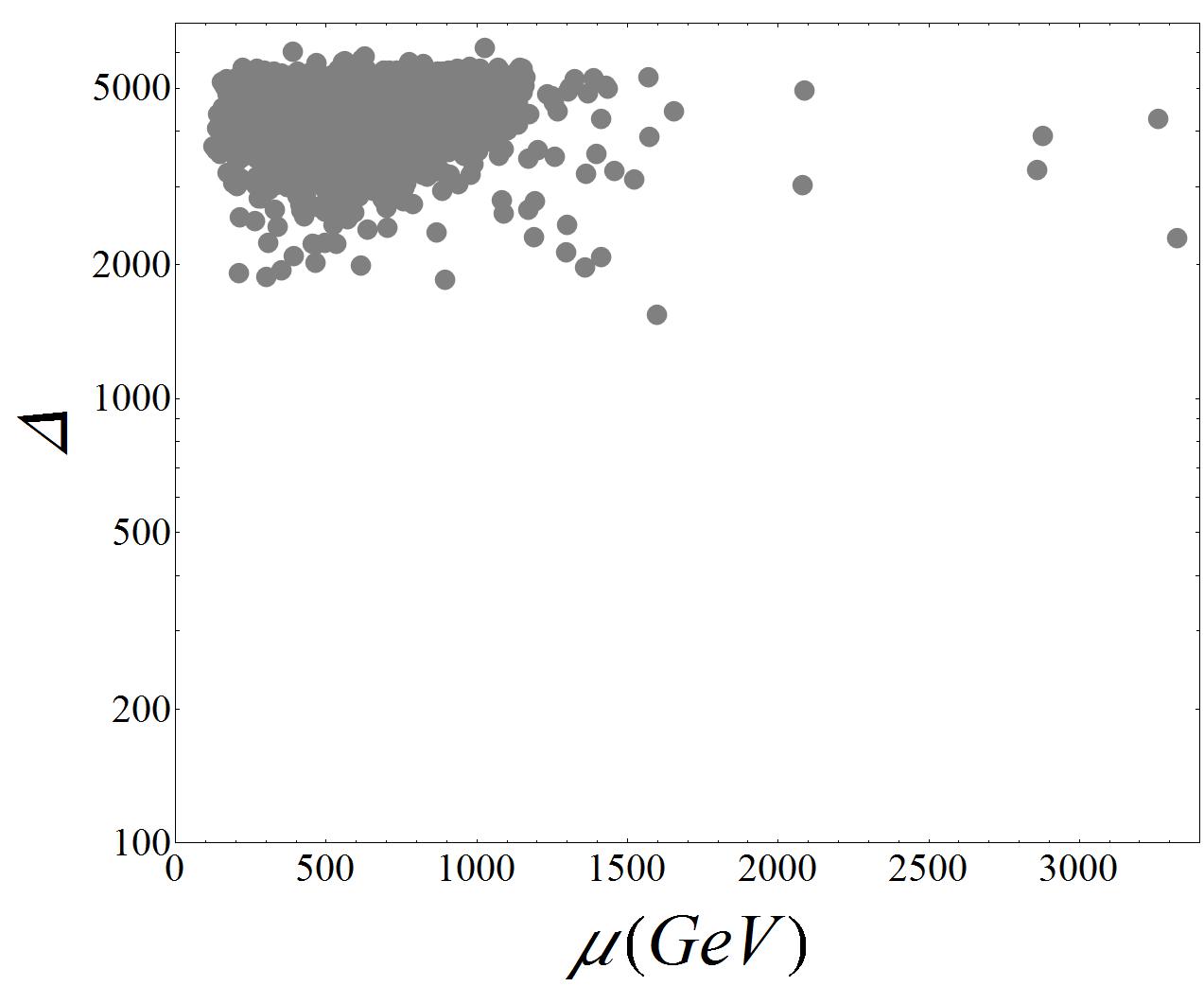}
\caption{\it Fine-tuning $\Delta$ compared to $\mu$ for universal gaugino mass scenarios. Note that the fine-tuning with respect to $\mu$ itself is {\em not} included in $\Delta$.}
\label{fig:UDeltas}
\end{figure}

We have seen that SO(10) GUT models with universal gaugino masses provide plenty of scenarios that survive the vacuum stability conditions and experimental constraints. However, we have not yet determined whether or not such scenarios require fine-tuning of the parameters to obtain the correct $Z$ boson mass. To evaluate this, we focus on parameters that provide the dominant contribution to $m_{H_u}^2$ such as the scalar masses $m_{16}$ and $m_{10+126}$, the D-term $g^2_{10} D$, the gaugino mass $M_{1/2}$, and the trilinear coupling $a_{10}$, and use SOFTSUSY's implementation of fine-tuning throughout. As one would expect, we find that the individual fine-tunings can be reduced by making the corresponding parameter smaller. However, we have no scenarios with small values of $M_{1/2}$ or $m_{10+126}$, so these fine-tunings remain sizable. The maximal fine-tuning for these scenarios is shown in Fig.~\ref{fig:UDeltas} in comparison to $\mu$, where we see that all viable scenarios have fine-tuning above 1500.

To summarise our findings for SO(10) inspired scenarios with universal gaugino masses, we find plenty of physically viable solutions in accordance with experimental constraints and the dark matter relic density. Furthermore, it seems considerably easier to find these solutions than in the less constrained SU(5) models. However, these scenarios suffer from unavoidable and unattractive fine-tuning.

\section{Non-Universal Gaugino Masses}
\label{sec:nugm}

We have been unable to find satisfactory solutions with universal gaugino masses, so now examine non-universal gaugino masses. We extend our parameter space by introducing $\rho_1 = M_1/M_3$ and $\rho_2=M_2/M_3$ (at the GUT scale), letting them vary in the interval $[-15, 15]$. To preserve notation, we identify $M_{1/2}$ with the value of $M_3$ at the GUT scale. For another analysis of SO(10) with non-universal gaugino masses with particular emphasis on dark matter constraints, see Ref.~\cite{Chakrabortty:2013voa}.

\subsection{An Inclusive Scan}
\label{subsec:IN}

\hspace{5mm} We first perform an inclusive scan over the parameter space to identify regions of interest. The number of initial scenarios is now 4,100,000; when we remove charged LSPs, apply stability constraints and impose the LHC and LUX bounds, we find 97,457 (2.3\%) of these survive, which is a sizable increase in the fraction of accepted points in comparison to SU(5). This fraction is lower than for universal gaugino masses due to the removal of scenarios with coloured dark matter in regions where \mbox{$M_{3} \ll M_{1,2}$}. Requiring ${ P}_{\rm tot}>10^{-3}$ leaves 59,833 scenarios of which 9200 have the preferred dark matter relic density. 

\begin{figure}[ht!]
\centering
\includegraphics[width=0.45\textwidth]{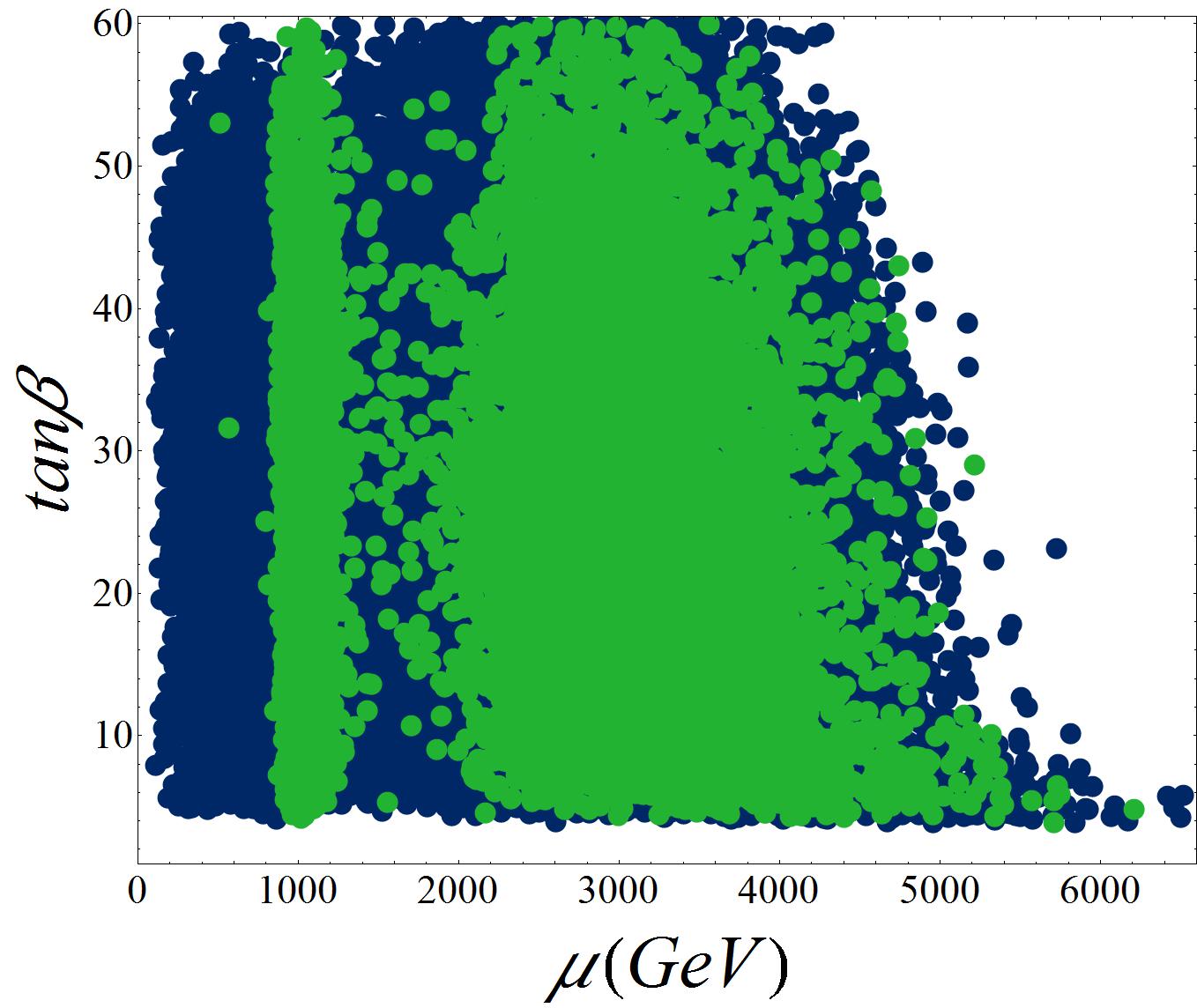}
\caption[\it Viable non-universal gaugino mass scenarios in the $\mu$-$\tan \beta$ plane.]{\it Viable non-universal gaugino mass scenarios in the $\mu$-$\tan \beta$ plane, with colours as in Fig.~\ref{fig:UmutanbGG}.}
\label{fig:NUmutanbGG}
\end{figure}

In Fig.~\ref{fig:NUmutanbGG} we show the surviving scenarios as a $\mu$-$\tan \beta$ projection. As well as the usual higgsino dark matter scenarios with the correct relic density around $1\,{\rm TeV}$, we now have many bino and wino dark matter scenarios with higher $\mu$. 

\begin{figure}[ht!]
\centering
\includegraphics[width=0.49\textwidth]{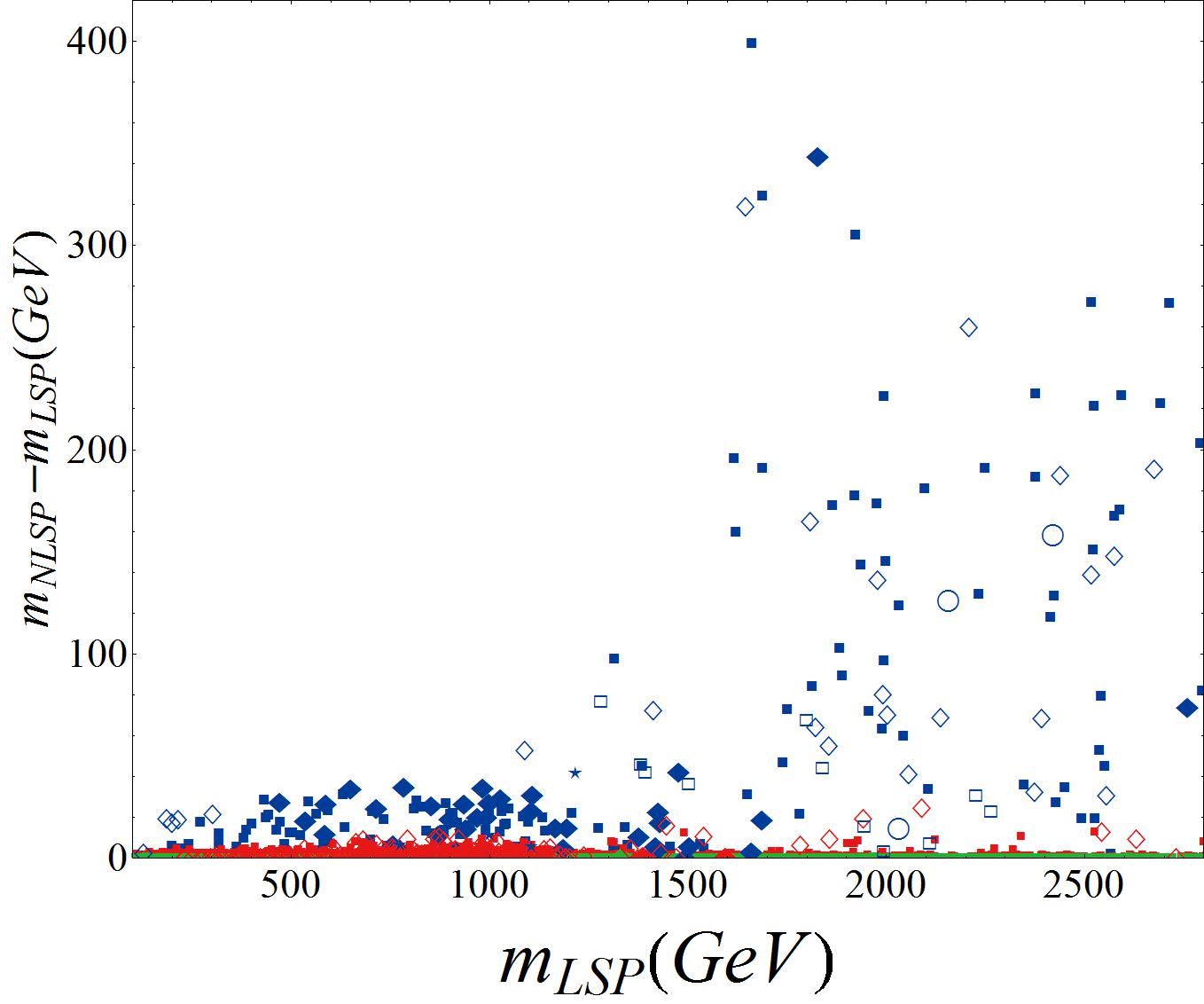}
\includegraphics[width=0.48\textwidth]{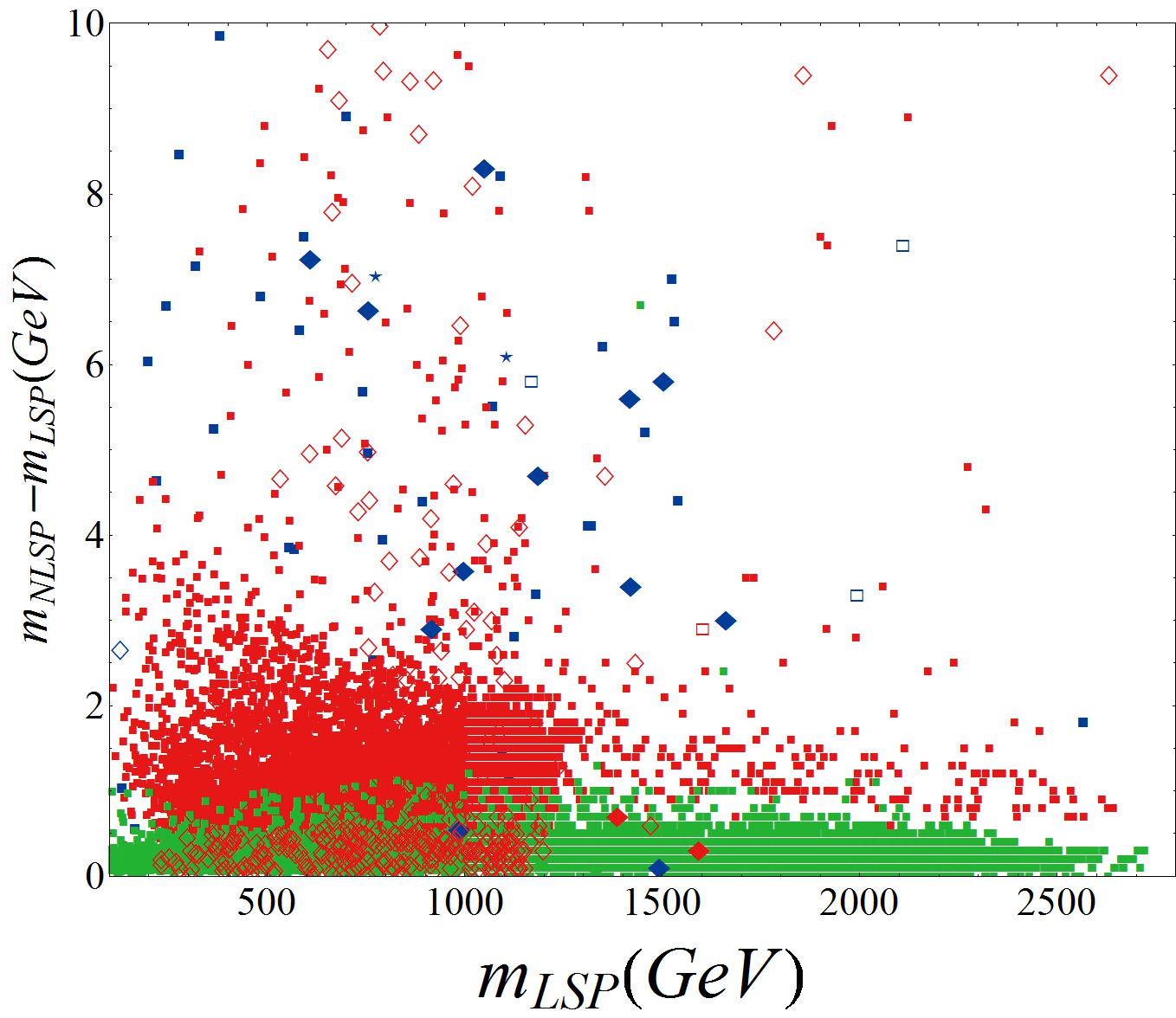}
\caption[\it Solutions in the plane of LSP mass vs.~the NLSP-LSP mass splitting for non-universal gaugino mass in SO(10) scenarios.]{\it Solutions in the plane of LSP mass vs.~the NLSP-LSP mass splitting for non-universal gaugino mass scenarios. The colour indicates the flavour of LSP, with red, blue and green denoting higgsino, bino and wino dominated dark matter respectively. The shape indicates the flavour of NLSP; filled squares, empty squares, filled diamonds, empty diamonds, circles and stars denote chargino, gluino, stop, neutralino, stau and sbottom NLSP respectively. The right-hand plot is a zoomed in version of the left-hand plot. Scenarios with too little dark matter are also shown.}
\label{fig:NUDMGG}
\end{figure}

The identity and mass splittings of the LSP and NLSP are shown in Fig.~\ref{fig:NUDMGG}, where we see many additional LSP-NLSP pairings, including wino dominated dark matter when $M_2 < 2 M_1$. Such solutions can provide the correct relic density for higher LSP masses. As before, most scenarios have the LSP and NLSP close in mass permitting co-annihilation, but we also have bino dominated dark matter with the NLSP as much as $400\,{\rm GeV}$ heavier than its LSP and dark matter annihilation via a heavy Higgs resonance. Note that in Fig.~\ref{fig:NUDMGG} we show all our surviving scenarios, including these with too little dark matter. 

Particularly interesting scenarios that we observe consist of bino-gluino and bino-wino co-annihilation regions, where the lack of supersymmetry searches at the LHC experiments make such scenarios rather unconstrained and relevant to explore at the $14~\rm{TeV}$ LHC. For the first case, the masses of the bino-gluino pairs range from $(1284, 1360)~\rm{GeV}$ (the leftmost blue empty square in the left panel of Fig.~\ref{fig:NUDMGG}), up to $(2267, 2290)~\rm{GeV}$  (the rightmost blue empty square in the left panel of Fig.~\ref{fig:NUDMGG}). Although the LSP is too heavy to be visible at the LHC, the gluino mass is within the LHC reach. We also find first and second generation squarks in a wide range from $2~\rm{TeV}$ up to $10~\rm{TeV}$. The third generation sfermions are approximately between $1.5-4~\rm{TeV}$ and the Higgs sector predicts heavy partners between $3$ and $6\,\rm{TeV}$. These scenarios are favored by models consisting of a constrained gaugino sector, where $M_{1/2}$ is within $400\--800~\rm{GeV}$, $\left| \rho_1 \right| \sim 7$ and prefer negative values of $a_{10}$. The other co-annihilation region consists of bino-wino (almost) degenerate states, where the lightest LSP-NLSP pair is found to have $130$ and $132\rm{GeV}$ neutralinos. The full set of solutions that respect this bino-wino degeneracy are represented by blue empty diamonds in Fig.~\ref{fig:NUDMGG} with a LSP mass always lighter than $305~\rm{GeV}$. In particular, we found a point which predicts a $125~\rm{GeV}$ Higgs boson and $189~\rm{GeV}$ binos which co-annihilate with $208~\rm{GeV}$ winos in such a way that the dark matter relic density saturates the WMAP bounds. The remaining low energy spectrum follows the same trend as the previous case. Again, models with a constrained gaugino sector and negative values of $a_{10}$ favour these class of solutions. In particular,  bino-wino degeneracy may be obtained if we set $0.4 \lesssim \rho_1 \lesssim 0.5 $ and $\lvert \rho_2 \rvert \sim 0.25$ with $M_{1/2}$ between $600$  and $1400~\rm{GeV}$. One could also speculate that bino-wino mixed dark matter is allowed by a rather constrained gaugino sector. While there is no \textit{a priori} reason to reject this argument, it would require full mass degeneracy among the $U(1)_Y$ and $SU(2)_L$ neutral gauginos at the low scale. However, as the boundary conditions are set at the GUT scale, finding identical masses for the lightest neutralinos is rather hard in our top-down approach, which involves numerical evolution of both $M_1$ and $M_2$.

\begin{figure}[ht!]
\centering
\includegraphics[width=0.48\textwidth]{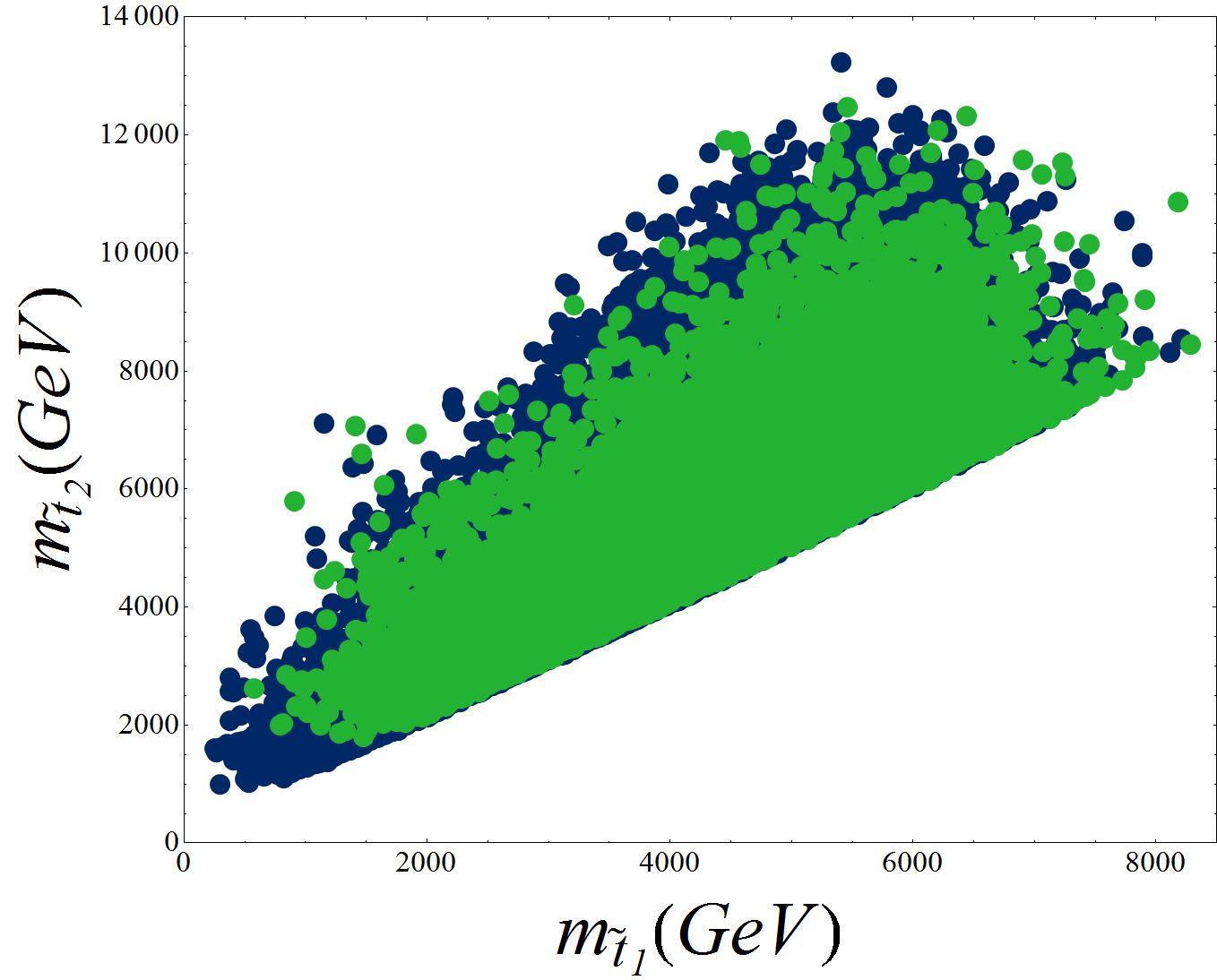}
\includegraphics[width=0.49\textwidth]{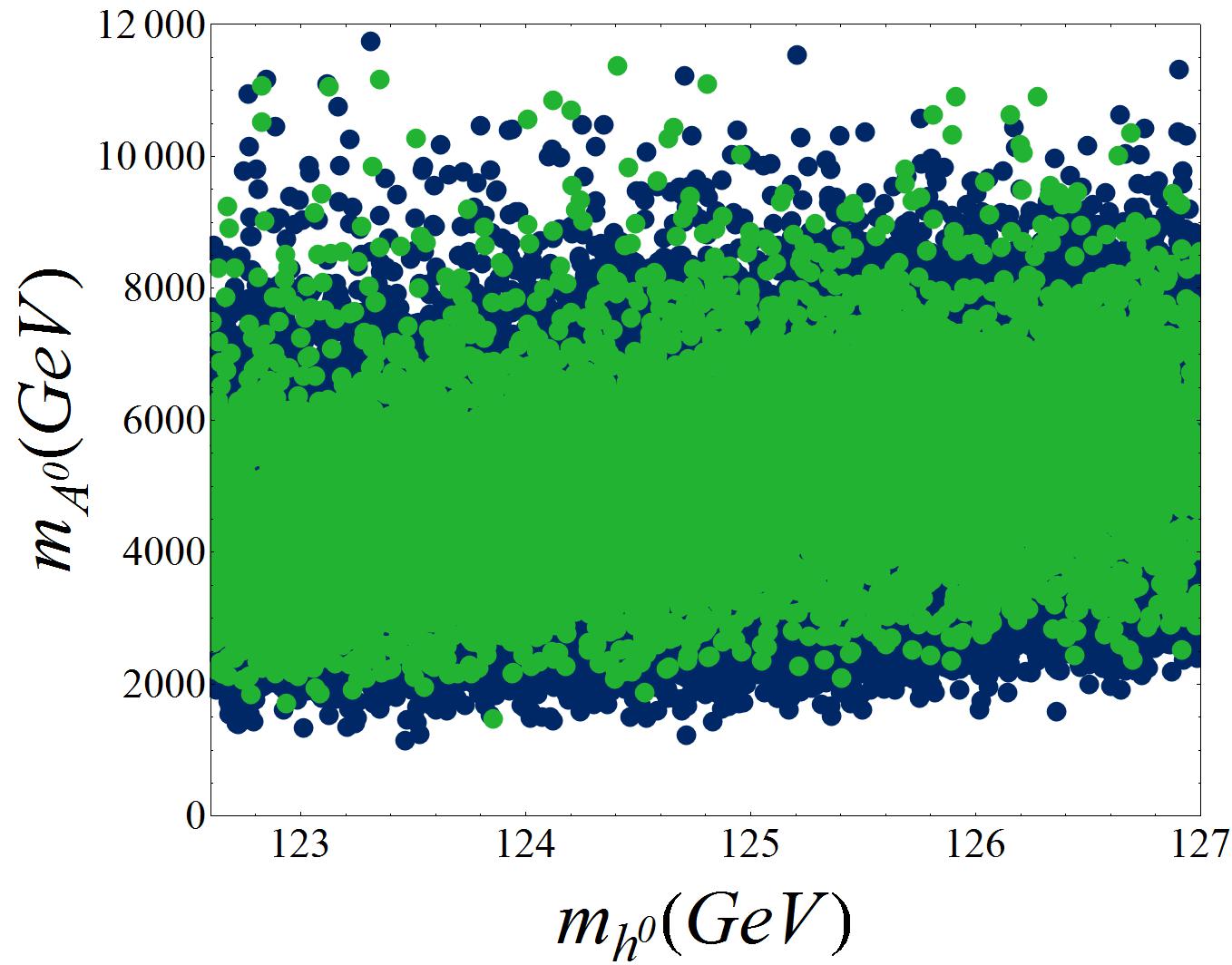}
\caption[\it Viable non-universal gaugino mass scenarios in the stop mass lightest scalar -- pseudoscalar mass planes]{\it Viable non-universal gaugino mass scenarios in the stop mass  (left) and the lightest scalar -- pseudoscalar mass  (right) planes, with colours as in Fig.~\ref{fig:UmutanbGG}.}
\label{fig:NUstophiggsGG}
\end{figure}

In general, the range of allowed stop masses is now much greater, as can be seen in Fig.~\ref{fig:NUstophiggsGG} (left), and we even find some relatively light stops. The lightest stop in a scenario with the correct relic density has a mass of $576\,\rm{GeV}$. If we require stop-neutralino degeneracy (filed diamonds on the right pane of Fig.~\ref{fig:NUDMGG}), the lightest stop we find is $759~\rm{GeV}$. This scenario gains further importance for the LHC14 as there are no experimental searches constraining models that predict degenerate stop-neutralinos states. As for the universal gaugino masses, we have no difficulty in achieving a sufficiently heavy Higgs boson, as shown in Fig.~\ref{fig:NUstophiggsGG} (right). The wider range of allowed masses is also reflected in our bottom squarks and staus (though we do not reproduce the plots here). The staus in particular can become very heavy due to allowing the region $M_{1,2} \gg M_3$ in our scan.

\begin{figure}[h!]
\centering
\includegraphics[width=0.5\textwidth]{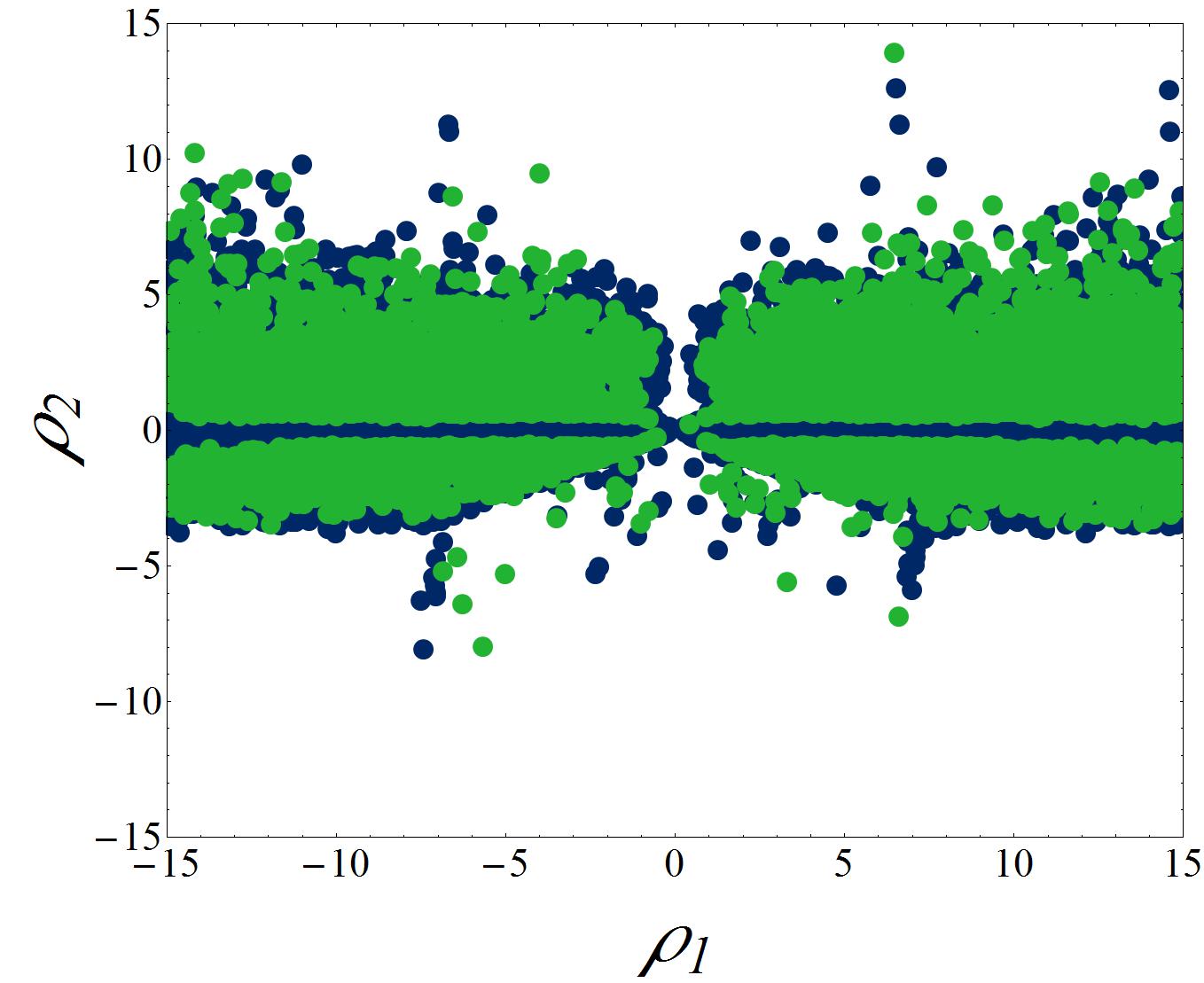}
\caption[\it Viable non-universal gaugino mass scenarios in the non-universal parameters $\rho{1,2}$.]{\it Viable non-universal gaugino mass scenarios in the non-universality parameters $\rho_{1,2}$, with colours as in Fig.~\ref{fig:UmutanbGG}.}
\label{fig:NUrhoGG}
\end{figure}

The allowed scenarios projected onto the $\rho_{1,2}$ plane are shown in Fig.~\ref{fig:NUrhoGG}. As was the case for SU(5) there are very few viable scenarios in the region corresponding to universal gaugino masses, $\rho_1=\rho_2=1$, which of course reflects the difficulty for finding viable scenarios in our universal gaugino study of Section~\ref{sec:ugmSO10}. The asymmetry with respect to the $\rho_1$ axis caused by choosing $\mu>0$ for this scan.

\begin{figure}[ht!]
\centering
\includegraphics[width=0.48\textwidth]{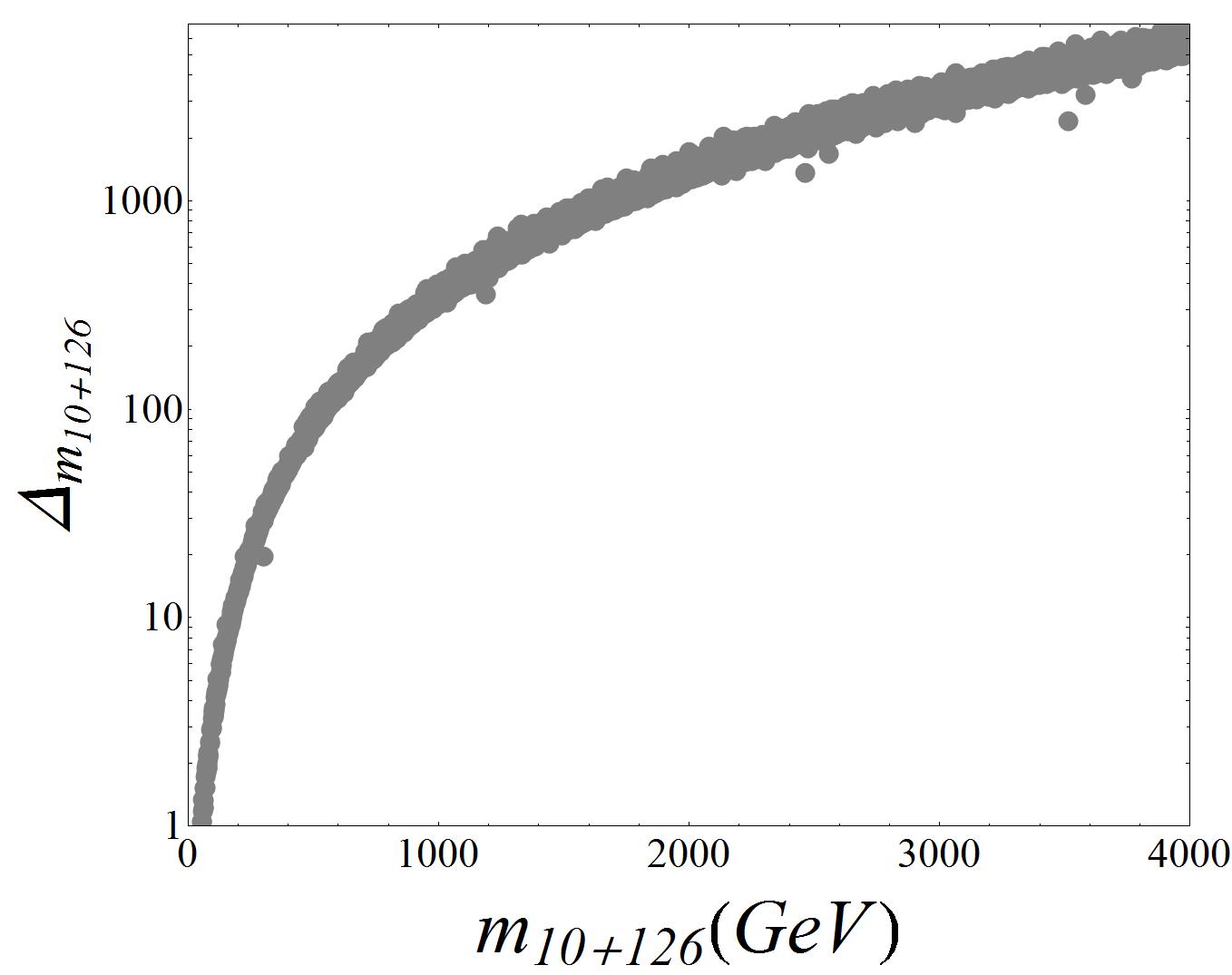}
\includegraphics[width=0.48\textwidth]{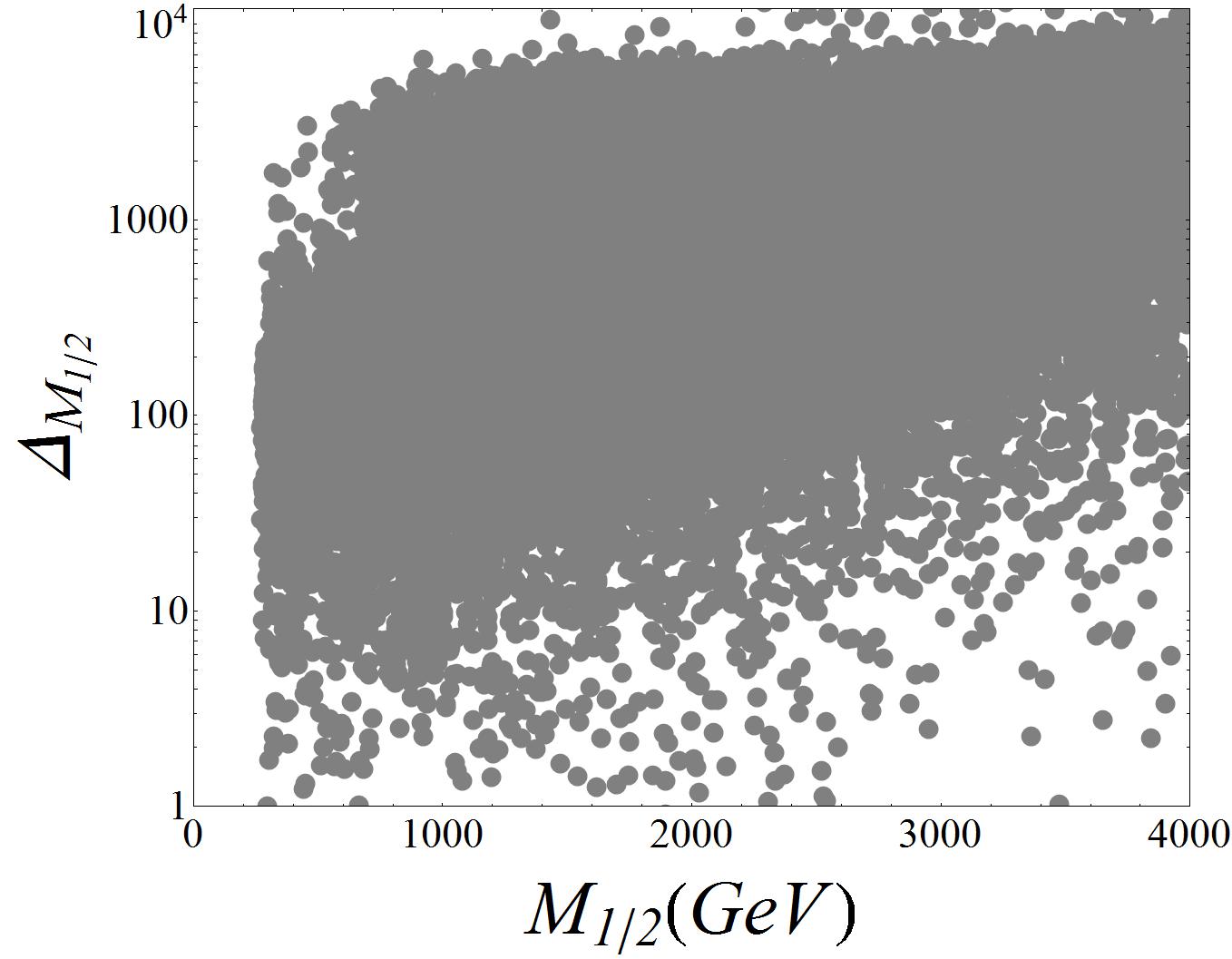}\\[4mm]
\includegraphics[width=0.48\textwidth,height=0.27\textheight]{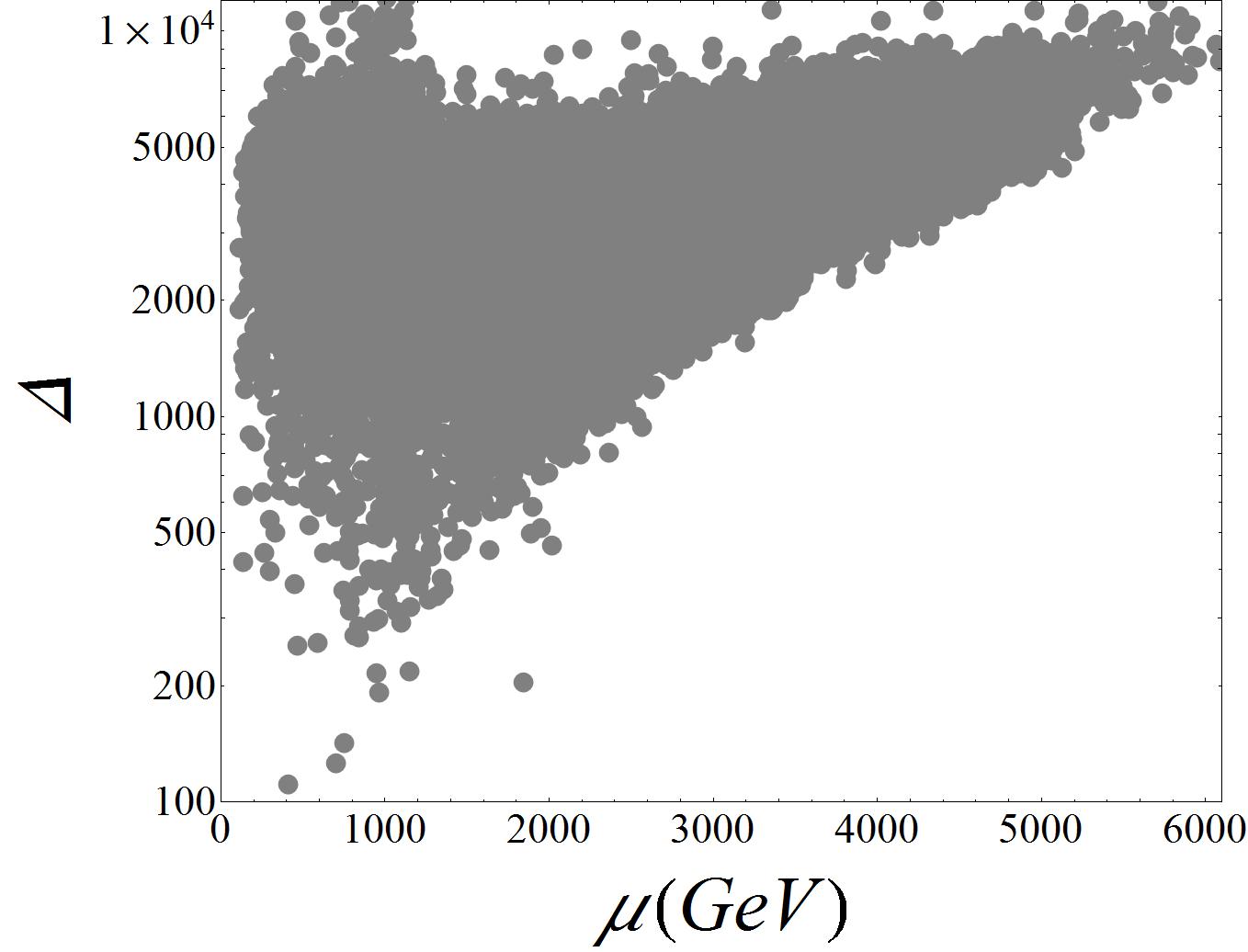}
\caption{\it Top: Fine-tuning in $M_Z$ with respect to the input parameters $m_{10+126}$ and $a_{10}$ and $g^2_{10} D$ for non-universal gaugino mass scenarios. Bottom: Fine-tuning $\Delta$ compared to $\mu$ for non-universal gaugino mass scenarios. Note that the fine-tuning with respect to $\mu$ itself is not included in $\Delta$.}
\label{fig:NUDeltasGG}
\end{figure}

The picture of fine-tuning is significantly different in the non-universal gaugino scan. As before, the fine-tuning due to $m_{16}$, $a_{10}$ and the D-term can be reduced by reducing the size of the individual parameters. However, in contrast to the universal gaugino scenarios, we now find scenarios with low values of $m_{10+126}$, shown in Fig.~\ref{fig:NUDeltasGG} (top left), thereby allowing us to relieve the fine-tuning associated with it. Furthermore, we see scenarios with low fine-tuning caused by $M_{1/2}$, Fig.~\ref{fig:NUDeltasGG} (top right), for a wide range of GUT scale gaugino masses, even rather large values. This is the same phenomenon we encountered in Ref.~\cite{Miller:2013jra} where we found that for particular choices of $M_{1/2}$, $m^2_{H_u}$ may sit close to a minimum, rendering it insensitive to fluctuations in $M_{1/2}$. While this will prove useful later in the paper, for now we note that we still find no points where these small fine-tunings occur simultaneously for all parameters and therefore the total fine-tuning $\Delta$, Fig.~\ref{fig:NUDeltasGG} (bottom), is always greater than $100$ for all our scenarios.

\subsection{An Enhanced Scan Over $M_{1/2}$, $\rho_1$ and $\rho_2$}
\label{subsec:EN}

\begin{figure}[ht]
\centering
\includegraphics[width=0.48\textwidth]{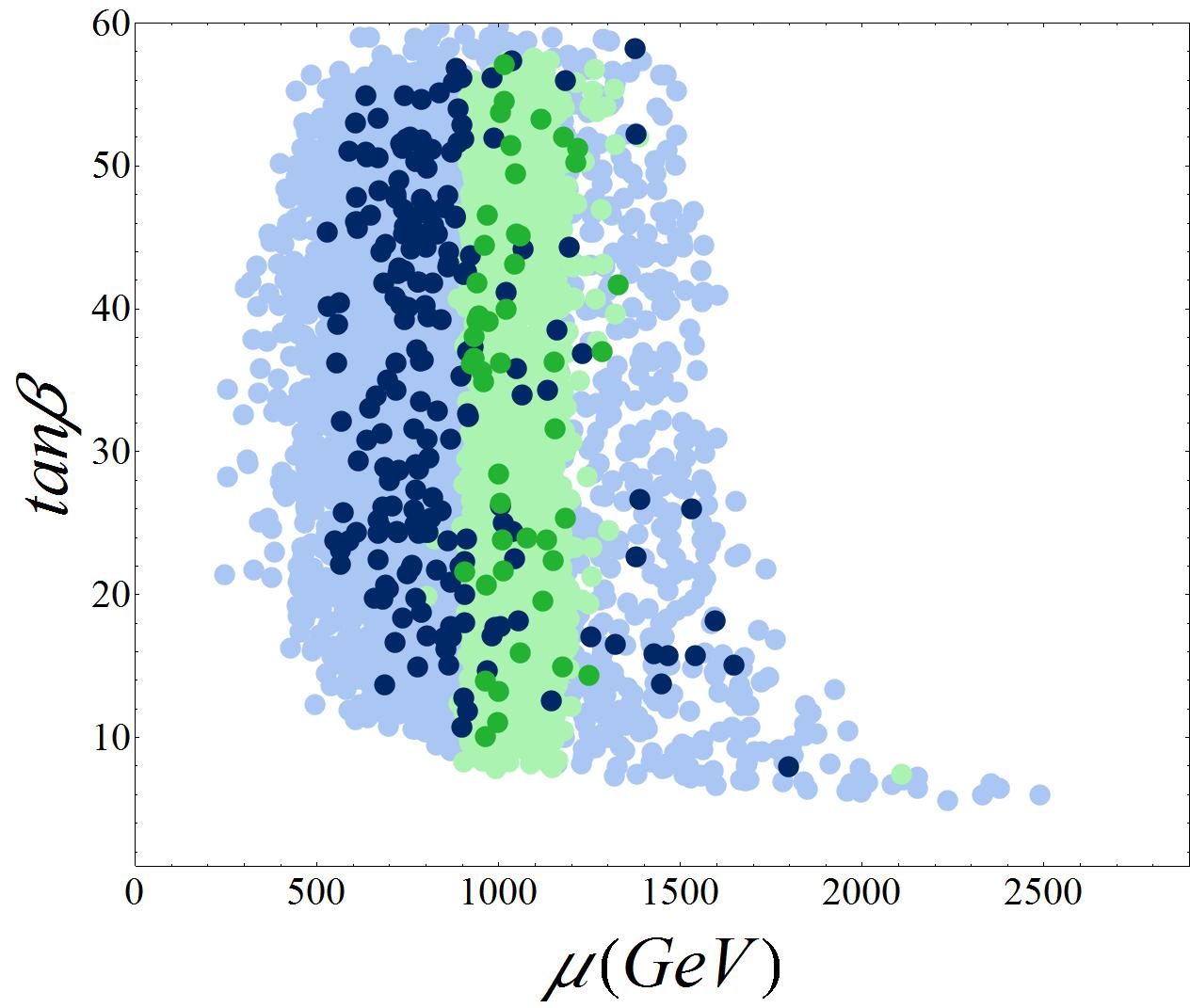}
\caption[\it Viable scenarios in the $\mu$-$\tan \beta$ plane for the enhanced scan with non-universal gaugino masses for SO(10).]{\it Viable scenarios in the $\mu$-$\tan \beta$ plane for the enhanced scan with non-universal gaugino masses. Points with the preferred dark matter relic density are shown in green, while those with a relic density below the bounds are in blue. Darker and lighter shades denote the fine-tuning: darker shades have fine-tuning $\Delta<10$ while lighter shades have $10<\Delta<100$.}
\label{fig:NUmutanbGG_e}
\end{figure}

Although our inclusive scan found no satisfactory points that evade experimental constraints, give a good dark matter relic density and have low fine-tuning, we may suspect that such points do exist and may be found by a more extended search. To guarantee that natural fluctuations of the scalar masses and trilinear couplings are small, we restrict \mbox{$0 < m_{scalar} < 150~\rm{GeV}$}, \mbox{$0 < \left|g^2_{10}D\right| < 150~\rm{GeV}$} and  \mbox{$0 < \left|a_{10}\right| < 150~\rm{GeV}$}. Using the experience of SU(5) studies, we allow $\rho_1$ and $\rho_2$ to vary over $\left[-13,13\right]$ and $\left[-3.5,3.5\right]$ respectively. Experimental and stability constraints are implemented as before, but we will also add a fine-tuning cut $\Delta<100$ (where again fine-tuning from $\mu$ is excluded). With approximately 2,000,000 initial attempts, 10,158 solutions survived the experimental and stability constraints, of which 5,760 were accepted by the probability and fine-tuning cuts. Such scenarios are shown in  the $\mu$-$\tan \beta$ plane in Fig.~\ref{fig:NUmutanbGG_e}. We now identify plenty of points with fine tuning less than 100 (lighter shades of green and blue) as well as several with fine-tuning less than 10 (darker shades of green and blue). The number of points that provide a good description of the full dark matter (green points) increased from 1,028 in SU(5) to 1,478 in SO(10). These scenarios are now restricted to be close to $1\,{\rm TeV}$ and we have lost the most of the scenarios with larger $\mu$ but a correct relic density that we saw in the inclusive scan. 

\begin{figure}[ht!]
\centering
\includegraphics[width=0.48\textwidth]{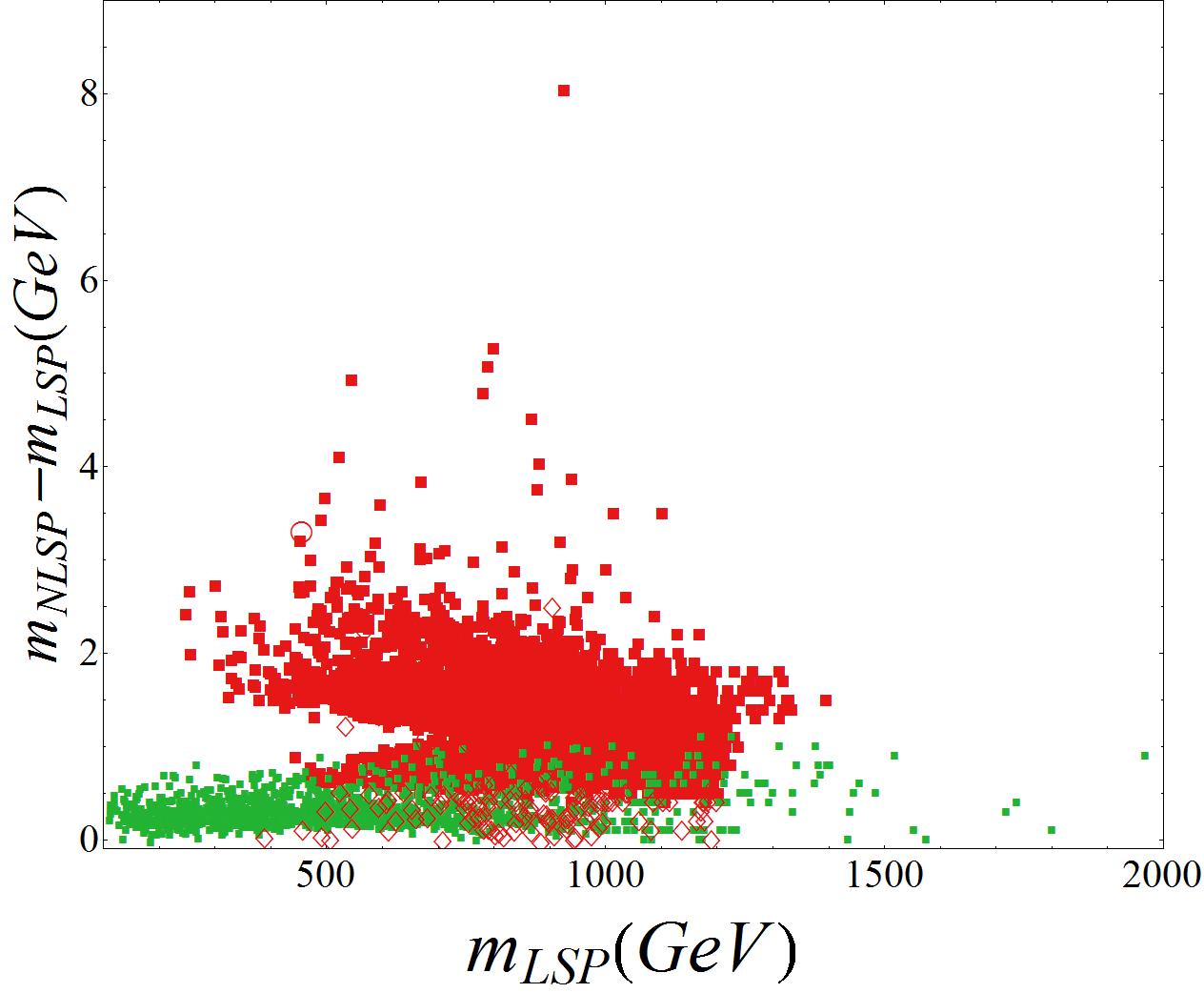}
\includegraphics[width=0.48\textwidth]{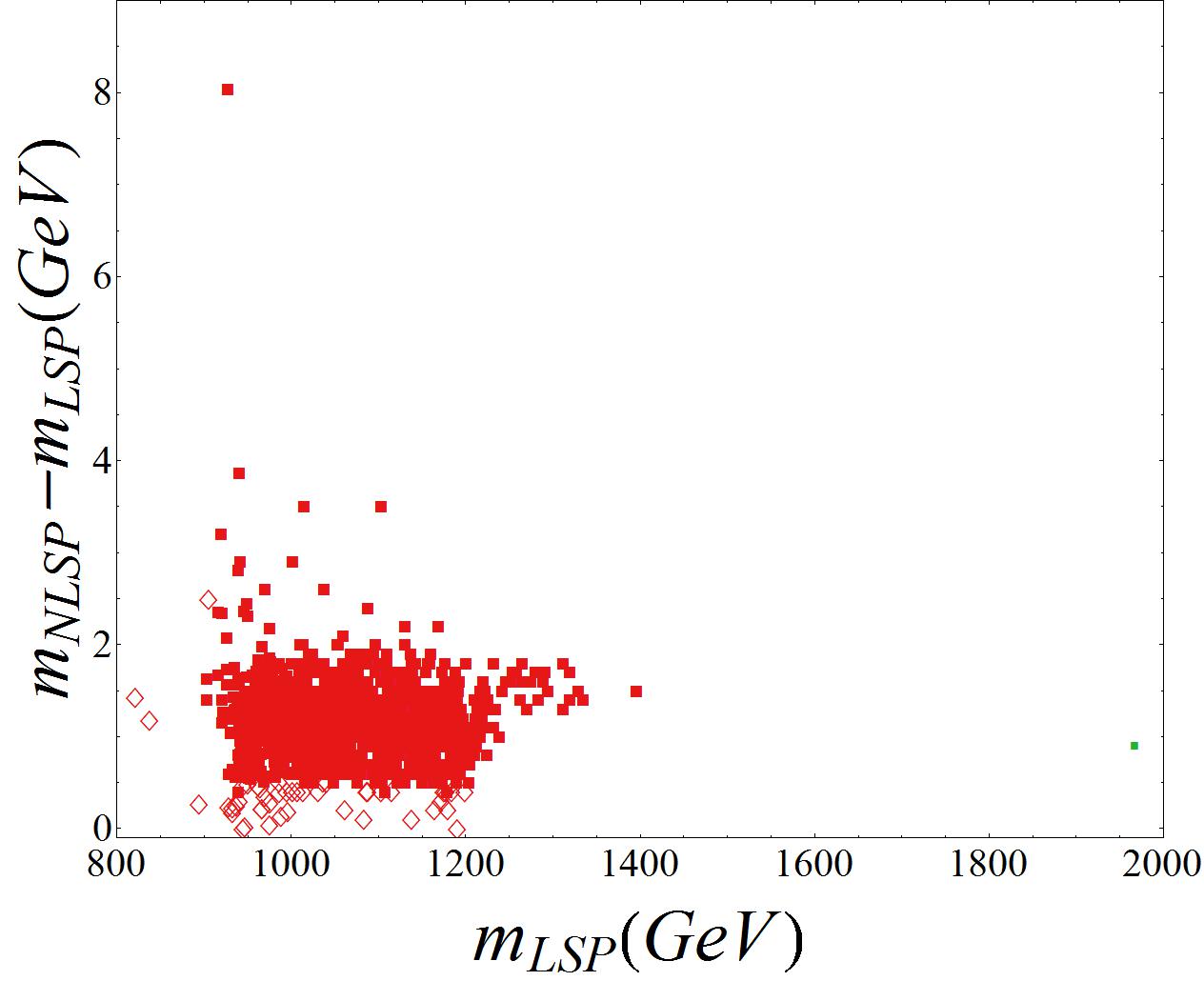}
\caption[\it Solutions in the plane of LSP mass vs.~the NLSP-LSP mass splitting for the enhanced scan over non-universal gaugino mass scenarios for SO(10).]{\it Solutions in the plane of LSP mass vs.~the NLSP-LSP mass splitting for the enhanced scan over non-universal gaugino mass scenarios. The colour indicates the flavour of LSP, with red, blue and green denoting higgsino, bino and wino dominated dark matter respectively. The shape indicates the flavour of NLSP; filled squares and empty diamonds denote chargino and neutralino NLSP respectively. The left-hand plot shows all scenarios with fine-tuning $\Delta<100$ while the right-hand plot restricts to scenarios with $\Delta<10$ and the preferred dark matter relic abundance.}
\label{fig:NUDMGG_e}
\end{figure}

The LSP and NLSP masses and nature is shown in Fig.~\ref{fig:NUDMGG_e}, where we show all surviving scenarios in the left panel. We now see no bino dominated dark matter at all. We do have wino dominated dark matter scenarios but when we restrict to the preferred relic density and $\Delta<10$ in the right panel, only higgsino dominated dark matter scenarios remain (as one might expect from Fig.~\ref{fig:NUmutanbGG_e}). 

The majority of the solutions have a chargino NLSPs though there are a few examples with a neutralino NLSP (empty diamonds). The mass of higgsino dominated neutralino and chargino is predominantly set by the $\mu$-parameter, while their mass splitting is set by $M_1$ and $M_2$. Since the $U(1)_Y$ gaugino mass term only contributes to the neutral components, the splitting between two neutralinos is typically larger. However, in regions where $M_1 \ll M_2$ (small $\rho_1$) the light neutralinos become degenerate. This provides for the empty diamonds in Fig.~\ref{fig:NUDMGG_e}, with $\rho_1$ constrained approximately to the interval $\left[-1.1,-2.3\right]$.

\begin{figure}[ht!]
\centering
\includegraphics[width=0.492\textwidth]{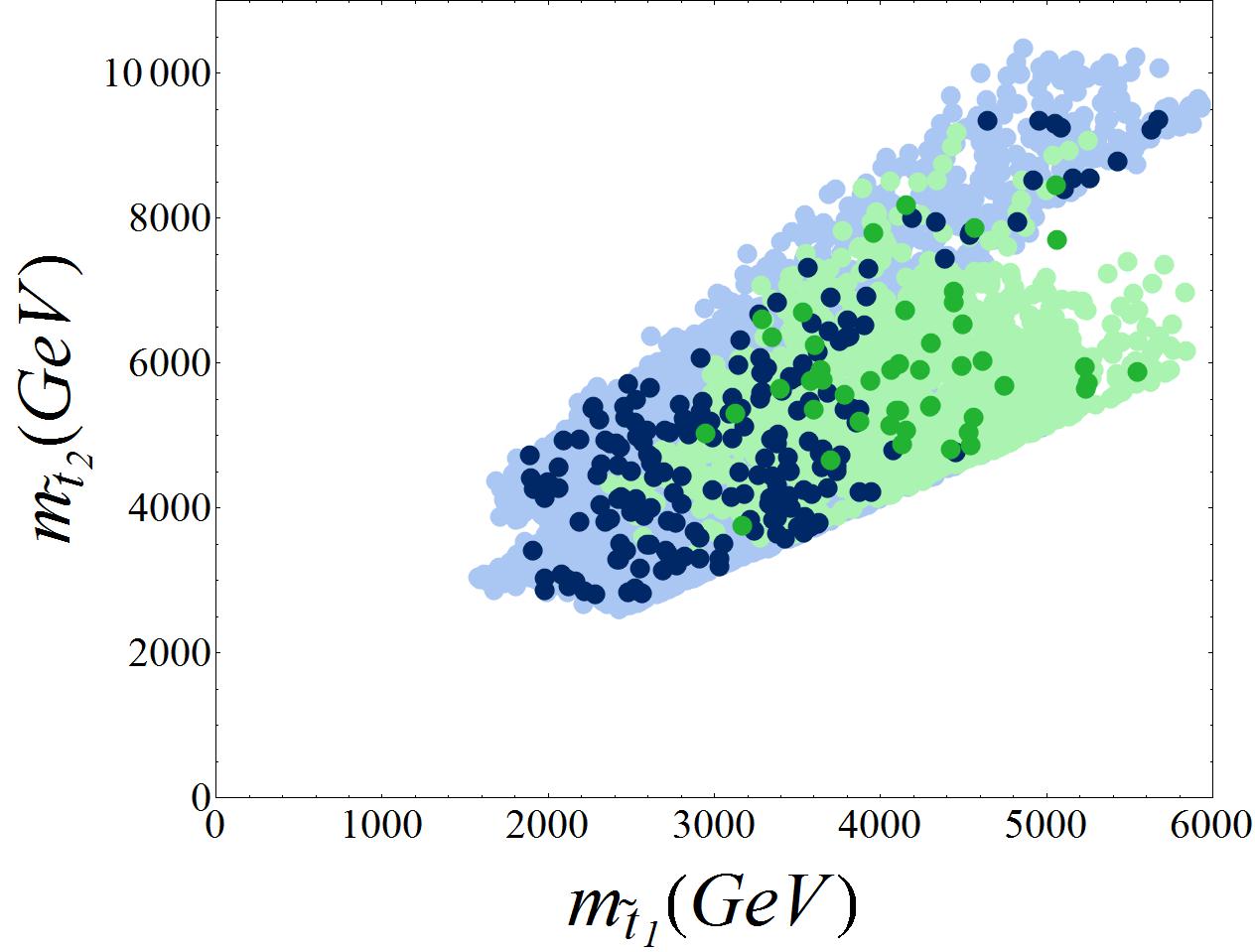}
\includegraphics[width=0.48\textwidth]{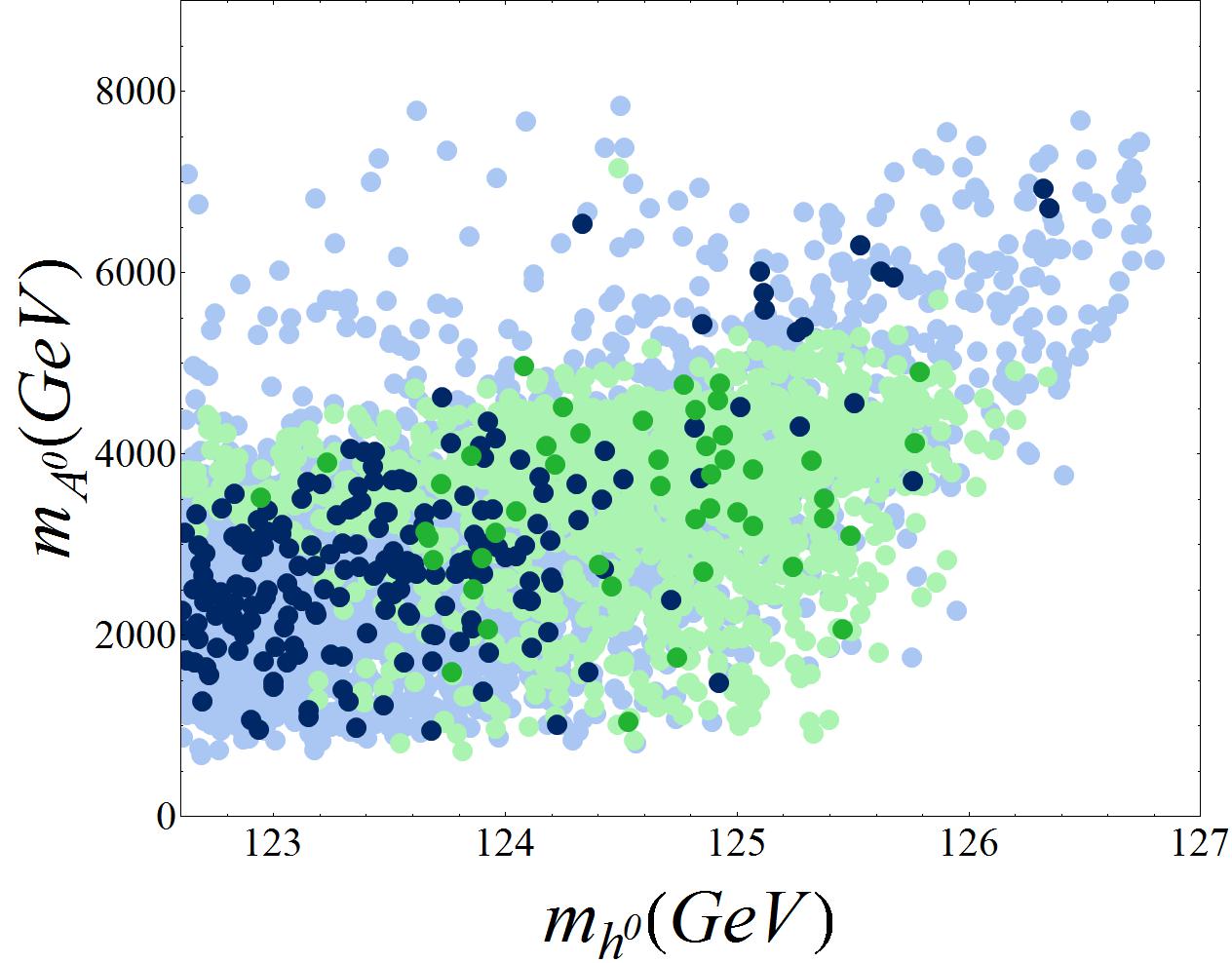} \\[4mm]
\includegraphics[width=0.465\textwidth]{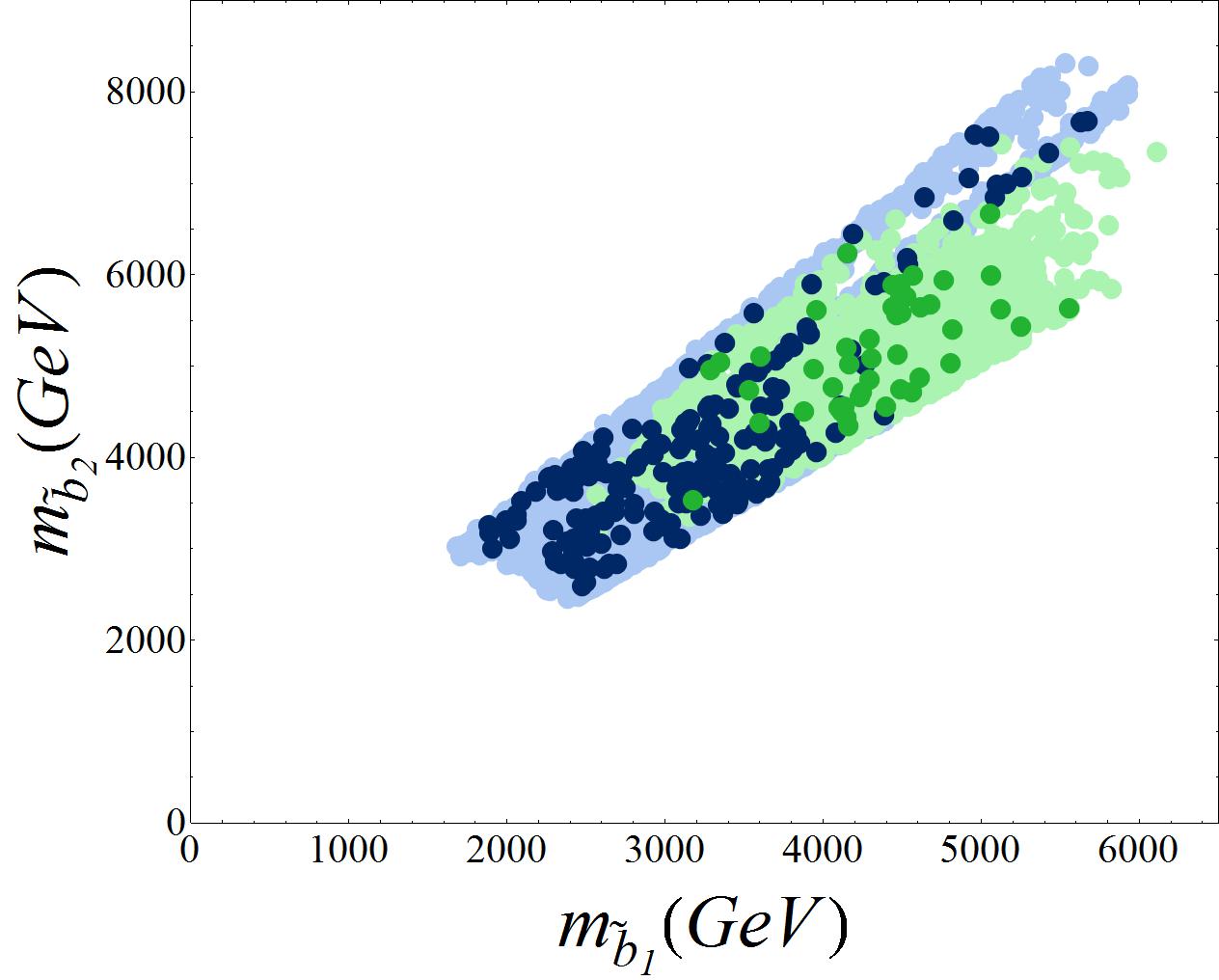}
\includegraphics[width=0.495\textwidth]{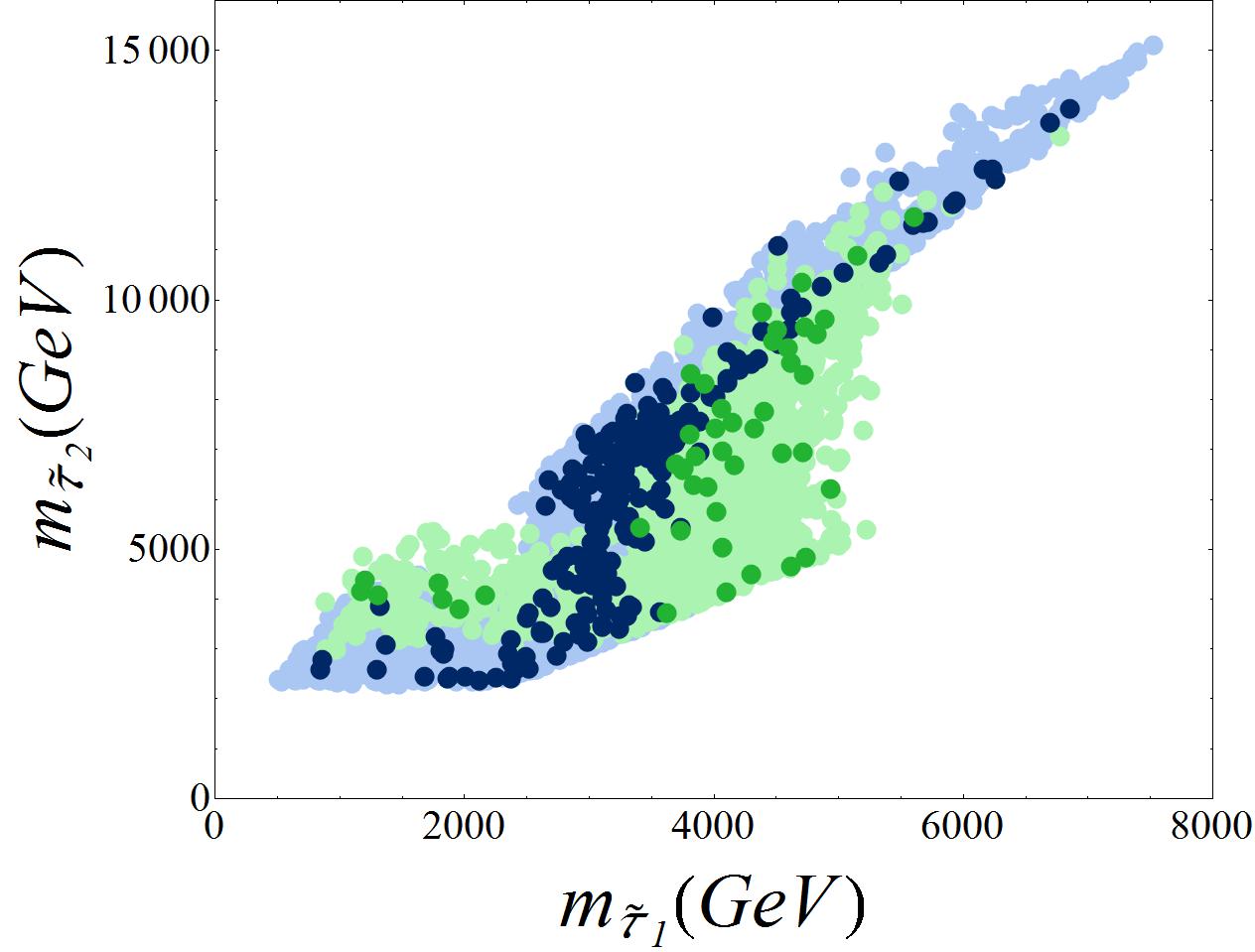}
\caption{\it Viable scenarios in the stop mass (top left), lightest scalar - pseudoscalar mass (top right) sbottom mass (bottom left) and stau mass (bottom right) planes for the enhanced scan with non-universal gaugino masses, with colours as in Fig.~\ref{fig:NUmutanbGG_e}.}
\label{fig:NUstophiggsGG_e}
\end{figure}

The masses of the top and bottom squarks, staus and Higgs bosons are shown in Fig.~\ref{fig:NUstophiggsGG_e}. The lightest top and bottom squarks are confined to $1.5$-$6\,{\rm TeV}$, lightest staus in the interval $0.5$-$6.0\,{\rm TeV}$ and the pseudoscalar Higgs boson mass can now vary over a wider region from $1$-$5\,{\rm TeV}$.

\begin{figure}[ht!]
\centering
\includegraphics[width=0.55\textwidth]{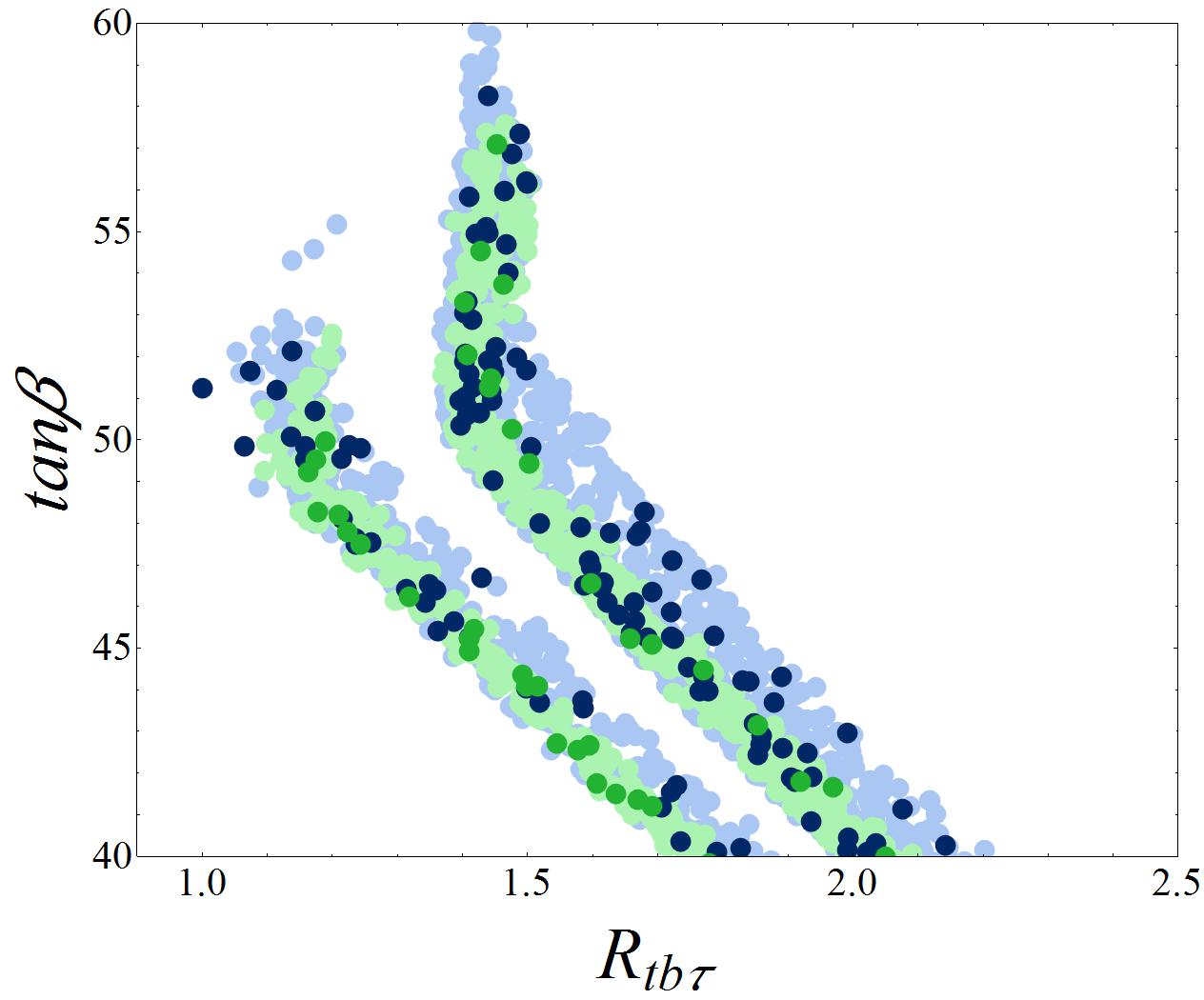}
\caption[\it Viable scenarios in the $R_{t b \tau}$-$\tan \beta$ plane for the enhanced scan with non-universal gaugino masses.]{\it  Viable scenarios in the $R_{t b \tau}$-$\tan \beta$ plane for the enhanced scan with non-universal gaugino masses, with colours as in Fig.~\ref{fig:NUmutanbGG_e}. The left (right) branch contains $\mu < 0$ ($\mu > 0$).}
\label{fig:NURtbtGG_e}
\end{figure}

It is also interesting to ask if these scenarios allow Yukawa coupling unification. We adopt the definition of Yukawa unification (YU) from \cite{Ajaib:2013uda}, and quantify how close to exact unification we are using the GUT scale ratio
\begin{eqnarray}
R_{t b \tau} = \frac{{\rm max}\left( y_t, y_b, y_{\tau} \right)}{{\rm min}\left( y_t, y_b, y_{\tau} \right)}.
\label{eq:Rtbt}
\end{eqnarray}
As discussed in Section~\ref{sec:int}, we will not require exact unification so do not throw away the scenarios with $R_{t b \tau} \neq 1$. Furthermore, as discussed in \cite{Antusch:2012gv, Antusch:2013rxa,Antusch:2014poa} (and references therein), GUT symmetry breaking may provide additional Clebsch-Gordan factors that alter the quark-lepton mass relations and provide new predictions for GUT scale Yukawa ratios. Particularly interesting is the prediction $y_{\tau}/y_{b} = 3/2$, calculated in the context of SU(5) and Pati-Salam unification. In Fig.~\ref{fig:NURtbtGG_e} we show solutions with $\tan \beta$ between $40$ and $60$, where the left branch corresponds to points with $\mu < 0$ and the right branch to $\mu > 0$. It is for negative values of $\mu$ that we get the best YU conditions with a point having $R_{t b \tau} = 1.07$ for $\tan \beta = 51.6$ (the dark blue point furthest to the left)\footnote{Threshold corrections to the Yukawa couplings at the GUT scale are dependent on the sign of $\mu$ as well as the sign of $M_{1,2,3}$ such that solutions with $\mu < 0$ may favour YU \cite{Ajaib:2013zha,Badziak:2013eda}.}. For the $\mu$ positive branch we do not find exact YU; the smallest ratio with the preferred relic density is $R_{t b \tau} = 1.37$ with $\tan \beta = 51.6$.  Refs.~\cite{Badziak:2011wm,Badziak:2011bn,Badziak:2012mm,Gogoladze:2011aa} find lower values of $R_{t b \tau}$ consistent with exact YU but here these are removed by our requirement $\Delta<100$.

\begin{figure}[ht!]
\centering
\includegraphics[width=\textwidth]{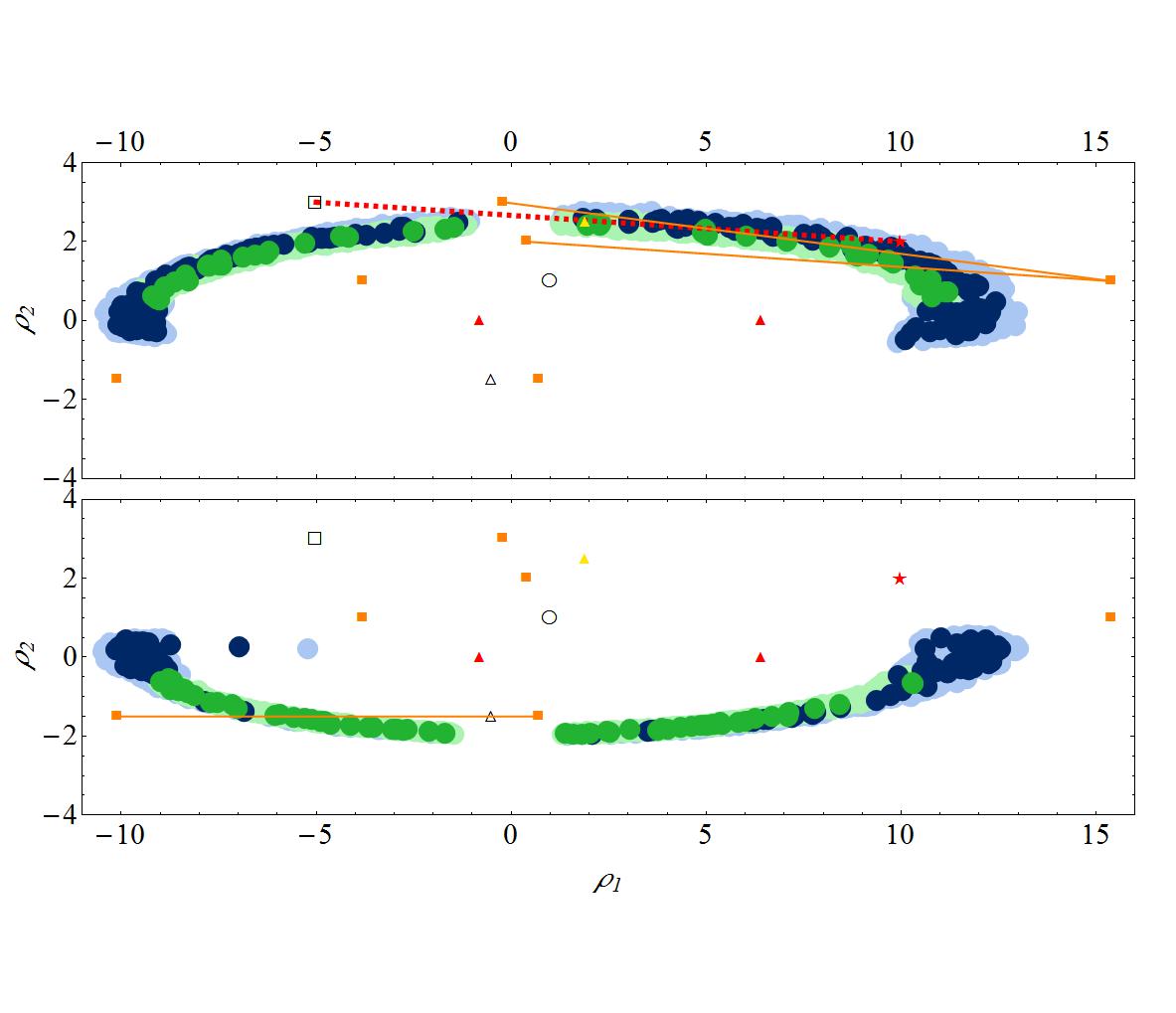}
\vspace*{-15mm} 
\caption[\it Viable scenarios in $\rho_1$-$\rho_2$ plane for the enhanced scan with non-universal gaugino masses in SO(10).]{\it Viable scenarios in $\rho_1$-$\rho_2$ plane for the enhanced scan with non-universal gaugino masses. Points with the preferred dark matter relic density are shown in green, while those with a relic density below the bounds are in blue. Darker and lighter shades denote the fine-tuning: darker shades have fine-tuning $\Delta<10$ while lighter shades have $10<\Delta<100$. The upper pane is for scenarios with $\mu>0$ while the lower pane is for $\mu<0$. The additional symbols represent particular gaugino mass ratios as predicted by the mechanisms described in tables \ref{table:GG}, \ref{table:FL} and \ref{table:PS}. Scenarios arising from embeddings in the $\mathbf{1}$, $\mathbf{54}$, $\mathbf{210}$, and $\mathbf{770}$ representations of SO(10) and transforming as a $\mathbf{1}$, $\mathbf{24}$, $\mathbf{75}$, and $\mathbf{200}$ of SU(5) with Georgi-Glashow embedding are shown by an empty circle, an empty triangle, an empty square and a red star respectively. Orange squares represent gaugino mass ratios when the proper maximal subgroup is SU(5) with flipped embedding. Red and yellow triangles coincide with ratios as predicted by transformations under $SU(4) \times SU(2)_R$. Scenarios with combinations of two representations lie along the straight lines: the $GG75+200$ lies on the red dashed line, whereas the $FL75+1$, $FL1+200$ and $FL24+24$ lie on the uppermost, middle and lowermost orange lines respectively.}
\label{fig:NUrhoGG_e}
\end{figure}

In Fig.~\ref{fig:NUrhoGG_e} we show the $\rho_{1,2}$ plane. We now see the surviving scenarios fall on an ellipse, reminiscent of that obtained in our SU(5) study~\cite{Miller:2013jra}. This ellipse is forced by our new fine-tuning requirement $\Delta<100$, in contrast to Fig.~\ref{fig:NUrhoGG}. The upper (lower) panel corresponds to $\mu>0$ ($\mu<0$) and the scenarios in the missing half of the ellipse are excluded due to the presence of a charged LSP. This ellipse is related to similar effects seen in \cite{Antusch:2012gv,Antusch:2013rxa,Abe:2007kf,Horton:2009ed,Kaminska:2013mya,Abe:2012xm}. We have also marked on this figure various gaugino mass ratios predicted by various breaking mechanisms that will be explained on the next section. For example, the model examined in \cite{Badziak:2011wm,Badziak:2011bn,Badziak:2012mm,Gogoladze:2011aa} is represented by an empty triangle in Fig.~\ref{fig:NUrhoGG_e} and is far from our ellipse.

\section{Scenarios with Fixed Gaugino Mass Ratios}
\label{sec:sfgmrSO10}

Although we have varied our gaugino masses in the above in order to form a complete scan over the parameter space, we implicitly assume that the gaugino mass ratios are fixed by some GUT or string inspired mechanism. One such possibility is the breaking of supersymmetry through a hidden sector field $\hat X$ in a representation (or combination of representations) belonging to the product \mbox{$\left(\mathbf{45} \times \mathbf{45}\right)_{symm} = \mathbf{1} + \mathbf{54} + \mathbf{210} + \mathbf{770}$}. Ref.~\cite{Martin:2009ad} presented the possible $M_1 : M_2 : M_3$ coefficients for the GG, FL and PS breaking routes, and in Tables \ref{table:GG}, \ref{table:FL} and \ref{table:PS} we summarise the fixed ratios that lie closest to our ellipse. The SU(5) route with GG embedding produces the same points that we have already studied in \cite{Miller:2013jra}. In particular, the $SU(5)_{\mathbf{200}}$ model is equivalent to a $\mathbf{770}$-$\hat X$ that transforms as a $\mathbf{200}$ under its maximal proper subgroup. 
\renewcommand{\arraystretch}{1.5}
\begin{table}[ht!]
\centering
\begin{tabular}{|>{\centering\arraybackslash}m{3.0cm}|>{\centering\arraybackslash}m{1.0cm}|>{\centering\arraybackslash}m{1.0cm}|>{\centering\arraybackslash}m{3.5cm}|}
\hline
\small{$SO(10) \rightarrow SU(5)$}&\small{$\rho_1$}&\small{$\rho_2$}&\small{Label in Fig.~\ref{fig:NUrhoGG_e}}\tabularnewline
\hline
\small{$\begin{aligned}  \mathbf{1} &\rightarrow \mathbf{1} \\  \mathbf{210}  &\rightarrow \mathbf{1} \\ \mathbf{770} &\rightarrow \mathbf{1}\end{aligned}$}  &  $1$ & $1$ &\small{empty circle}\tabularnewline
\hline
\small{$ \begin{aligned}  \mathbf{54} &\rightarrow \mathbf{24} \\  \mathbf{210}  &\rightarrow \mathbf{24} \\ \mathbf{770} &\rightarrow \mathbf{24}\end{aligned}$}  &  $ -\frac{1}{2}$ & $ -\frac{3}{2}$ &\small{empty triangle}\tabularnewline
\hline
\small{$ \begin{aligned}  \mathbf{210} &\rightarrow \mathbf{75} \\  \mathbf{770}  &\rightarrow \mathbf{75}\end{aligned}$}  &  $ -5$ & $3$ &\small{empty square}\tabularnewline
\hline
\small{$  \mathbf{770} \rightarrow \mathbf{200}$}  &  \small{$10$} & \small{$2$} &\small{red star}\tabularnewline
\hline
\end{tabular}
\caption[\it Fixed gaugino mass ratios for hidden sector chiral superfield $\hat X$ in representations of $SU(5) \subset SO(10)$ with the Georgi-Glashow embedding.]{\it Fixed gaugino mass ratios for hidden sector chiral superfield $\hat X$ in representations of $SU(5) \subset SO(10)$ with the Georgi-Glashow embedding.}
\label{table:GG}
\end{table}
\renewcommand{\arraystretch}{1.5}
\begin{table}[ht!]
\centering
\begin{tabular}{|>{\centering\arraybackslash}m{3.0cm}|>{\centering\arraybackslash}m{1.0cm}|>{\centering\arraybackslash}m{1.0cm}|}
\hline
\small{$SO(10) \rightarrow SU(5)^{\prime}$}&\small{$\rho_1$}&\small{$\rho_2$}\tabularnewline
\hline
\small{$\mathbf{210} \rightarrow \mathbf{1}$}  &  $ -\frac{19}{5}$ & $1$\tabularnewline
\hline
\small{$\mathbf{210} \rightarrow \mathbf{24}$}  &  $ \frac{7}{10}$ & $-\frac{3}{2}$\tabularnewline
\hline
\small{$ \begin{aligned}  \mathbf{210} &\rightarrow \mathbf{75} \\  \mathbf{770}  &\rightarrow \mathbf{75}\end{aligned}$}  &  $ -\frac{1}{5}$ & $3$\tabularnewline
\hline
\small{$  \mathbf{770} \rightarrow \mathbf{1}$}  &  $\frac{77}{5}$ & $1$ \tabularnewline
\hline
\small{$  \mathbf{770} \rightarrow \mathbf{24}$}  &  $-\frac{101}{10}$ & $-\frac{3}{2}$ \tabularnewline
\hline
\small{$  \mathbf{770} \rightarrow \mathbf{200}$}  &  $\frac{2}{5}$ & ${2}$ \tabularnewline
\hline
\end{tabular}
\caption[\it Fixed gaugino mass ratios for hidden sector chiral superfield $\hat X$ in representations of $SU(5) \subset SO(10)$ with the flipped embedding.]{\it Fixed gaugino mass ratios for hidden sector chiral superfield $\hat X$ in representations of $SU(5) \subset SO(10)$ with the flipped embedding. All the ratios in this table are labeled by filled orange squares in Fig.~\ref{fig:NUrhoGG_e}.}
\label{table:FL}
\end{table}
\renewcommand{\arraystretch}{1.5}
\begin{table}[ht!]
\centering
\begin{tabular}{|>{\centering\arraybackslash}m{4.0cm}|>{\centering\arraybackslash}m{1.0cm}|>{\centering\arraybackslash}m{1.0cm}|>{\centering\arraybackslash}m{3.5cm}|}
\hline
\small{$SO(10) \rightarrow SU(4) \times SU(2)_R$}&\small{$\rho_1$}&\small{$\rho_2$} & \small{Label in Fig.~\ref{fig:NUrhoGG_e}}\tabularnewline
\hline
\small{$\mathbf{210} \rightarrow \mathbf{(15,1)}$}  &  $ -\frac{4}{5}$ & $0$ &\small{red triangle}\tabularnewline
\hline
\small{$\mathbf{770} \rightarrow \mathbf{(1,1)}$}  &  $ \frac{19}{10}$ & $\frac{5}{2}$ &\small{yellow triangle}\tabularnewline
\hline
\small{$  \mathbf{770} \rightarrow \mathbf{(84,1)}$}  &  $\frac{32}{5}$ & ${0}$ &\small{red triangle}\tabularnewline
\hline
\end{tabular}
\caption[\it Fixed gaugino mass ratios for hidden sector chiral superfield $\hat X$ in representations of $SU(4) \times SU(2)_R \times SU(2)_L \subset SO(10)$ with the flipped embedding.]{\it Fixed gaugino mass ratios for hidden sector chiral superfield $\hat X$ in representations of \mbox{$SU(4) \times SU(2)_L \times SU(2)_R \subset SO(10)$}. All the ratios in this table are labeled by filled triangles in Fig.~\ref{fig:NUrhoGG_e}.}
\label{table:PS}
\end{table}

We note that there is only one additional viable model in SO(10) that is not already present in SU(5) - the yellow triangle on the ellipse with $\hat X$ transforming as a singlet under $SU(4) \times SU(2)_R$. However, we may extend our analysis to also allow mixing among the representations of $\hat X$, providing models that lie on lines joining the points. Let $\mathbf{R}$ and $\mathbf{R}^{\prime}$ be two of those irreps and $\theta_{RR'}$ a mixing angle. The gaugino masses at the input scale are
\begin{eqnarray}
M_1 &=& M_{1/2} (\rho^R_1 \cos \theta_{RR'} + \rho^{R'}_1 \sin \theta_{RR'}) \label{eq:M1RR}, \\
M_2 &=& M_{1/2} (\rho^R_2 \cos \theta_{RR'} + \rho^{R'}_2 \sin \theta_{RR'}) \label{eq:M2RR}, \\
M_3 &=& M_{1/2} (\cos \theta_{RR'} + \sin \theta_{RR'}) \label{eq:M3RR}, 
\end{eqnarray}
such that we recover the standard form when either $\theta_{RR'} = 0$ ($\hat X \in \mathbf{R}$), or $\theta_{RR'} = \pi/2$ ($\hat X \in \mathbf{R}^{\prime}$). In eqs.~(\ref{eq:M1RR} - \ref{eq:M2RR}), $\rho^{R,R'}_{1,2}$ are the usual gaugino mass ratios fixed by the representation $\mathbf{R}, \,\mathbf{R^{\prime}}$ and the transformation properties of the hidden sector fields under the maximal subgroups. Note that now $M_{3}$ is no longer $M_{1/2}$ at the GUT scale, unless the mixing angle is zero or $\pi/2$. The gauino mass ratios are now
\begin{eqnarray}
\rho_1 &=& \frac{M_1}{M_3} = \frac{ \rho^R_1 \cos \theta_{RR'} + \rho^{R'}_1 \sin \theta_{RR'} }{\cos \theta_{RR'} + \sin \theta_{RR'}} \label{eq:r1RR}, \\
\rho_2 &=& \frac{ M_2}{M_3} =  \frac{ \rho^R_2 \cos \theta_{RR'} + \rho^{R'}_2 \sin \theta_{RR'} }{\cos \theta_{RR'} + \sin \theta_{RR'}}\label{eq:r2RR}.
\end{eqnarray}
Alternatively, we may eliminate $\theta_{RR'}$ 
\begin{equation}
\tan \theta_{RR'} =   \frac{\rho_1 - \rho^R_{1}}{\rho^{R'}_1 - \rho_1}  \label{eq:thRR}, 
\end{equation}
to provide a general expression relating $\rho_{1}$ with $\rho_2$,
\begin{equation}
\rho_{2} =\frac{\left( \rho^{R'}_2 - \rho^R_2 \right) \rho_1 + \rho^R_2 \rho^{R'}_1 - \rho^R_1 \rho^{R'}_2}{\rho^{R'}_1 - \rho^R_1} \label{eq:r12},
\end{equation}
which defines a line in the $\rho_1 \-- \rho_2$ plane with end points corresponding to the original fixed ratios $\rho^{R,R'}_{1,2}$.

As we are interested in models with low fine tuning, we give preference to scenarios that best overlap with the ellipse. These models are:

\begin{enumerate}
\item The hidden sector fields $\hat X$ are embedded in the combinations $\mathbf{R} + \mathbf{R^{\prime}} = \mathbf{210} + \mathbf{770}$ or $\mathbf{770} + \mathbf{770^{\prime}}$ transforming as $\mathbf{75} + \mathbf{200}$ under SU(5) with GG embedding. The respective fixed ratios to provide the end points are in Table \ref{table:GG} (the last two rows) giving
\begin{equation}
\rho_2 = -\frac{1}{15} \rho_1 + \frac{8}{3}.
\label{eq:GG75+200}
\end{equation} 
This model is identified by the red dashed line in Fig.~\ref{fig:NUrhoGG_e} and we refer to it as $GG75+200$.

\item We also consider the same combinations of SO(10) irreps as in 1.~but transform them as $\mathbf{75} + \mathbf{1}$ of SU(5) with flipped embedding. The endpoints are in Table \ref{table:FL} (third and fourth rows), giving the line
\begin{equation}
\rho_2 = -\frac{5}{39} \rho_1 + \frac{116}{39},
\label{eq:FL75+1}
\end{equation} 
the upper orange line in the upper pane of Fig.~\ref{fig:NUrhoGG_e}. We refer to this as $FL75+1$.

\item The third model mixes two $770$-dimensional irreps,  $\mathbf{R} + \mathbf{R^{\prime}} = \mathbf{770} + \mathbf{770^{\prime}}$ which transform as $\mathbf{1} + \mathbf{200}$ under flipped SU(5). The gaugino mass ratios are related by
\begin{equation}
\rho_2 = -\frac{1}{15} \rho_1 + \frac{152}{75},
\label{eq:FL1+200}
\end{equation} 
which corresponds to the lower orange line in the upper pane of Fig.~\ref{fig:NUrhoGG_e}. We name this model $FL1+200$. 

\item The fourth model contains fields in \mbox{$\mathbf{R} + \mathbf{R^{\prime}} = \mathbf{210} + \mathbf{770}$}, which under SU(5) transforms as  $\mathbf{24} + \mathbf{24^{\prime}}$ with flipped embedding. The gaugino mass ratios fix
\begin{equation}
\rho_2 = -\frac{3}{2},
\label{eq:FL24+24}
\end{equation} 
while leaving $\rho_1$ unconstrained. This is the orange line in the lower pane of Fig.~\ref{fig:NUrhoGG_e}. This model is denoted $FL24+24$.

\item Finally, we also consider an interesting model without any mixing corresponding to a single $\mathbf{770}$ which transforms as a $\mathbf{(1,1)}$ under the Pati-Salam maximal subgroup. This is identified by the yellow triangle in the upper pane of Fig.~\ref{fig:NUrhoGG_e} corresponding to $\rho_1 = 19/10$ and $\rho_2 = 5/2$. We denote this model $PS$. 
\end{enumerate}

We note that by allowing mixed representations, we have enlarged the parameter space to include the mixing angle $\theta_{RR'}$ and potentially have a new source of fine-tuning. Indeed, since the low energy value of $m^2_{H_u}$ depends on the gaugino masses, and therefore $\theta_{RR'}$, this additional fine-tuning destroys the vast majority of the scenarios, giving $\Delta$ of several hundreds or even thousands. However, this dependence is parabolic so there is always a point along the line where $m^2_{H_u}$ reaches a minimum and is insensitive to fluctuations in $\theta_{RR'}$. For the $FL75+1$ model, if we set $\tan \beta = 31.9$, $M_{1/2} = 2456.8~\rm{GeV}$, $a_{10} = -2.3~\rm{GeV}$, $m_{16} = 49.3~\rm{GeV}$, $m_{10} = 75.9~\rm{GeV}$, $\sqrt{g^2_{10}D} = 75.9~\rm{GeV}$ and $K_{16} = 12.2$, we find a minimum for $\theta_{RR'} \approx 0.65$.

As we observed in Section~\ref{subsec:IN}, in order to have low fine-tuning with respect to $M_{1/2}$ we also need to to sit close to a minimum of $m^2_{H_u}$, and indeed the solutions that we found on the $\rho_1 \-- \rho_2$ ellipse manifest this behaviour. Now for scenarios with mixed representations we need to have a minimum of $m^2_{H_u}$ with respect to $M_{1/2}$ and $\theta_{RR'}$ simultaneously. For most choices they do not coincide and the scenario becomes fine-tuned, but sometimes these minima are rather close and fine-tuning is small. This is indeed the case for the $FL75+1$ point described above; when $\Delta_{\theta_{RR'}} \approx 0$ we have $\Delta_{M_{1/2}} = 5$.  

However, one should be careful interpreting this fine-tuning and remember that this behaviour results from a two-loop evolution of $m^2_{H_u}$ using the $\overline{\rm{DR}}$ scheme in \mbox{SOFTSUSY 3.3.0.} One might expect these minima to shift somewhat with any change of treatment, such as inclusion of higher orders, a change in renormalisation prescription, or an different implementation of threshold effects. 

To examine this we attempted to reproduce the same behaviour using SPheno 3.2.4 \cite{Porod:2003um}. We first reproduced the supersymmetric particle spectrum for a variety of representative points with low fine-tuning, and found that both SOFTSUSY and SPheno are in agreement to within about $1.0~\rm{to}~3.5\%$. We also performed $1\%$ shifts in $M_{3}$ keeping $\rho_1$ and $\rho_2$ (and the soft parameters) fixed and observed fluctuations in this spectrum of less than $1\%$, suggesting that the scenarios are indeed stable. To investigate the stability of $M_Z$, we varied $M_{3}$ by hand (since SPheno contains no fine-tuning algorithm) and examined the behaviour of $m_{H_u}^2$. Unfortunately we found a fine-tuning of about $10^3$ indicating that fine-tuning behaviour is sensitive to the details of the calculation.  If we allow a small shift in $\rho_1$ and/or $\rho_2$ we can always again find a point where $m_{H_u}^2$ is a minimum with respect to $M_3$, indicating that the ellipse of stability is slightly shifted in comparison to SOFTSUSY due to the theoretical uncertainties. Indeed this effect may lead to some points on the $\rho_1-\rho_2$ plane moving on or off the ellipse. The scenarios examined in Refs.~\cite{Gogoladze:2011aa,Gogoladze:2013wva}, which were found to have low fine-tuning (``natural'' supersymmetry) when using ISAJET~\cite{Paige:2003mg}, are represented by the empty and filled squares to the top of Fig.~\ref{fig:NUrhoGG_e}. This is an effect that we will not study further in this paper, but note that some gaugino mass scenarios that have good fine-tuning properties in SOFTSUSY may not have such good behaviour in SPheno or ISAJET, and vice versa. Consequently we will abandon scenarios with mixed representations since the choice of a particular $\rho_1$ and $\rho_2$ become unmotivated. 

\subsection{The $\mathbf{PS}$ Model}
\label{sec:ps1}

We now examine the $PS$ model and investigate how this may be restricted by low energy constraints. For this model, the gaugino mass ratios are entirely determined by a single representation of the  $\hat X$ superfields. This is analogous to the $SU(5)_{200}$ model studied in \cite{Miller:2013jra}, but now $\rho_1 = 19/10$ and $\rho_2 = 5/2$. The scan is performed using the same range for the input parameters as in Section~\ref{subsec:EN}. 

\begin{figure}[ht]
\centering
\includegraphics[width=0.48\textwidth]{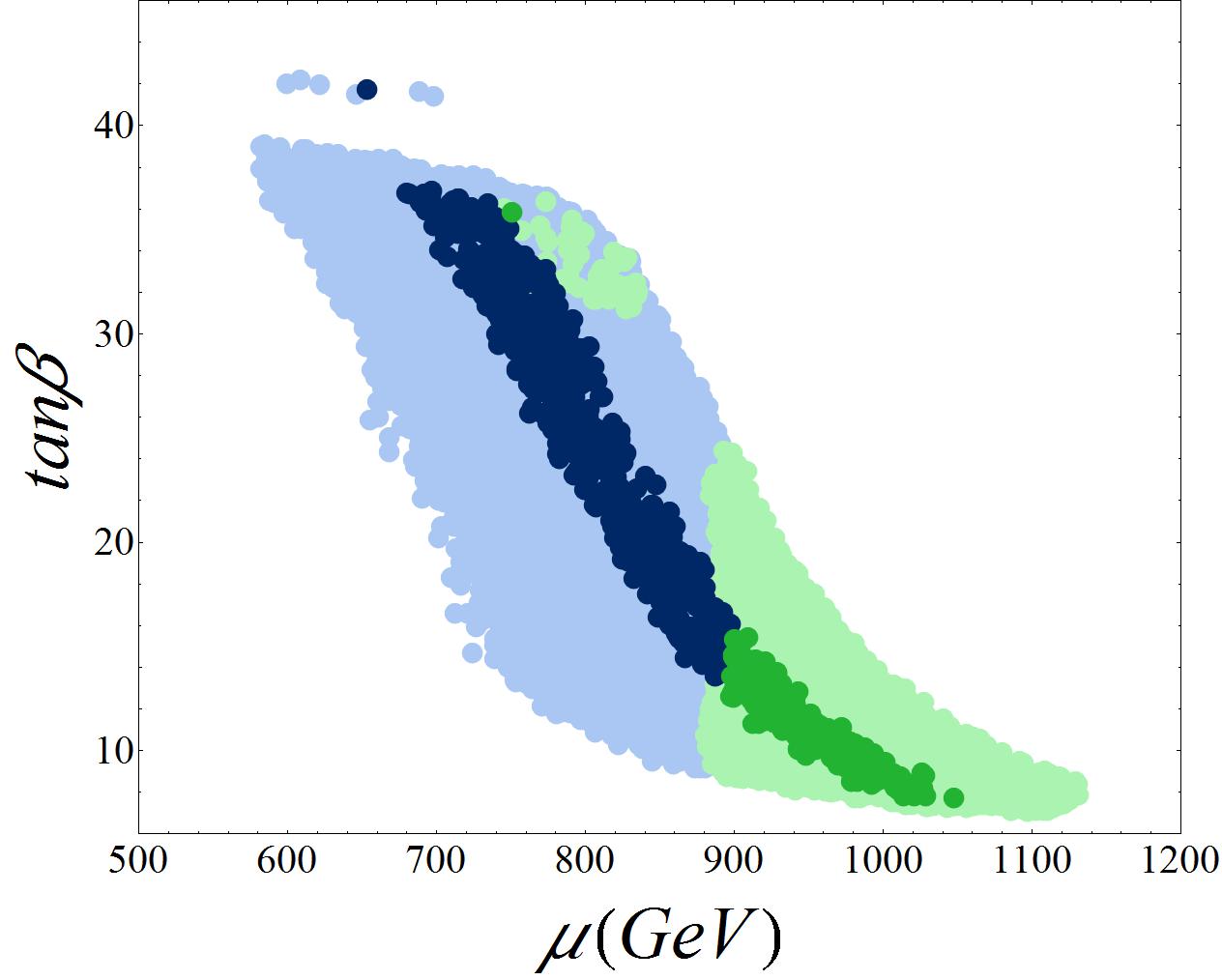}
\caption[\it Viable scenarios in the $\mu \-- \tan \beta$ and $R_{t b \tau} \-- \tan \beta$ planes for the $FF75+1$ model.]{\it Viable scenarios in the $\mu \-- \tan \beta$ (left) and $R_{t b \tau} \-- \tan \beta$ (right) planes for the $PS1$ model, with colours as in Fig.~\ref{fig:NUmutanbGG_e}.}
\label{fig:PS1mutanb}
\end{figure}

We first note in Fig.~\ref{fig:PS1mutanb} that $\tan \beta$ takes values from $7$ up to $40$, with some scattered solutions at $42$. This range becomes slightly restricted, $8 \--38$, if we insist $\Delta < 10 $. The viable scenarios have moderate to large values of $\mu$, $0.6 \-- 1.1~\rm{TeV}$, with the preferred relic density when $\mu$ is around a TeV. We also see a region when the correct relic abundance for $\tan \beta$ around $35$ and $\mu$ between $750\--850~\rm{GeV}$. For all of these points, the LSP is predominantly a higgsino and the NLSP is (higgsino-dominated) chargino, just $2\--4.5~\rm{GeV}$ heavier. Despite only one solution with large $\tan \beta$ and $\Delta < 10$ (the isolated dark green point), we observe several other light green points close to the dark blue band, where the fine-tuning is still not large. The points between the two green areas are also only just below the $2\sigma$ dark matter bounds. 

\begin{figure}[t!]
\centering
\includegraphics[width=0.48\textwidth]{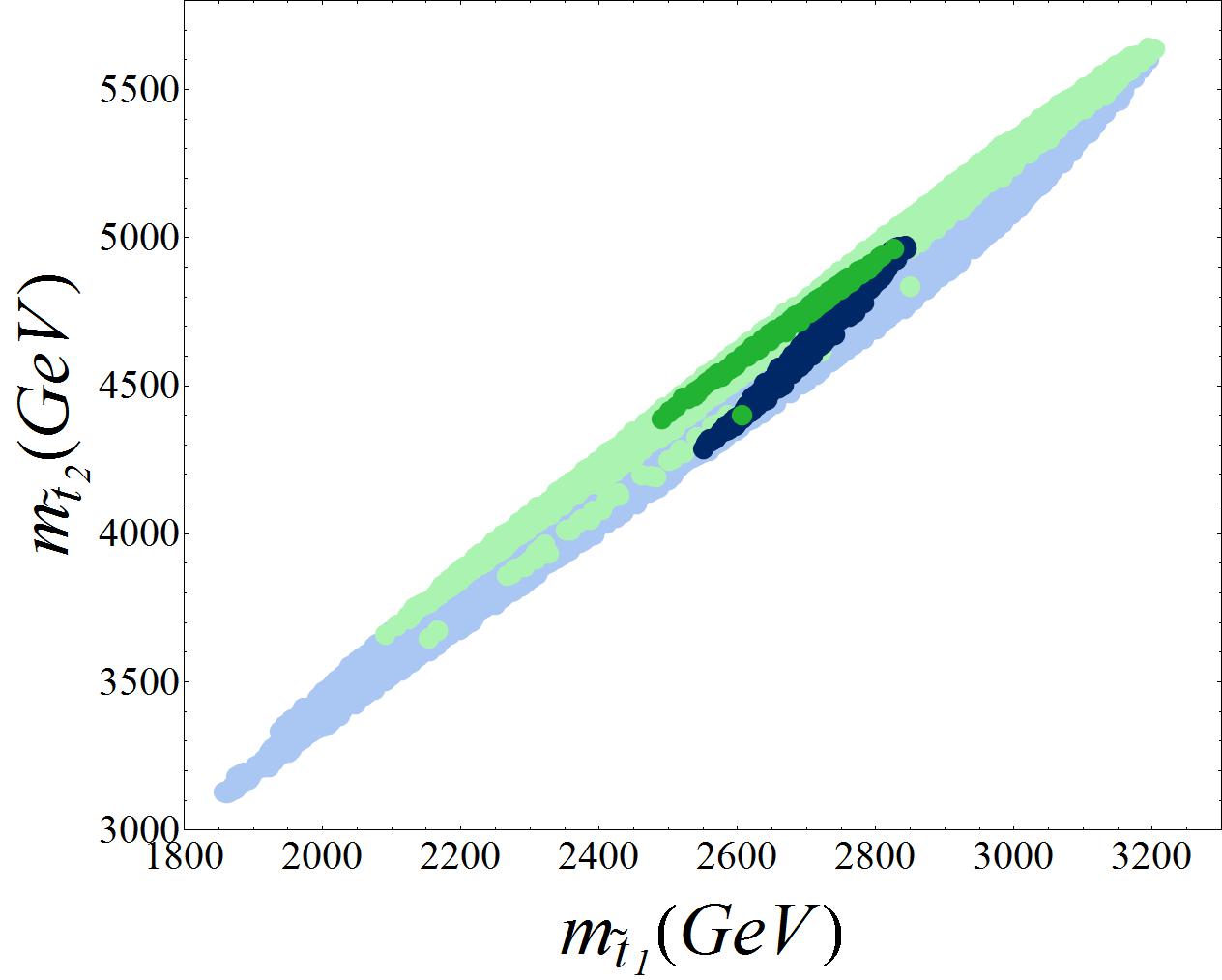}
\includegraphics[width=0.50\textwidth]{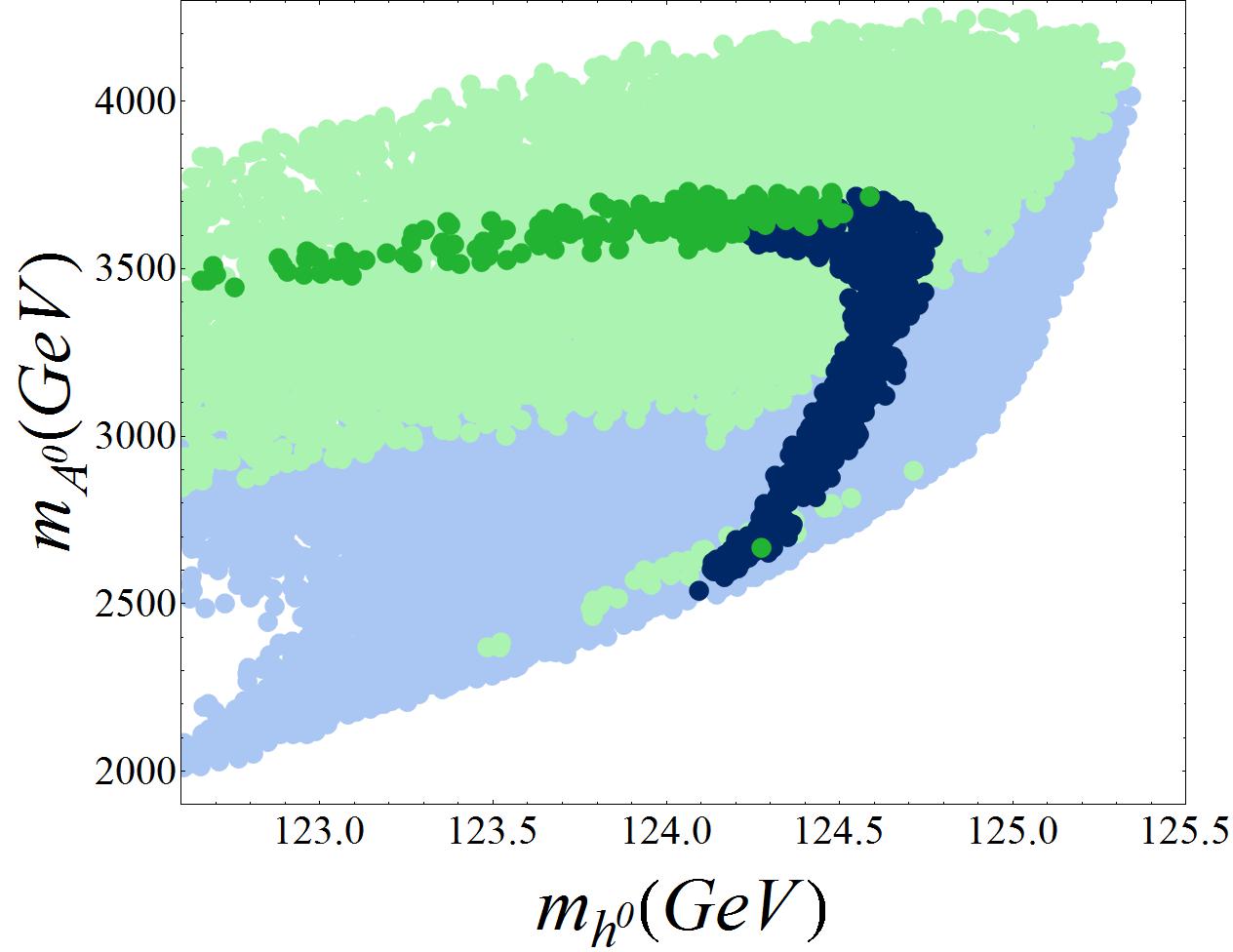}
\includegraphics[width=0.49\textwidth]{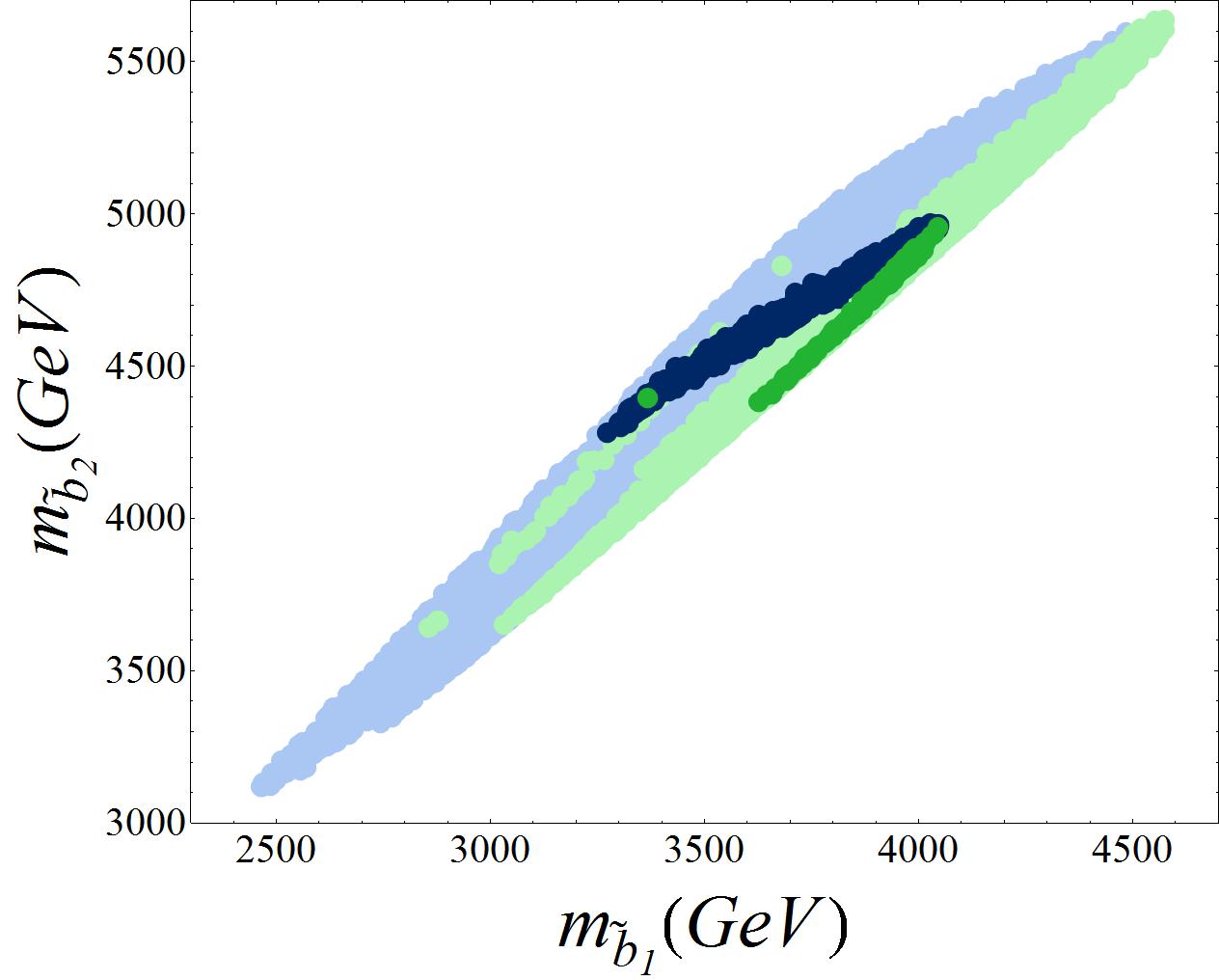}
\includegraphics[width=0.49\textwidth]{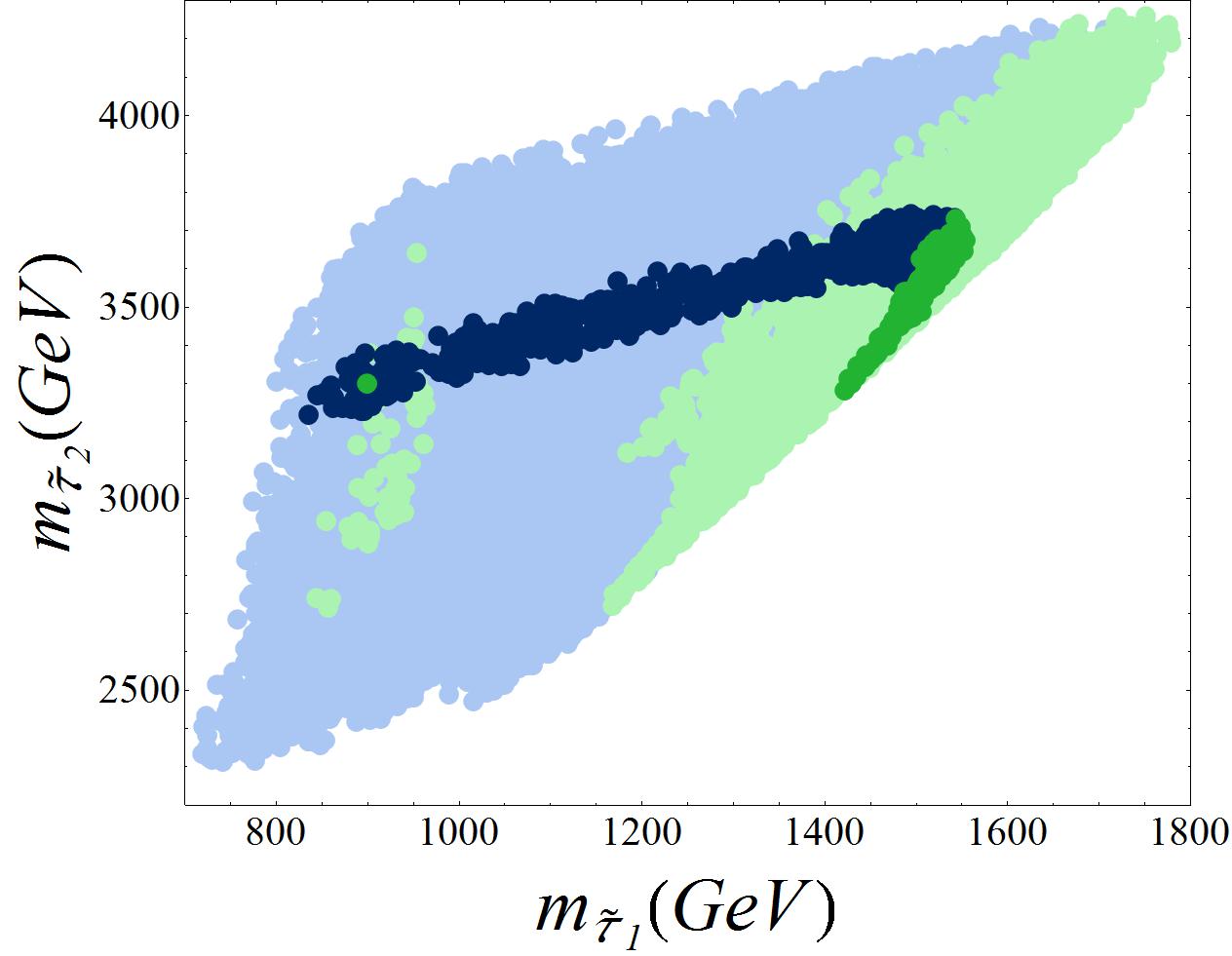}
\caption{\it Viable scenarios in the stop mass (top left), lightest scalar - pseudoscalar mass (top right), sbottom mass (bottom left) and stau mass (bottom right) planes for the $PS1$ model, with colours as in Fig.~\ref{fig:NUmutanbGG_e}.}
\label{fig:PS1stophiggs}
\end{figure}

The stop, sbottom, stau and Higgs masses are shown in Fig.~\ref{fig:PS1stophiggs}, where we see stops, sbottoms and staus above $1.8~\rm{TeV}$, $2.5~\rm{TeV}$  and $750~\rm{GeV}$ respectively. However, these lowest mass solutions have fine-tuning $10 < \Delta < 100$ and predict too little dark matter density. Insisting on low fine-tuning and the correct relic density resigns us to a much heavier spectrum. It is interesting to note that for the higher $\tan \beta$ region we also observe relatively light staus when $M_1$ is comparatively small; the tau-Yukawa contribution to the RGE for right-handed stau dominates $M_1$ for large $\tan \beta$, resulting in a light mainly right-handed $\tilde{\tau}_1$. The other stau, mainly left-handed, remains heavy due to $M_2$.

\begin{figure}[ht!]
\centering
\includegraphics[width=0.48\textwidth]{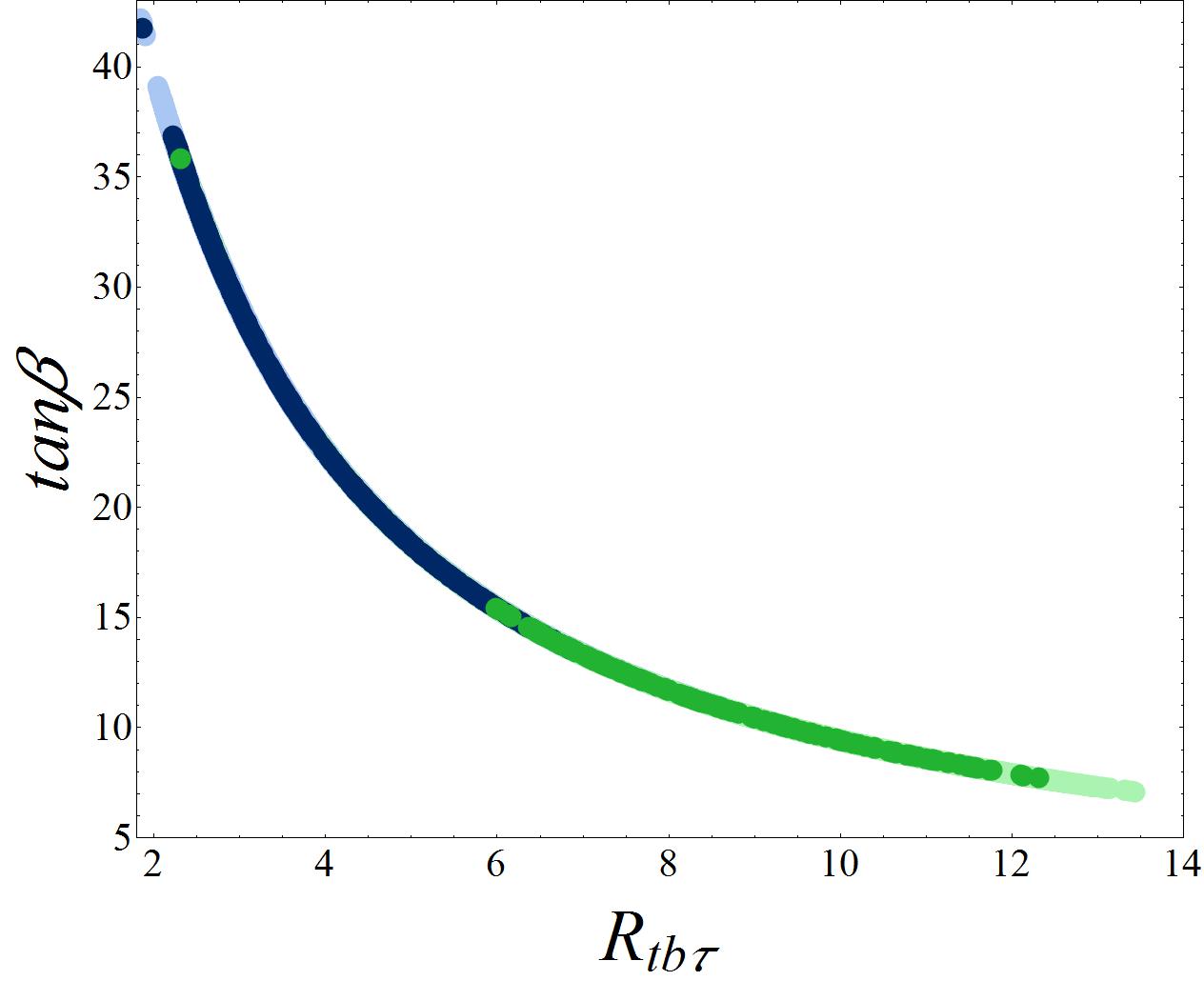}
\includegraphics[width=0.48\textwidth]{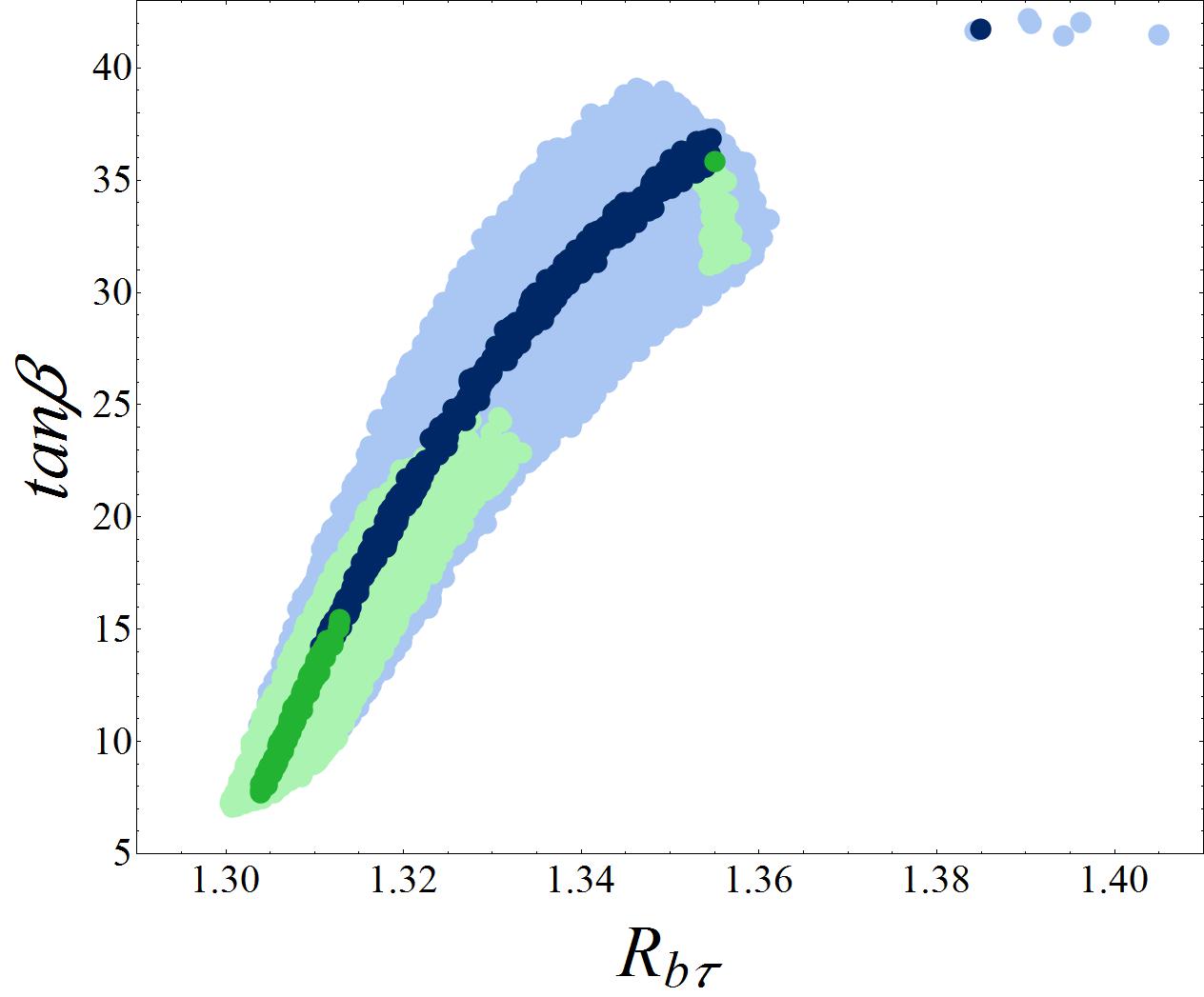}
\caption{\it Viable scenarios in the $R_{t b \tau} \-- \tan \beta$ (left) and  $R_{b \tau} \-- \tan \beta$ (right) planes for the $PS1$ model, with colours as in Fig.~\ref{fig:NUmutanbGG_e}.}
\label{fig:PS1R}
\end{figure}

The values of $\tan \beta$ here are not sufficiently large to reach the region preferred for Yukawa coupling unification, as seen in Fig.~\ref{fig:PS1R}. Therefore the $PS$ model is not a good candidate for exact top-bottom-tau unification, and can only provide quasi-unification. Also, the bottom-tau ratio, $1.30 < R_{b \tau} < 1.41$, is much closer to $3/2$ than an exact $y_b = y_{\tau}$ unification.

\begin{figure}[h!]
\centering
\includegraphics[width=0.48\textwidth]{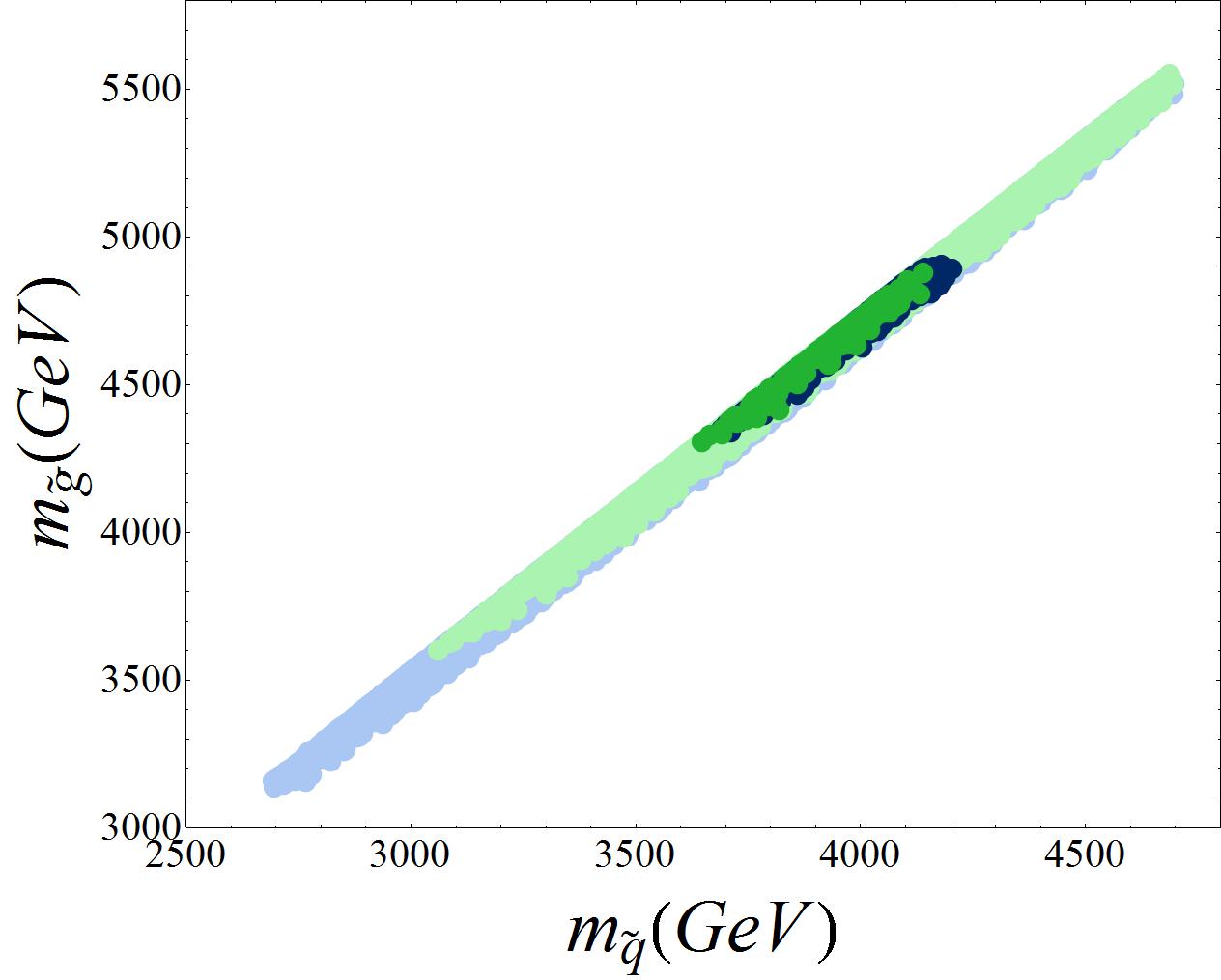}
\caption{\it The lightest squark mass and the gluino mass for $PS1$ model, with colours as in Fig.~\ref{fig:NUmutanbGG_e}.}
\label{fig:so10gluino}
\end{figure}

The gluino and the lightest first and second generation squark masses are shown in Fig.~\ref{fig:so10gluino}, where we see a correlation between the gluino and the lighest squark masses, reminiscent of that for SU(5) models seen in \cite{Miller:2013jra}. Using the boundary condition (\ref{eq:5scalarBCGG}), the $\tilde{d}_R$ squark mass, typically the lightest, takes the approximate form (see Refs~\cite{Miller:2012vn,Ananthanarayan:2003ca,Ananthanarayan:2004dm,Ananthanarayan:2007fj} for deeper discussions on the first and second generation squark masses)
\begin{equation}
m^2_{\tilde d_R}(t) = K_{\mathbf{16}} \left( m^2_{\mathbf{16}} - 3 g^2_{10}D \right)
+ M_3^2 (t) \left[ 0.78 + 0.002 \,\rho_1^2 \right],
\label{eq:approxdr}
\end{equation}
where, we have ignored all two loop contributions. Since we keep $m_{\tilde{d}_R} (0)$ small, the dominant contribution arises from the gluino mass term. For the $PS1$, \mbox{$\rho_1 = 1.9$}, so \mbox{$m_{\tilde d_R} \approx 0.9 m_{\tilde{g}}$}. 

As for the $SU(5)_{200}$ and the $O$-$II$ models studied in \cite{Miller:2013jra}, we find squarks and gluinos accessible to the $14~\rm{TeV}$ LHC. However, these solutions predict a relic density below the preferred bounds and have $10 < \Delta < 100$. Although it is possible to satisfy both the low fine-tuning and preferred relic density requirements simultaneously, the lightest squarks would possibly escape the region reachable by the LHC, with masses around $3.6 \-- 4.1 ~\rm{TeV}$.

We present two example benchmarks for viable PS scenarios with non-universal masses that may be interesting to consider at either the $14\,{\rm TeV}$ LHC or the energy-upgraded Super-LHC with \mbox{$\sqrt{s} = 28\,{\rm TeV}$}. These two benchmarks $PS_1$ and $PS_2$, predict relatively light staus as a consequence of the smallness of $\rho_1$. In particular, $PS_1$ corresponds to the isolated dark green point in Figs.~\ref{fig:PS1mutanb} and \ref{fig:PS1stophiggs}.

The GUT scale parameters can be found in Tab.~\ref{tab:InputsSO10}. The masses of the first and third generation sfermions, as well as the gauginos, are shown in Tab.~\ref{tab:1st3rdSO10}. The second generation sfermions are assumed degenerate with the first. Finally in Tab.~\ref{tab:parsSO10} we present the Higgs masses together with $\mu$, $\tan \beta$, the Yukawa coupling ratios $R_{t b \tau}$ and $R_{b \tau}$, the fine-tuning $\Delta$, the fine-tuning from $\mu$ alone and the predicted relic density of dark matter. In both scenarios, the LSP is predominantly dominated by the Higgsino component.  

\begin{table}[h!]
\centering
\begin{tabular}{ccc}
 & $\rm{PS}_1$ & $\rm{PS}_2$ \\
\hline \\[-4mm]
$m_{\mathbf{16}}$  & 147.5 & 113.8 \\
$K_{\mathbf{16}}$  & 12.79 & 12.3 \\
$m_{\mathbf{10+126}}$  & 130.6 & 132.5 \\
$g^2_{10} D$  & 4065 & -6674 \\
$a_{\mathbf{10}}$ & -38.43 & -116.7 \\
$ M_{1/2}$ & 2105 & 2471 \\
$ \rho_1 $ & 1.90 & 1.90 \\
$\rho_2$ & 2.50 & 2.50 \\
$\theta_{RR'}$ & 0 & 0 \\
\hline
\end{tabular}
     \caption{\it GUT scale parameters for our two scenarios. Masses and trilinear couplings are in GeV. }\label{tab:InputsSO10}
\end{table}

\begin{table}[h!]
\centering
\begin{tabular}{ccc}
 & $\rm{PS}_1$ & $\rm{PS}_2$ \tabularnewline
\hline \\[-4mm]
$m_{\tilde{u}_L}$   & 4997 & 5785 \tabularnewline
$m_{\tilde{u}_R}$  & 3898 & 4481 \tabularnewline
$m_{\tilde{d}_L}$  & 4998 & 5786 \tabularnewline
$m_{\tilde{d}_R}$  & 3786 & 4417 \tabularnewline
$m_{\tilde{e}_L}$  & 3424 & 4036 \tabularnewline
$m_{\tilde{e}_R}$  & 1594 & 1765 \tabularnewline
$m_{\tilde{\nu}^1}$ & 3423 & 4035 \tabularnewline
\hline
\end{tabular} ~~~~~
\begin{tabular}{ccc}
 & $\rm{PS}_1$ & $\rm{PS}_2$ \tabularnewline
\hline \\[-4mm]
$m_{\tilde{t}_1}$   & 2606 & 2987  \tabularnewline
$m_{\tilde{t}_2}$  & 4401 & 5243 \tabularnewline
$m_{\tilde{b}_1}$  & 3366 & 4240 \tabularnewline
$m_{\tilde{b}_2}$  & 4396 & 5239 \tabularnewline
$m_{\tilde{\tau}_1}$  & 900.0 & 1577 \tabularnewline
$m_{\tilde{\tau}_2}$  & 3302 & 3955 \tabularnewline
$m_{\tilde{\nu}^3}$   & 3300 & 3954 \tabularnewline
\hline  
\end{tabular} ~~~~~
\begin{tabular}{ccc}
 & $\rm{PS}_1$ & $\rm{PS}_2$ \tabularnewline
\hline \\[-4mm]
$M_{\tilde{g}}$   & 4450 & 5175 \tabularnewline
$M_{\tilde{\chi}^0_1}$  & 794.8 & 949.4 \tabularnewline
$M_{\tilde{\chi}^0_2}$  & 798.0 & 952.2 \tabularnewline
$M_{\tilde{\chi}^0_3}$  & 1740 & 2050 \tabularnewline
$M_{\tilde{\chi}^0_4}$  & 4288 & 5040 \tabularnewline
$M_{\tilde{\chi}^{\pm}_1}$  & 796.9 & 951.3 \tabularnewline
$M_{\tilde{\chi}^{\pm}_2}$  & 4288 & 5040 \tabularnewline
\hline
\end{tabular}
     \caption{\it First and third generation sfermion masses (we assume the first and second generation sfermions are degenerate), and Gaugino masses for the two scenarios. All masses are in GeV.}
\label{tab:1st3rdSO10}
\end{table}

\begin{table}[h!]
\centering
\begin{tabular}{ccc}
 & $\rm{PS}_1$ & $\rm{PS}_2$ \tabularnewline
\hline \\[-4mm]
$m_{h^0}$  & 124.3 & 125.0 \tabularnewline
$m_{A^0}$  & 2667 & 3842 \tabularnewline
$m_{H^0}$  & 2667 & 3842 \tabularnewline
$m_{H^{\pm}}$  & 2668 & 3843 \tabularnewline
$\mu$   & 751.1 & 907.5 \tabularnewline
$\tan \beta$  & 35.83 & 19.13 \tabularnewline
\hline
\end{tabular} ~~~~~
\begin{tabular}{ccc}
 & $\rm{PS}_1$ & $\rm{PS}_2$ \tabularnewline
\hline \\[-4mm]
$R_{t b \tau}$  & 2.31 & 4.76 \tabularnewline
$R_{b \tau}$  & 1.36 & 1.32 \tabularnewline
$\Delta$  & 9.83 & 33.62 \tabularnewline
$\Delta_{\mu}$  & 302.3 & 453.5 \tabularnewline
$\Omega_c h^2$  & 0.0944 & 0.0934 \tabularnewline
\hline
\phantom{xx} & \phantom{yy} & \phantom{zz} \\
\end{tabular}
\caption{\it Higgs masses, $\mu$ (all in GeV) and $\tan \beta$ for the two scenarios. $tb\tau$ and $b\tau$ unification ratios, the fine-tuning $\Delta$ (not including $\mu$), fine-tuning from $\mu$ alone, and the predicted relic density are also shown.}
\label{tab:parsSO10}
\end{table}

\section{Discussion and Conclusion}\label{sec:conc}

\hspace{5mm} We have investigated the low energy spectrum of Grand Unification with SO(10) boundary conditions considering both universal and non-universal gaugino masses, using SOFTSUSY. We confronted our results with low energy measurements such as the Higgs boson mass, $b \to s \gamma$, $B_S \to \mu^+ \mu^-$, $B \to \tau \nu_\tau$, as well as $g-2$ of the muon. Such scenarios are also consistent with the so far negative searches for supersymmetry at the LHC and the LUX direct dark matter searches. We also insist in scenarios with a stable vacuum at low energies, as well as a dark matter relic density within or below the experimental bounds of the WMAP and Plank satellites.

For both the universal and non-universal gaugino masses, phenomenologically viable scenarios suffer from considerable fine-tuning, in part due to their high value of $\mu$. Since the fine-tuning in $\mu$ seems unavoidable, we instead look for scenarios that minimise the fine-tuning from the soft parameters. [We stress again that fine-tuning in $\mu$ remains an unsolved problem for these scenarios.] We saw that setting small values of $m_{16}$, $m_{10+126}$, $a_{10}$ and $g^2_{10} D$ at the GUT scale reduces their individual tunings, leaving only fine-tuning from $M_{1/2}$. We therefore preformed a dedicated scan with small GUT scale scalar masses, trilinear couplings and D-term splittings, allowing the first two to become sizeable at the electroweak scale due to the contribution of $M_{1/2}$ in the renormalization group flow. The high scale $|M_{1/2}|$ is set to beyond a TeV.

Several scenarios with low fine-tuning, $\Delta < 10$, emerge from this scan, lying on an ellipse in the $\rho_1$-$\rho_2$  (or equivalently $M_1$-$M_2$) plane. We have confronted this ellipse with models of non-universal gaugino masses that make concrete predictions for $\rho_1,\,\rho_2$. In particular, we examined SO(10) models where supersymmetry is broken by hidden sector fields belonging to (possibly combinations of) $\mathbf{1}$, $\mathbf{54}$, $\mathbf{210}$ and $\mathbf{770}$ irreps. By including two such irreps, we introduce a mixing angle as an extra parameter, which also contributes to fine-tuning and must be included in $\Delta$. We found several scenarios for which $\Delta$ is small. We then examined these scenarios in SPheno in order to ensure that their phenomenology is stable to changes in theoretical treatment. While we find the low energy supersymmetric spectrum is unchanged, SPheno finds considerably higher fine-tuning caused by a shift in the ellipse. Consequently the motivation for choosing particular scenarios with multiple hidden sector irreps becomes weak and we do not study these further. Nevertheless, we do present results based on a single hidden sector field that transforms as a singlet under the Pati-Salam maximal subgroup. 

A scan dedicated to this model was performed and proved to be rather restrictive. We find this model is accessible to the $14\,{\rm TeV}$ LHC with squarks lighter than $3\,{\rm TeV}$ only if we allow a low dark matter relic density and moderate fine-tuning. The preferred relic density and $\Delta <10$ requires a heavier spectrum beyond the expected $14\,{\rm TeV}$ reach. Exact top-bottom-tau Yukawa unification is also not achieved but it is possible to get \mbox{$1.30<R_{b \tau} < 1.41$,} close to the $y_{\tau}/y_b = 3/2$ ratio. Finally, as with all scenarios that keep the GUT scale soft scalar masses small, we find that the scalar masses are dominated by the gaugino contribution, which results in the prediction that the first and second generation squarks are approximately degenerate with the gluino. We believe this model is interesting for consideration at future colliders, so present the spectra of two representative benchmark scenarios.

The models discussed here are by no means unique. There is no reason why hidden sector fields should not belong to combinations of two or more representations. Allowing this would in principle cover a much larger region on the $\rho_1$-$\rho_2$ plane, but the addition of further mixing angles would make it challenging to find scenarios with low fine-tuning. It would also be interesting to study Grand Unification models based on other gauge groups (such as E$_6$ or trinification models $[SU(3)]^3$) following a similar philosophy. We also note that none of these models fully solves some of the persistent shortcomings of conventional GUT models, such as proton stability, doublet-triplet splitting and the $\mu$-problem. Although we may invoke string inspired mechanisms as possible explanations, additional solutions are also desirable and new ideas from less traditional perspectives are required.

The Higgs mass at the LHC suggests rather heavy superpartner masses, so it is not surprising that we have not yet observed evidence of supersymmtry. Indeed, we have shown that the majority of SO(10) inspired supersymmetric spectra will only be accessible to an energy upgraded super-LHC. It will therefore be exciting to see if the scenarios discussed here can be found at the LHC or its successor colliders.

\section*{Acknowledgements}

D.J.M.~acknowledges partial support from the STFC Consolidated Grant ST/G00059X/1. A.P.M.~would like to acknowledge FCT for the doctoral grant SFRH/BD/62203/2009 under which part of this work was developed. A.P.M. also acknowledges FCT for the post-doctoral grant SFRH/BPD/97126/2013 and partial support by the grant PTDC/FIS/116625/2010, which permitted to further develop the research that is discussed in this paper. A.P.M.~and D.J.M.~would also like to thank Dr.~David Sutherland for his constant support, criticism and fruitful discussions in the realisation of the work presented in this paper.

\end{document}